\documentclass[pre,onecolumn,12pt,showpacs,superscriptaddress,notitlepage]{revtex4-1}
\pdfoutput=1
\usepackage{amsmath,amssymb,mathtools}
\usepackage[xdvi]{graphicx}
\usepackage{color}
\usepackage{bm}

\usepackage[utf8]{inputenc}
\usepackage[T1]{fontenc}
\usepackage{ae,aecompl}

\usepackage{color}
\definecolor{darkblue}{rgb}{0,0,0.6}
\definecolor{darkred}{rgb}{0.6,0,0}
\definecolor{darkgreen}{rgb}{0,0.6,0}

\usepackage[colorlinks=true,urlcolor=darkblue,citecolor=darkblue,linkcolor=darkred,hyperfootnotes=false]{hyperref}

\begin{document}

\title{Dynamical phase transitions in the current distribution\\ of driven diffusive channels}

\author{Yongjoo Baek}
\email{y.baek@damtp.cam.ac.uk}
\affiliation{Department of Physics, Technion, Haifa 32000, Israel}
\affiliation{DAMTP, Centre for Mathematical Sciences, University of Cambridge, Cambridge CB3 0WA, United Kingdom}

\author{Yariv Kafri}
\affiliation{Department of Physics, Technion, Haifa 32000, Israel}

\author{Vivien Lecomte}
\affiliation{LIPhy, Universit\'{e} Grenoble Alpes and CNRS, F-38042 Grenoble, France}

\date{\today}

\begin{abstract}
We study singularities in the large deviation function of the time-averaged current of diffusive systems connected to two reservoirs.
A set of conditions for the occurrence of phase transitions, both first and second order, are obtained by deriving Landau theories.
First-order transitions occur in the absence of a particle-hole symmetry, while second-order occur in its presence and are associated with a symmetry breaking.
The analysis is done in two distinct statistical ensembles, shedding light on previous results.
In addition, we also provide an exact solution of a model exhibiting a second-order symmetry-breaking transition.
\end{abstract}

\maketitle

\tableofcontents

\section{Introduction}

In recent years there has been an ongoing effort to understand full distribution functions of time-averaged currents in a host of scenarios, including both quantum~\cite{levitov1993pis,levitov_electron_1996,pilgram_stochastic_2003,jordan_fluctuation_2004,Esposito2009,flindt_trajectory_2013,genway_trajectory_2014} and classical contexts~\cite{Derrida2007,derrida_universal_1999,bodineau_distribution_2005,Bodineau:2007iq,prolhac_current_2008,appert-rolland_universal_2008,prolhac_cumulants_2009,baek_large_n_2016,hurtado_test_2009,prados_large_2011,de_gier_large_2011,lazarescu_exact_2011,hurtado_spontaneous_2011,derrida_microscopic_2011,gorissen_current_2012,gorissen_exact_2012,krapivsky_fluctuations_2012,meerson_extreme_2013,meerson_extreme_2014,znidaric_exact_2014,Hurtado:2014bn,Lazarescu2015,Shpielberg2016,Zarfaty:2016dv,hirschberg_zrp_2015,Shpielberg2017}. Since the time-averaged current is a history-dependent observable, its distribution depends on dynamical aspects and not only on the density of states. This makes the problem nontrivial even for equilibrium systems which fall into the Boltzmann--Gibbs framework. Nevertheless, a lot of information has been obtained about long-time properties of the distribution, which are encoded in the current {\em large deviation function} (LDF)~\cite{touchette_large_2009}. For various low-dimensional current-bearing systems, the LDFs have been derived using both microscopic models~\cite{derrida_exact_1998,prolhac_cumulants_2009,lazarescu_exact_2011,mallick_exact_2011,lazarescu_matrix_2013,lazarescu_matrix_2013,Lazarescu2015,ayyer_full_2015} and a hydrodynamic approach~\cite{spohn_long_1983,Spohn_Book}. In the latter case, the system is described by a small number of transport coefficients, and the LDF is obtained using the {\em macroscopic fluctuation theory} (MFT)~\cite{bertini_macroscopic_2002,jordan_fluctuation_2004,bertini_macroscopic_2015}. This approach has shed light on many interesting properties of driven diffusive systems~\cite{derrida_current_2004,bodineau_current_2004,bertini_current_2005,bertini_non_2006,Harris2005,imparato_equilibriumlike_2009,Lecomte2010,Shpielberg2016,bodineau_vortices_2008,Akkermans2013}.

One of the most intriguing discoveries of these studies is that the LDF can be singular even when the underlying hydrodynamic equations have smooth coefficients. In the context of current LDFs, these singularities are referred to as {\em dynamical phase transitions} (DPTs)~\cite{*[{For singularities appearing in different contexts, see }] [{ and the references therein.}] baek_singularities_2015}. When a DPT occurs, there is a singular change in the way the system sustains a given value of the current. This leads to an enhanced probability of observing values of the current beyond the transition point. Various kinds of DPTs have been reported to date~\cite{derrida_lebowitz_speer_2002,hirschberg_zrp_2015,bertini_current_2005,bodineau_distribution_2005,Bodineau:2007iq,lecomte_thermodynamic_2007,garrahan_dynamical_2007,bunin_non-differentiable_2012,baek_singularities_2015,tizon-escamilla_order_2016,shpielberg_geometrical_2017}. In particular, early studies identify DPTs in driven diffusive systems with periodic boundaries~\cite{bertini_current_2005,bodineau_distribution_2005,Bodineau:2007iq,appert-rolland_universal_2008,tizon-escamilla_order_2016}. In these systems, small values of the current are sustained by {\em time-independent} configurations. In contrast, when the transition point is crossed, they are realized by configurations which are {\em periodic in time}. This second-order DPT, which involves a breaking of time-translation symmetry, is said to be originating from a violation of the {\em additivity principle}~\cite{bodineau_current_2004}. Numerical verifications of the phenomena can be found in~\cite{Hurtado2011,Espigares:2013dx,Hurtado:2014bn}.

Until very recently, much less has been known about DPTs in systems coupled to two reservoirs. A criterion for the occurrence of DPTs, originating from the breaking of the additivity principle, was given in~\cite{Shpielberg2016}. More recently~\cite{baek_dynamical_2016}, we found that such systems can have DPTs which are not associated with the breaking of the time-translation symmetry. Based on Landau theories, we showed that when the transition occurs the presence of particle--hole symmetry leads to second-order DPTs, while in the absence of a particle--hole symmetry the transition is first-order. We also identified microscopic models (\emph{e.g.}~Katz--Lebowitz--Spohn model~\cite{katz_nonequilibrium_1983}) and suggested experimental systems (\emph{e.g.}~a graphene channel~\cite{Adam2007,*Tan2007,*Chen2008}) which realize these DPTs. In this paper we discuss the results of~\cite{baek_dynamical_2016} in detail and extended them. Specifically, we present an in-depth analysis of the correspondence between the different path ensembles used in the calculations and discuss the precise nature of the {\em phase coexistence} at first-order DPTs. Moreover, we present a simple model which can be exactly solved for arbitrary values of the control parameters. This result goes well beyond the perturbative treatment presented previously. 

This paper is organized as follows. In Sec.~\ref{sec:driven_diffusive}, we introduce a fluctuating hydrodynamics description of driven diffusive systems and define a pair of conjugate path ensembles, namely the {\em $J$-ensemble} and the {\em $\lambda$-ensemble}. In Sec.~\ref{sec:lambda_ens}, we describe DPTs in the $\lambda$-ensemble. In Sec.~\ref{sec:J_ens}, we describe DPTs in the $J$-ensemble and compare them to those of the $\lambda$-ensemble. In Sec.~\ref{sec:exact_model}, we study the symmetry-breaking DPTs in an exactly solvable model, which provides a non-perturbative verification of our general results. In Sec.~\ref{sec:conclusions}, we conclude with a summary of our findings and possible extensions.

\section{Driven diffusive systems and path ensembles}
\label{sec:driven_diffusive}

\subsection{Driven diffusive systems}

\begin{figure}[b]
\includegraphics[width=0.4\columnwidth]{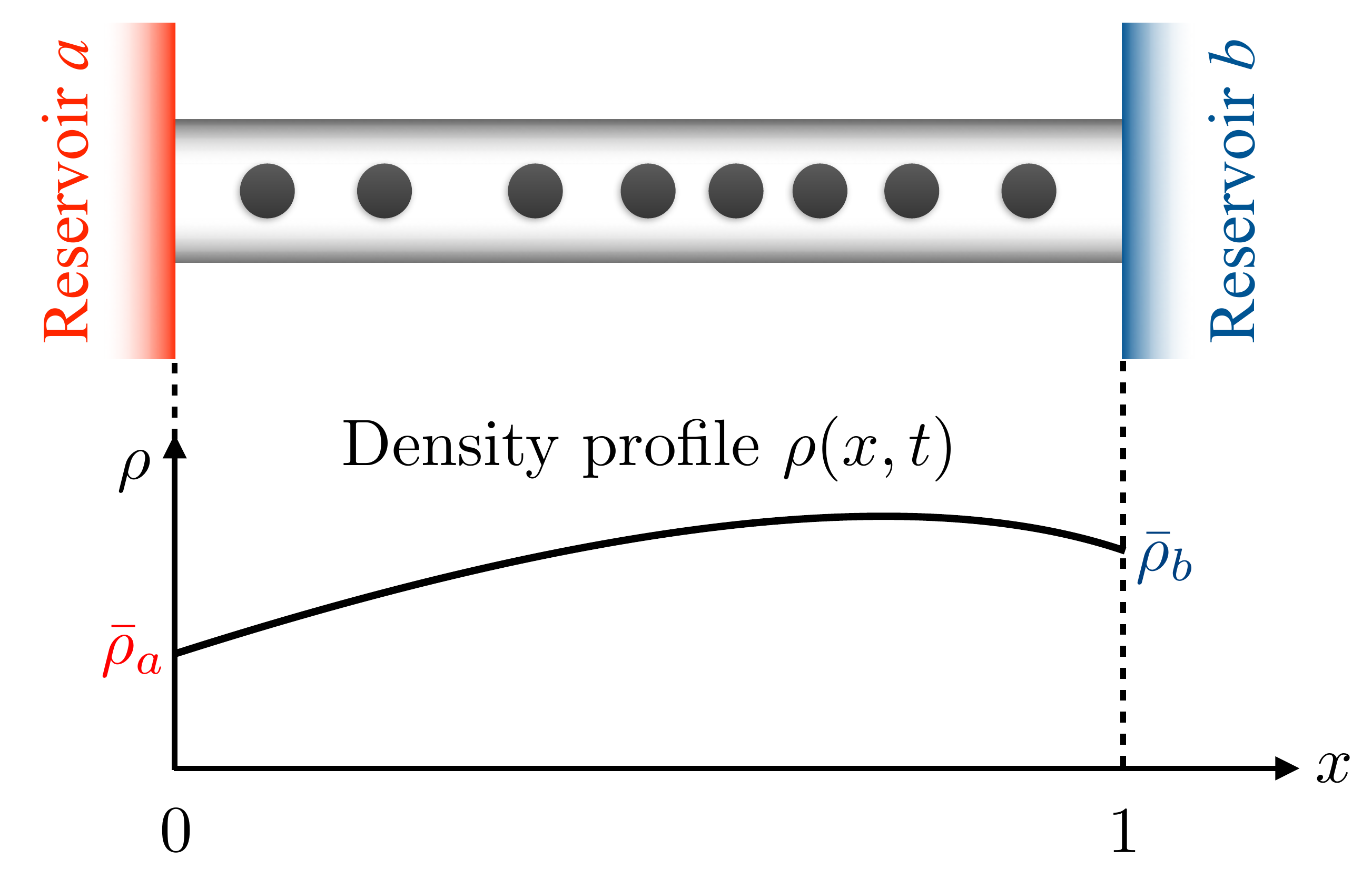}
\caption{\label{fig:fig1} A schematic diagram of a driven diffusive system with open boundaries.}
\end{figure}

We consider a one-dimensional (1D) channel of length $L \gg 1$ coupling a pair of particle reservoirs (see Fig.~\ref{fig:fig1}). The diffusive channel holds a large number of locally conserved and mutually interacting particles. Using the standard formalism of fluctuating hydrodynamics~\cite{spohn_long_1983,Spohn_Book}, after rescaling the space coordinate $x$ by $L$ (so that $0 \le x \le 1$) and the time coordinate $t$ by $L^2$, the transport is described by the continuity equation
\begin{align} \label{eq:continuity}
	\partial_t \rho(x,t) + \partial_x j(x,t) = 0\,,
\end{align}
where the density profile $\rho(x,t)$ is subject to boundary conditions
\begin{align} \label{eq:r_bcs}
\rho(0,t) = \bar\rho_a\,, \quad \rho(1,t) = \bar\rho_b\,,
\end{align}
and the current density $j(x,t)$ is given by
\begin{align} \label{eq:j}
j(x,t) = -D(\rho)\partial_{x}\rho + \sigma(\rho)E + \sqrt{\sigma(\rho)}\eta(x,t)\,.
\end{align}
The terms on the rhs of Eq.~\eqref{eq:j} represent contributions from Fick's law, the response to a bulk field $E$, and the noise, respectively. The diffusivity $D(\rho)$ and the mobility $\sigma(\rho)$, which are determined by the local particle density $\rho$, satisfy the Einstein relation
\begin{align} \label{eq:fdt}
\frac{2D(\rho)}{\sigma(\rho)} = f''(\rho)\,,
\end{align}
where $f(\rho)$ is the equilibrium free energy density. We assume that the transport coefficients $D(\rho)$ and $\sigma(\rho)$ are smooth, so that there are no phase transitions stemming trivially from the singularities of $f(\rho)$. Finally, denoting the average over all histories by $\langle\cdot\rangle$, the Gaussian noise $\eta(x,t)$ satisfies $\langle \eta(x,t) \rangle = 0$ and
\begin{align} \label{eq:eta_corr}
\langle \eta(x,t)\eta(x',t') \rangle = \frac{1}{L}\,\delta(x-x')\delta(t-t') \,.
\end{align}

\subsection{Large deviations and path ensembles}

We aim to calculate the statistics of the time-averaged current
\begin{align} \label{eq:J_def}
J \equiv \frac{1}{T} \int_0^T dt \int_0^1 dx \, j(x,t)\,,
\end{align}
where $T$ denotes the duration of observation. It is well known, and shown below, that for diffusive systems the distribution of $J$ satisfies the large deviation principle
\begin{align} \label{eq:PJ_LDP}
P(J) \sim \exp \left[-TL\Phi(J)\right]
	\quad \text{for $T \gg 1$ and $L \gg 1$}\,,
\end{align}
where $\Phi(J) = 0$ only at $J = \langle J \rangle$, and $\Phi(J) > 0$ otherwise. We discuss how the large-$T$ and the large-$L$ limits are taken in more detail later. The function $\Phi(J)$, which quantifies the rarity of nonzero deviations from the average current $J - \langle J \rangle$ for large $T$ and $L$, is called the large deviation function (LDF) of $J$.

The statistics of $J$ are also encoded in the scaled cumulant generating function (CGF), which is defined by
\begin{align} \label{eq:Psi_def}
\Psi(\lambda) \simeq \frac{1}{TL} \ln \left\langle e^{TL\lambda J} \right\rangle
	\quad \text{for $T \gg 1$ and $L \gg 1$}\,.
\end{align}
Provided that the large deviation principle~\eqref{eq:PJ_LDP} is valid, we have
\begin{align} \label{eq:cgf_J_int}
\left\langle e^{TL\lambda J} \right\rangle
	= \int dJ\, e^{TL\lambda J} P(J) \sim \int dJ\, e^{TL[\lambda J-\Phi(J)]}\,.
\end{align}
For large $T$ and $L$, saddle-point asymptotics yields
\begin{align} \label{eq:phi_to_psi}
\Psi(\lambda) = \sup_J \left[\lambda J - \Phi(J) \right]\,,
\end{align}
which implies that $\Psi(\lambda)$, being the Legendre transform of $\Phi(J)$, is a convex function. This transform amounts to changing the path ensemble from the $J$-ensemble to the one whose probabilities are biased by $e^{TL\lambda J}$, which we call the $\lambda$-ensemble. If $\Psi'(\lambda)$ is well defined, it relates $\lambda$ and $J$ by
\begin{align} \label{eq:lambda_to_Javg}
	\Psi'(\lambda) \overset{\eqref{eq:Psi_def},\,\eqref{eq:cgf_J_int}}{=\joinrel=\joinrel=} \frac{\left\langle Je^{TL\lambda J}\right\rangle}{\left\langle e^{TL\lambda J}\right\rangle} \equiv \langle J \rangle_\lambda\,,
\end{align}
where $\langle J \rangle_\lambda$ denotes the mean current for the given value of $\lambda$. Noting that $\left\langle e^{TL\lambda J}\right\rangle$ is an analog of the partition function of a canonical ensemble, we can regard $\Psi(\lambda)$ as an analog of a free energy density. Thus singularities of $\Psi(\lambda)$ represent DPTs in the $\lambda$-ensemble. 

\section{Dynamical phase transitions in the $\lambda$-ensemble}
\label{sec:lambda_ens}

\subsection{Hamiltonian formalism}

Using the standard Martin--Siggia--Rose formalism~\cite{martin_statistical_1973,dominicis_technics_1976,janssen_lagrangean_1976}, the calculation of the scaled CGF $\Psi(\lambda)$ can be reduced to solving a system of Hamiltonian field equations. For completeness, we briefly review how these equations are derived.

By Eq.~\eqref{eq:Psi_def}, $\Psi(\lambda)$ is calculated from the ensemble average $\left\langle e^{TL\lambda J} \right\rangle$. The latter can be expressed in a path-integral form
\begin{align} \label{eq:integ_rje}
\left\langle e^{TL\lambda J} \right\rangle
= \int \mathcal{D}[\rho,j,\eta]\,
	 &\Bigg\{\exp\!\left[-L\int_0^{T}dt\,\int_0^1dx\, \left(\frac{\eta^2}{2} - \lambda j\right)\right]\nonumber\\ 
&\quad\times\delta\Big[\partial_t{\rho}+\partial_x j\Big] \,
	\delta\Big[j + D(\rho)\partial_x \rho - \sigma(\rho) E - \sqrt{\sigma(\rho)}\eta \Big]\Bigg\}\,,
\end{align}
where the two delta functionals account for Eqs.~\eqref{eq:continuity} and \eqref{eq:j}, respectively. The first delta functional can be replaced with the Fourier transform
\begin{align} \label{eq:fourier_rh}
\delta\Big[\partial_t\rho + \partial_x j\Big]
= \int \mathcal{D}\hat\rho \; \exp\!\left[-L \int_0^T dt \, \int_0^1 dx \, \hat\rho\,(\partial_t\rho + \partial_x j)\right]\,,
\end{align}
where the field $\hat\rho = \hat\rho(x,t)$ is integrated along the whole imaginary axis. Since $\rho$ is fixed at the boundaries, $\hat\rho$ satisfies the boundary conditions (see~\cite{Tailleur2008} for a more detailed discussion)
\begin{align} \label{eq:rh_bcs}
\hat\rho(0,t) = 0\,,\qquad \hat\rho(1,t) = 0\,.
\end{align}
After using Eq.~\eqref{eq:fourier_rh} in Eq.~\eqref{eq:integ_rje}, one can evaluate the integral over $j$ and $\eta$ to obtain
\begin{align} \label{eq:integ_rrh}
\left\langle e^{TL\lambda J} \right\rangle
= \int \mathcal{D}[\rho,\hat\rho]\,
	\exp\!\Bigg\{-L \int_0^T dt\, 
	\int_0^1 dx\,
	&\bigg[\hat\rho\,\partial_t\rho + D(\rho)(\partial_x\rho)(\lambda + \partial_x\hat\rho)\nonumber\\
	&\quad-\frac{\sigma(\rho)}{2}(\lambda + \partial_x\hat\rho)(\lambda + \partial_x\hat\rho + 2E) \bigg]\Bigg\}\,.
\end{align}
For convenience, we introduce a change of variables
\begin{align} \label{eq:rh2rhl}
\hat\rho(x,t) \to \hat\rho_\lambda(x,t) - \lambda x\,,
\end{align}
which replaces the boundary conditions in Eq.~\eqref{eq:rh_bcs} with
\begin{align} \label{eq:rhl_bcs}
\hat\rho_\lambda(0,t) = 0\,,\qquad \hat\rho_\lambda(1,t) = \lambda\,.
\end{align}
Then Eq.~\eqref{eq:integ_rrh} changes to
\begin{align} \label{eq:integ_rrhl}
\left\langle e^{TL\lambda J} \right\rangle
= \int \mathcal{D}[\rho,\hat\rho_\lambda]\,
	\exp\!\left\{-LS_T[\rho,\hat\rho_\lambda] \right\}\,,
\end{align}
where the action functional $S_T[\rho,\hat\rho_\lambda]$ is defined as
\begin{align} \label{eq:action}
S_T[\rho,\hat\rho_\lambda] \equiv -\lambda\int_0^1 dx\, x\left[\rho(x,T)-\rho(x,0)\right] + \int_0^T dt\, 
	\int_0^1 dx\,
	\left[\hat\rho_\lambda\partial_t\rho - H(\rho,\hat\rho_\lambda) \right]
\end{align}
with
\begin{align} \label{eq:hamiltonian}
H(\rho,\hat\rho_\lambda) \equiv -D(\rho)(\partial_x\rho) (\partial_x\hat\rho_\lambda)
	+\frac{\sigma(\rho)}{2}(\partial_x\hat\rho_\lambda)(\partial_x\hat\rho_\lambda + 2E)\,.
\end{align}
When $L$ is large, the path integral in Eq.~\eqref{eq:integ_rrhl} can be evaluated by saddle-point asymptotics. Thus the calculation of $\Psi(\lambda)$ is simplified to a minimization problem
\begin{align} \label{eq:Psi_least_action}
\Psi(\lambda) = -\lim_{T \to \infty} \frac{1}{T} \inf_{\rho,\,\hat\rho_\lambda} S_T[\rho,\hat\rho_\lambda]
= -\lim_{T \to \infty} \frac{1}{T} \inf_{\rho,\,\hat\rho_\lambda} \int_0^T dt\, 
	\int_0^1 dx\,
	\left[\hat\rho_\lambda\partial_t\rho - H(\rho,\hat\rho_\lambda) \right]\,,
\end{align}
where the minimum is found among the histories of $\rho$ and $\hat\rho_\lambda$ in the complex plane satisfying the boundary conditions given by Eqs.~\eqref{eq:r_bcs} and \eqref{eq:rhl_bcs}. The second identity of Eq.~\eqref{eq:Psi_least_action} holds because the first term of $S_T[\rho,\hat\rho_\lambda]$, shown in Eq.~\eqref{eq:action}, becomes negligible for $T\to\infty$. One easily observes that Eq.~\eqref{eq:Psi_least_action} has the form of a least action principle, with $H(\rho,\hat\rho_\lambda)$ corresponding to a Hamiltonian density which is a function of a ``position'' field $\rho$ and a ``momentum'' field $\hat\rho_\lambda$. The {\em optimal histories}, which minimize the action and determine $\Psi(\lambda)$ by Eq.~\eqref{eq:Psi_least_action}, therefore satisfy the equations
\begin{align} \label{eq:rrhl_hist}
\partial_t\rho &= \frac{\delta}{\delta\hat\rho_\lambda}\int_0^1dx\,H(\rho,\hat\rho_\lambda) = \partial_x\left[D(\rho)\partial_x\rho - \sigma(\rho)(\partial_x\hat\rho_\lambda +E)\right]\,, \nonumber\\
\partial_t\hat\rho_\lambda &= -\frac{\delta}{\delta\rho}\int_0^1dx\,H(\rho,\hat\rho_\lambda) = - D(\rho) \partial_x^2 \hat\rho_\lambda - \frac{1}{2}\sigma'(\rho)(\partial_x\hat\rho_\lambda)(\partial_x\hat\rho_\lambda + 2E)\,,
\end{align}
which have real-valued solutions. Note that, by comparing the first equation with Eqs.~\eqref{eq:continuity} and \eqref{eq:j}, the real-valued $\sqrt{\sigma(\rho)}\partial_x\hat\rho_\lambda$ of such solutions can be interpreted as an optimal realization of the noise (up to a sign).

\subsection{Particle--hole symmetry}

In general, finding optimal histories from the nonlinear Eq.~\eqref{eq:rrhl_hist} is a difficult task. To make progress we consider systems with a particle--hole symmetry. A system is defined to be particle--hole symmetric when its dynamics is invariant under the transformation
\begin{align} \label{eq:par2hole}
	x \to 1-x\,, \quad \rho-\bar\rho \to \bar\rho-\rho\,,
	\quad \hat\rho \to \lambda - \hat\rho\,, \quad E \to -E\,.
\end{align}
Namely, if we define $\rho = \bar\rho$ as a baseline distinguishing `particles' and `holes', the dynamics is described by the same set of equations after an exchange of particles flowing to the right (left) and holes flowing to the left (right).

By looking at the fluctuating hydrodynamics given by Eqs.~\eqref{eq:continuity}, \eqref{eq:r_bcs}, and \eqref{eq:j}, it is clear that this symmetry holds only if the transport coefficients are even about $\bar\rho$, that is,
\begin{align} \label{eq:D_s_even}
	D(\rho) = D(2\bar\rho - \rho)\,, \qquad \sigma(\rho) = \sigma(2\bar\rho - \rho)
\end{align}
for any $\rho$. When $D(\rho)$ and $\sigma(\rho)$ are smooth functions of $\rho$, this evenness condition implies that all odd-order derivatives of the transport coefficients vanish at $\rho = \bar\rho$. In other words, introducing the notations
\begin{align} \label{eq:bar_notations}
	\bar X' \equiv X'(\bar\rho)\,, \quad \bar X'' \equiv X''(\bar\rho)\,,
	\quad \bar X^{(n)} \equiv X^{(n)}(\bar\rho)\,,
\end{align}
Eq.~\eqref{eq:D_s_even} implies
\begin{align} \label{eq:D_s_odd_zero}
	\bar D^{(2n+1)} = \bar \sigma^{(2n+1)} = 0
\end{align}
for any nonnegative integer $n$.

In most of the analysis that follows, we focus on the case when the boundary conditions satisfy
\begin{align} \label{eq:equal_ra_rb}
\bar\rho_a = \bar\rho_b = \bar\rho\,,
\end{align}
so that, if Eq.~\eqref{eq:D_s_odd_zero} holds, Eq.~\eqref{eq:rrhl_hist} has a time-independent linear solution
\begin{align} \label{eq:lin_profs}
	\rho(x,t) = \bar\rho\,, \qquad
	\hat\rho_\lambda(x,t) = \lambda x\,.
\end{align}
If $\lambda = 0$, the time-independent history~\eqref{eq:lin_profs} is clearly optimal since it corresponds to the mean behavior, which is the most probable. By continuity, if Eqs.~\eqref{eq:D_s_odd_zero} and \eqref{eq:equal_ra_rb} are satisfied, Eq.~\eqref{eq:lin_profs} gives the optimal history (or rather the {\em optimal profile} due to its time independence) for $\lambda$ sufficiently close to zero. The simplicity of this solution allows us to make much progress in the analysis of DPTs.
%

\subsection{Symmetry-breaking transitions at equilibrium}
\label{sec:eq_sym}

In what follows we analyze DPTs in equilibrium systems with
\begin{align} \label{eq:eq_conditions}
	\bar\rho_a = \bar\rho_b = \bar\rho\,, \qquad E = 0\,.
\end{align}
We first show that, when $\lambda$ reaches a critical value $\lambda_c$, the linear solution~\eqref{eq:lin_profs} becomes unstable. Moreover, we prove that for $\lambda^2 > \lambda_c^2$ there are two new time-independent solutions of Eq.~\eqref{eq:rrhl_hist} which minimize the action. Using these results, we then develop a Landau theory which describes the DPT, namely a second-order singularity of $\Psi(\lambda)$ and the associated critical behavior.

\subsubsection{Derivation of the transition point}
\label{sec:der-tr-point}

To show that the linear solution~\eqref{eq:lin_profs} becomes unstable at some value of $\lambda$, we look at the Gaussian space-time fluctuations of the action functional around the solution. Using Eq.~\eqref{eq:action}, the fluctuations are given by 
\begin{align}\label{eq:delta_action}
\Delta S_T[\varphi,\hat\varphi;\lambda]
&\equiv S_T[\bar\rho+\varphi,\lambda x+i\hat\varphi] - S_T[\bar\rho,\lambda x]\nonumber\\
&\simeq \int_0^T dt\,\int_0^1 dx\, \left[i\hat\varphi\,\partial_t\varphi + i\bar D(\partial_x\varphi)(\partial_x\hat\varphi) + \frac{\bar\sigma}{2}(\partial_x\hat\varphi)^2-\frac{\bar\sigma''\lambda^2}{4}\varphi^2\right] \,,
\end{align}
where the real-valued fields $\varphi = \varphi(x,t)$ and $\hat\varphi = \hat\varphi(x,t)$ satisfy the boundary conditions
\begin{align}
\varphi(0,t) = \varphi(1,t) = \hat\varphi(0,t) = \hat\varphi(1,t) = 0\,.
\end{align}
Note that in Eq.~\eqref{eq:delta_action} the momentum field fluctuations are written as $i\hat\varphi$ since $\hat\rho_\lambda$ is integrated along the imaginary direction in Eq.~\eqref{eq:fourier_rh}. Using the Fourier transforms
\begin{align} \label{eq:phi_fourier}
\varphi(x,t) = 2\sum_{n=1}^\infty \int_{-\infty}^\infty \frac{d\omega}{2\pi} \,
	\varphi_{n,\omega} \, e^{i \omega t} \sin(n\pi x) \,, \quad
\hat\varphi(x,t) = 2\sum_{n=1}^\infty \int_{-\infty}^\infty \frac{d\omega}{2\pi} \,
	\hat\varphi_{n,\omega} \, e^{i \omega t} \sin(n\pi x) \,,
\end{align}
Eq.~\eqref{eq:delta_action} becomes
\begin{align} \label{eq:delta_action_fourier}
\Delta S_T[\varphi,\hat\varphi;\lambda]
&= 2\sum_{n=1}^\infty \int_{-\infty}^\infty \frac{d\omega}{2\pi}\,
\begin{bmatrix}
\hat\varphi_{n,\omega} & \varphi_{n,\omega}
\end{bmatrix}
\mathbb{B}_{n,\omega,\lambda}
\begin{bmatrix}
\hat\varphi_{n,-\omega}\\
\varphi_{n,-\omega}
\end{bmatrix},
\end{align}
where $\mathbb{B}_{n,\omega,\lambda}$ is a two-by-two matrix given by
\begin{align}
\mathbb{B}_{n,\omega,\lambda} \equiv \begin{bmatrix}
\frac{n^2\pi^2\bar\sigma}{2} & \frac{in^2\pi^2\bar D+\omega}{2}\\
\frac{in^2\pi^2\bar D-\omega}{2} & -\frac{\bar\sigma''\lambda^2}{4}
\end{bmatrix}.
\end{align}
The linear profiles are unstable when $\Delta S_T < 0$ for some $\varphi_{n,\pm\omega}$ and $\hat\varphi_{n,\pm\omega}$, which is in turn possible when $\mathbb{B}_{n,\omega,\lambda}$ is not positive semidefinite. The eigenvalues of $\mathbb{B}_{n,\omega,\lambda}$, denoted by $b^{\pm}_{n,\omega,\lambda}$, are obtained as
\begin{align}
b^{\pm}_{n,\omega,\lambda} = \frac{2n^2\pi^2\bar\sigma-\bar\sigma''\lambda^2}{8} \pm \left[\left(\frac{2n^2\pi^2\bar\sigma-\bar\sigma''\lambda^2}{8}\right)^2+\frac{n^2\pi^2\bar\sigma\bar\sigma''}{8}\left(\lambda^2-\lambda_{n,\omega}^2\right)\right]^{1/2}\,,
\end{align}
which are both positive for $\lambda^2$ smaller than
\begin{align} \label{eq:lambda_sym_break}
\lambda_{n,\omega}^2 \equiv \frac{2 \left(n^4 \pi^4 \bar D^2 + \omega^2\right)}{n^2 \pi^2 \bar\sigma \bar\sigma''}\,.
\end{align}
When $\bar\sigma'' > 0$ and $\lambda^2 > \lambda_{n,\omega}^2$, a negative eigenvalue $b^-_{n,\omega,\lambda} < 0$ appears, which implies that $\mathbb{B}_{n,\omega,\lambda}$ is no longer positive semidefinite. These results imply that $\lambda_c^2$ is given by the smallest $\lambda_{n,\omega}^2$
\begin{align} \label{eq:lambda_c_def}
	\lambda_c^2 = \lambda_{1,0}^2 = \frac{2\pi^2\bar D^2}{\bar\sigma\bar\sigma''}\,.
\end{align}
Therefore a DPT occurs due to a time-independent ($\omega = 0$) mode with the longest wavelength ($n = 1$, corresponding to a wavelength of twice the system size), which breaks the particle--hole symmetry. Note that this scenario is different from that found for DPTs in periodic systems, where the unstable mode has a nonzero frequency $\omega$ and breaks the time-translation symmetry~\cite{bodineau_distribution_2005,Bodineau:2007iq,appert-rolland_universal_2008,bertini_current_2005,tizon-escamilla_order_2016}.
In the latter case, the additivity principle, which assumes the optimal profile to be time-independent, underestimates $\Psi(\lambda)$ beyond the transition. In contrast, in the former case the additivity principle correctly predicts $\Psi(\lambda)$ both below and above the transition.

\subsubsection{Derivation of the Landau theory}

With the above result, we now develop a Landau theory to describe the transition induced by the unstable mode. Specifically, we show that for $\lambda$ close to $\lambda_c$ the scaled CGF can be expressed as
\begin{align} \label{eq:Psi_landau}
\Psi(\lambda) = \int_0^1 \mathrm{d}x\, H(\bar\rho,\lambda x)
	- \inf_m {\cal L}_\lambda(m) = \frac{\bar\sigma\lambda^2}{2} - \inf_m {\cal L}_\lambda(m)\,,
\end{align}
where ${\cal L}_\lambda$ is a Landau theory of the form
\begin{align} \label{eq:landau_formal}
	\mathcal{L}_\lambda(m) = -a_2 \epsilon_\lambda m^2 + a_4 m^4 + O\!\left(m^6\right) \quad \text{with $a_2 > 0$ and $a_4 > 0$}\,,
\end{align}
whose minimization determines the value of the order parameter $m = m_\lambda$ as a function of the rescaled distance from the transition point $\epsilon_\lambda \equiv (\lambda-\lambda_c)/\lambda_c$. 

From our previous discussion, we know that a symmetry-breaking DPT occurs due to a zero-frequency mode. Thus Eq.~\eqref{eq:Psi_least_action} can be replaced with a simpler, time-independent version
\begin{align} \label{eq:Psi_least_action_time_indept}
\Psi(\lambda) = \sup_{\rho,\,\hat\rho_\lambda} \int_0^1 \mathrm{d}x\, H(\rho,\hat\rho_\lambda)\,,
\end{align}
where the extremum is found among time-independent solutions of Hamiltonian field equations~\eqref{eq:rrhl_hist} with the boundary conditions~\eqref{eq:r_bcs}, \eqref{eq:rhl_bcs}, and \eqref{eq:equal_ra_rb}.

We also know that the DPT is induced at the leading order by a sinusoidal mode of the longest possible wavelength, namely $\varphi(x)\sim\sin(\pi x)$. Thus the amplitude of $\sin(\pi x)$ can be naturally interpreted as an order parameter $m$. With this in mind, the deviations from the linear profiles can be expanded as
\begin{align}
\varphi^m(x) &\equiv m \sin (\pi x) + \sum_{l=2}^\infty m^l \varphi_l(x)\,, \label{eq:phi_m}\\
\hat\varphi^m(x) &\equiv \sum_{l=1}^\infty m^l \hat\varphi_l(x)\,,
\end{align}
with the boundary conditions for each $l$ given by
\begin{align} \label{eq:phi_m_bcs}
\varphi_l(0) = \varphi_l(1) = \hat\varphi_l(0) = \hat\varphi_l(1) = 0\,,
\end{align}
where the higher-order components $\varphi_2,\,\varphi_3,\,\ldots$ are chosen to be orthogonal to $\sin(\pi x)$, so that $m$ is \emph{exactly} the amplitude of $\sin(\pi x)$. Based on this expansion, we define
\begin{align} \label{eq:landau_lambda_def}
{\cal L}_\lambda(m) \equiv \int_0^1 \mathrm{d}x\, \left[
	H(\bar\rho,\lambda x)
	- H(\bar\rho + \varphi^m,\lambda x + \hat\varphi^m) \right]\,.
\end{align}
The relation~\eqref{eq:Psi_landau} between $\Psi(\lambda)$ and $\mathcal{L}_\lambda(m)$ is obtained from this definition and Eq.~\eqref{eq:Psi_least_action_time_indept}.

To proceed further, we need to obtain the functions $\varphi^m$ and $\hat\varphi^m$ order by order by solving the Hamiltonian field equations \eqref{eq:rrhl_hist} with $\rho = \bar\rho+\varphi^m$, $\hat\rho = \lambda x + \hat\varphi^m$, and $E = 0$ for the time-independent state. This can be done when at the leading order $\epsilon_\lambda$ satisfies
\begin{align} \label{eq:el_scaling}
\epsilon_\lambda \simeq c_\lambda \, m^2\,,
\end{align}
so that $\lambda$ is sufficiently close to $\lambda_c$. At order $m$, we have
\begin{align} \label{eq:ode_10}
	\partial_x^2 \hat\varphi_1  = -\frac{\pi^2\bar D}{\bar\sigma} \sin(\pi x)\,,
\end{align}
which is solved by
\begin{align} \label{eq:sol_10}
	\hat\varphi_1(x) = \frac{\bar D}{\bar\sigma}\sin(\pi x)\,.
\end{align}
At order $m^2$, we find
\begin{align} \label{eq:ode_20}
\partial_x^2 \varphi_2 = -\pi^2 \varphi_2\,, \qquad
\partial_x^2 \hat\varphi_2 = \frac{\bar D}{\bar\sigma} \partial_x^2 \varphi_2 - \frac{\pi\bar\sigma''\lambda_c}{2\bar\sigma}\sin(2\pi x)\,.
\end{align}
Choosing $\varphi_2$ to be orthogonal to $\sin(\pi x)$, the solution is
\begin{align} \label{eq:sol_20}
\varphi_2(x) = 0\,, \quad
\hat\varphi_2(x) = \frac{\bar\sigma''\lambda_c}{8\pi \bar\sigma}\sin(2\pi x)\,.
\end{align}
At order $m^3$, one has
\begin{align} \label{eq:ode_30_1}
\partial_x^2 \varphi_3 = -\pi^2 \varphi_3 + \left(\frac{\bar D''}{2\bar D} - \frac{\bar\sigma^{(4)}}{8\bar\sigma''} - 2c_\lambda\right) \pi^2 \sin (\pi x)
	-\left(\frac{\bar D''}{2\bar D}-\frac{\bar\sigma^{(4)}}{24\bar\sigma''}\right) \pi^2 \sin (3 \pi x)
\end{align}
and
\begin{align} \label{eq:ode_30_2}
\partial_x^2 \hat\varphi_3 &= \frac{\bar D}{\bar\sigma} \partial_x^2 \varphi_3
	+\frac{\pi ^2 \left(\bar D \bar\sigma''-\bar\sigma \bar D''\right)}{8 \bar\sigma^2} \left[\sin (\pi  x) - 3\sin (3\pi x)\right]\,.
\end{align}
The differential equation \eqref{eq:ode_30_1} has a solution with $\varphi_3(0) = \varphi_3(1) = 0$ if and only if
\begin{align} \label{eq:ode_30_const}
c_\lambda = \frac{1}{\bar\sigma''}\left(\frac{\bar D''}{4\bar D} - \frac{\bar\sigma^{(4)}}{16\bar\sigma''}\right)\,,
\end{align}
which implies
\begin{align} \label{eq:sol_30}
\varphi_3(x) = \left(\frac{\bar D''}{16\bar D}-\frac{\bar\sigma^{(4)}}{192\bar\sigma''}\right)\sin (3 \pi  x)
\end{align}
and
\begin{align}
\hat\varphi_3(x) = - \frac{\bar D \bar\sigma''-\bar D''\bar\sigma}{8\bar\sigma^2}\sin (\pi  x)
	+\left[\frac{\bar D}{\bar\sigma}\left(\frac{\bar D''}{16\bar D}-\frac{\bar\sigma^{(4)}}{192\bar\sigma''}\right) + \frac{\bar D \bar\sigma''-\bar D'' \bar\sigma}{24\bar\sigma^2}\right]\sin (3 \pi  x) \,.
\end{align}

Using these results in Eq.~\eqref{eq:landau_lambda_def}, we finally obtain
\begin{align} \label{eq:landau_eq_sym}
{\cal L}_\lambda(m) &= 
	-\frac{\pi^2 \bar D^2}{2\bar\sigma} \,\epsilon_\lambda \, m^2
	+\frac{\pi^2\bar D\left(4\bar D'' \bar\sigma'' - \bar D \bar\sigma^{(4)}\right)}{64\bar\sigma\bar\sigma''}\,m^4 + O(m^6) \,,
\end{align}
which confirms Eq.~\eqref{eq:landau_formal}. It is notable that $\varphi_l$ and $\hat\varphi_l$ with $l \ge 3$ do not contribute to $\mathcal{L}_\lambda(m)$ at this order.

\begin{figure}
\includegraphics[width=0.5\columnwidth]{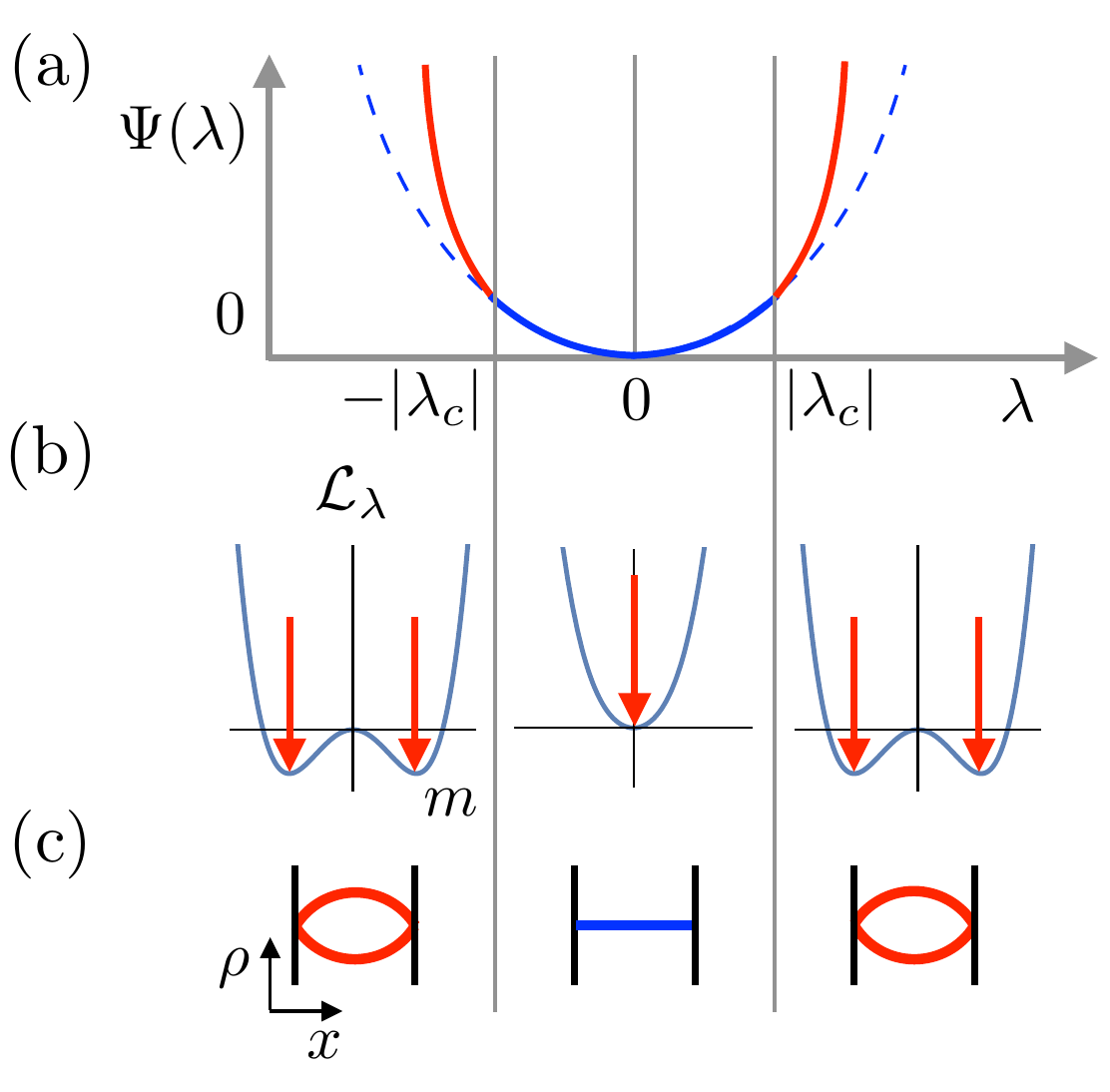}
\caption{\label{fig:fig2} (a) A schematic illustration of the scaled CGF $\Psi(\lambda)$ exhibiting second-order DPTs associated with particle--hole symmetry breaking. The branch dominated by the symmetric (symmetry-breaking) profile(s) is marked with solid blue (red) lines. The dashed blue lines indicate the action of the symmetric profile when it is no longer optimal. (b) The Landau theory $\mathcal{L}_\lambda(m)$ in each regime of $\lambda$. (c) The optimal density profile(s) in each regime of $\lambda$.}
\end{figure}

\subsubsection{The nature of the transition}

We now discuss the implications of the Landau theory on the singular behaviors at $\lambda = \lambda_c$. If the transport coefficients satisfy 
\begin{align} \label{eq:sym_break_criterion}
	\bar\sigma'' > 0\,,\qquad 4\bar D''\bar\sigma'' - \bar D\bar\sigma^{(4)} > 0\,,
\end{align}
the minimization of $\mathcal{L}_\lambda$ implies that for $\epsilon_\lambda > 0$ there are two oppositely signed optimal values of $m$, which take the form
\begin{align}             
	m_\lambda^\pm \simeq \pm\left(\frac{16\bar D \bar\sigma'' \epsilon_\lambda}{4\bar D''\bar\sigma''-\bar D\bar\sigma^{(4)}}\right)^{1/2}\,.
\end{align}
Using $m = m_\lambda^\pm$ in Eqs.~\eqref{eq:Psi_landau} and \eqref{eq:landau_eq_sym}, we obtain
\begin{align} \label{eq:Psi_eq_sym}
\Psi(\lambda) = \begin{cases}
 \frac{\bar\sigma \lambda^2}{2} &\text{ if $\epsilon_\lambda < 0$\,,}\\
 \frac{\bar\sigma \lambda^2}{2} + \frac{4\pi^2\bar D^3\bar\sigma''}{\bar\sigma\left(4\bar D''\bar\sigma'' - \bar D\bar\sigma^{(4)}\right)}\epsilon_\lambda^2 &\text{ if $\epsilon_\lambda \ge 0$\,.}	
 \end{cases}
\end{align}
Therefore the second-order derivative $\Psi''(\lambda)$ has a jump discontinuity at $\epsilon_\lambda = 0$, which is given by
\begin{align} \label{eq:Psi_2nd_jump}
\Delta\!\left(\Psi''\right) \equiv \lim_{\epsilon_\lambda \downarrow 0}\Psi'' - \lim_{\epsilon_\lambda \uparrow 0}\Psi''
=
\frac{4\bar D\bar\sigma''^2}{4\bar D''\bar\sigma''-\bar D\bar\sigma^{(4)}}\,.
\end{align}
See Fig.~\ref{fig:fig2} for a schematic illustration of $\Psi(\lambda)$ showing such singularities. Note that these singular structures imply  $\Delta\!\left(\Psi''\right) \sim \epsilon_\lambda^\alpha$ and $m_\lambda \sim \epsilon_\lambda^\beta$ with Ising mean-field exponents $\alpha = 0$ and $\beta = 1/2$.

Given these singular behaviors, one may ask whether $\lambda = \lambda_c$ is indeed ``critical'' in the sense that there exists a diverging scale. Using Eq.~\eqref{eq:delta_action_fourier}, we obtain the marginal distribution of the unstable mode
\begin{align} \label{eq:prob_ph1}
	P_\lambda[\varphi_{1,\pm\omega}] &= \int \mathcal{D}[\varphi_{2,\pm\omega},\varphi_{3,\pm\omega},\cdots;\hat\varphi]\, e^{-L\Delta S_T[\varphi,\hat\varphi;\lambda]} \nonumber\\
	&\sim \exp\left[-L\int\frac{d\omega}{2\pi}\,\frac{2\omega^2 - \pi^2\bar\sigma\bar\sigma''(\lambda^2-\lambda_c^2)}{2\pi^2\bar\sigma}\varphi_{1,\omega}\varphi_{1,-\omega}\right]
	\quad \text{for $\lambda^2 < \lambda_c^2$}\,,
\end{align}
where saddle-point asymptotics has been used to calculate the integral. Thus the density-density correlations satisfy
\begin{align}
\left\langle\varphi_{1,\omega}\varphi_{1,\omega'}\right\rangle = \frac{2\pi^3\bar\sigma}{L[2\omega^2 - \pi^2\bar\sigma\bar\sigma''(\lambda^2-\lambda_c^2)]}\,\delta(\omega+\omega')\,,
\end{align}
whose inverse Fourier transform gives
\begin{align}
\left\langle\varphi_1(t)\varphi_1(t')\right\rangle = \frac{\bar\pi^2\bar\sigma\tau_\lambda}{4L}e^{-|t-t'|/\tau_\lambda}\,,
\end{align}
with the correlation time
\begin{align}
\label{eq:correltimetaulamba}
\tau_\lambda \equiv \frac{1}{\bar D\pi^2}\left(1-\frac{\lambda^2}{\lambda_c^2}\right)^{-1/2} \quad \text{for $\lambda^2 < \lambda_c^2$}\,.
\end{align}
This time scale diverges to infinity as $\tau_\lambda \sim |\epsilon_\lambda|^{-\nu}$ for $\lambda \to \lambda_c$,
with a mean-field correlation exponent $\nu = 1/2$.

In spite of the low dimensionality of the system, the critical behaviors at the DPT are well described by a mean-field theory because the weak-noise limit imposed by Eq.~\eqref{eq:eta_corr} keeps the effects of fluctuations negligible. In order to see this more clearly, we consider the contribution of the unstable mode $\varphi_1$ to the jump discontinuity of $\Psi''(\lambda)$ at $\lambda = \lambda_c$. From Eq.~\eqref{eq:prob_ph1}, the leading correction to $\Psi(\lambda)$ from $\varphi_1$ is obtained as
\begin{align}
\delta\Psi(\lambda) \equiv -\frac{1}{TL}\int d\omega\, \ln \left[\omega^2 + \tau_\lambda^{-2}\right] \quad \text{for $\lambda^2 < \lambda_c^2$}\,.
\end{align}
This modifies $\Psi''(\lambda)$ by
\begin{align}
\delta\Psi''(\lambda) = \frac{\pi^2\bar\sigma\bar\sigma''}{TL}\int d\omega\, \frac{\omega^2+2\pi^2\bar D^2}{\left(\omega^2+\tau_\lambda^{-2}\right)^2} \sim \frac{\tau_\lambda^3}{L} \sim \frac{\epsilon_\lambda^{-3/2}}{L}\,,
\end{align}
where $T^{-1}$ is canceled by the IR cut-off of the frequency range, and $\tau_\lambda^3$ is extracted from the low-frequency behavior of the integrand. Although the magnitude of the correction becomes larger as $\epsilon_\lambda$ approaches zero, the large $L$ keeps it much smaller than the jump discontinuity shown in Eq.~\eqref{eq:Psi_2nd_jump}. 
Indeed the thermodynamic limit ensures $L\gg |\lambda-\lambda_c|^{-3/2}$. For finite system size, one thus expects a rounding of the transition in the region $|\lambda-\lambda_c| \lesssim L^{-2/3}$. Similarly, for finite time, Eq.~\eqref{eq:correltimetaulamba} implies that a rounding should occur for  $|\lambda-\lambda_c|\sim T^{-2}$.
In sum, for any finite $\epsilon_\lambda$ in the infinite-time and infinite-size limits, the mean-field exponent $\alpha = 0$ correctly describes the second-order singularity of $\Psi(\lambda)$.

\subsubsection{$T\to\infty$ then $L\to\infty$~~vs.~~$L\to\infty$ then $T\to\infty$}
\label{sec:TLLTinflims}
The MFT predicts the existence of two solutions for $\epsilon_\lambda > 0$. However, one might imagine that an instanton connecting one solution to another could allow the system to switch between the two solutions. The description of the corresponding time-dependent trajectories falls beyond the scope of the MFT; however, we now give simple arguments to analyze such a possibility.
As we show, the order of the limits $T\to\infty$ and $L\to\infty$, and how they are taken, are both important.

To analyze whether a transition between profiles is possible, one needs to calculate the cost in action of the instanton connecting them.
We do this using a heuristic argument which also applies to the cost of a domain wall in a Ginzburg-Landau theory of, say, an Ising model.
The Landau theory developed above implies that the action per unit time scales in the symmetry-broken phase as $m_\lambda^4$.
In Eq.~(\ref{eq:correltimetaulamba}) we showed that the time correlation $\tau_\lambda$ of fluctuations decays as $\tau_\lambda\sim |\epsilon_\lambda|^{-1/2}$.
Therefore the instanton connecting the two solutions is expected to extend over a duration scaling as $|\epsilon_\lambda|^{-1/2}$. 
Since $m_\lambda$ scales as $|\epsilon_\lambda|^{1/2}$, the cost of the instanton then scales as
\begin{align} \label{eq:SDW_4}
\Delta S_\lambda^\text{DW} \sim |m_\lambda|^4 \tau_\lambda \sim \epsilon_\lambda^{3/2}\,.
\end{align}
Thus the typical time between the occurrences of instantons behaves as $\tau_\text{dom} \sim e^{cL|\epsilon_\lambda|^{3/2}}$ with $c > 0$.
This implies that an optimal history develops domain walls if $T$ is much greater than the typical duration  $\tau_\text{dom}$ between instantons.
Otherwise, only one of the two optimal profiles is observed with equal probability during the entire optimal history. These considerations show that the order of limits $T\to\infty$ and $L\to\infty$ plays an important role in determining the optimal history.

\subsection{Effects of weak particle--hole asymmetry at equilibrium}
\label{ssec:lambda_eq_asym}

We turn to the case when the odd-order derivatives of $D(\rho)$ and $\sigma(\rho)$ are nonzero at $\rho = \bar\rho$. Again $\bar\rho_a=\bar\rho_b=\bar\rho$, and $\bar\rho$ is near a point where $\sigma'(\rho)=0$. Although the system is then no longer particle--hole symmetric, the linear solution~\eqref{eq:lin_profs} is still the optimal profile for $\lambda = 0$. In what follows, we treat the odd-order derivatives as perturbative parameters to explore how the optimal profiles depend on $\lambda$. We show below that a weak asymmetry between particles and holes either destroys the DPTs altogether or induces first-order DPTs.

\subsubsection{Derivation of the Landau theory}

The Landau theory, which was derived above for the symmetry-breaking DPTs, can be generalized to systems with a weak particle--hole asymmetry such that odd-order derivatives $\bar D^{(2n+1)}$ and $\bar\sigma^{(2n+1)}$ are nonzero. 
This can be carried out in a consistent manner when,  in addition to Eq.~\eqref{eq:el_scaling}, we take the odd-order derivatives $\bar D'$, $\bar\sigma'$, and $\bar\sigma^{(3)}$ to scale as
\begin{align} \label{eq:scalings_asym}
\bar D' \simeq c_D m\,,\quad
\bar\sigma' \simeq c_1 m^3\,,\quad
\bar\sigma^{(3)} \simeq c_3 m\,.
\end{align}
Then we again solve the Hamiltonian field equations~\eqref{eq:rrhl_hist} order by order for time-independent profiles of the form $\rho = \bar\rho + \varphi^m$ and $\hat\rho = \lambda x + \hat\varphi^m$, where $\varphi^m$ and $\hat\varphi^m$ satisfy Eqs.~\eqref{eq:phi_m} and~\eqref{eq:phi_m_bcs}. Following Section~\ref{sec:der-tr-point}, we find that when the coefficient $c_\lambda$ in Eq.~\eqref{eq:el_scaling} is
\begin{align}
c_\lambda = \frac{1}{\bar\sigma''}\left(\frac{\bar D''}{4\bar D} - \frac{\bar\sigma^{(4)}}{16\bar\sigma''}\right) + \frac{2}{\pi}\left(\frac{c_D}{\bar D}-\frac{3 c_1 + c_3}{3\bar\sigma''}\right)\,,
\end{align}
a nonzero solution for $\varphi^m$ and $\hat\varphi^m$ can be obtained up to order $m^3$. Using this solution and Eq.~\eqref{eq:landau_lambda_def}, the Landau theory is obtained as
\begin{align} \label{eq:landau_eq_asym}
{\cal L}_\lambda(m) &=
	-\frac{2\pi\bar D^2}{\bar\sigma\bar\sigma''}\,\bar\sigma'\,m
	-\frac{\pi^2 \bar D^2}{2\bar\sigma} \,\epsilon_\lambda \, m^2
	-\frac{2\pi \bar D(\bar D \bar\sigma^{(3)}-3\bar D'\bar\sigma'')}{9\bar\sigma\bar\sigma''}
	\,m^3 \nonumber\\
	&\quad + \frac{\pi^2\bar D\left(4\bar D'' \bar\sigma'' - \bar D \bar\sigma^{(4)}\right)}{64\bar\sigma\bar\sigma''}\, m^4 + O(m^5)\,.
\end{align}

The Landau theory implies that
when $\bar\sigma' \neq 0$, $\mathcal{L}_\lambda$ contains a linear term in $m$, which destroys the DPT in the vicinity of $\lambda_c$. Note that other DPTs might appear for larger values of $|\lambda|$ where the perturbative approach presented here is not valid.

\begin{figure}
\includegraphics[width=0.5\columnwidth]{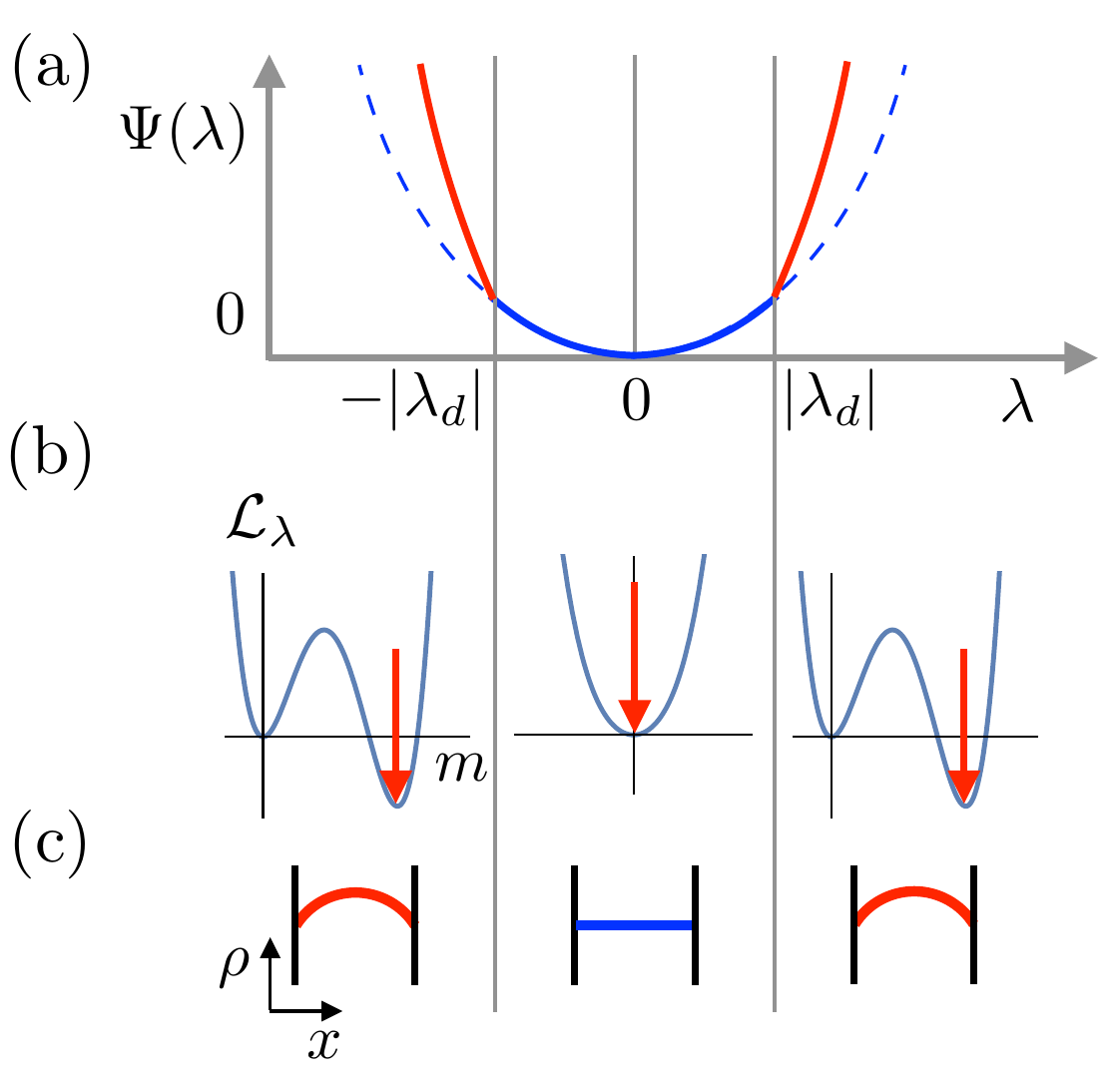}
\caption{\label{fig:fig3} (a) A schematic illustration of the scaled CGF $\Psi(\lambda)$ exhibiting first-order DPTs for $\bar D\bar\sigma^{(3)} > 3\bar D'\bar\sigma''$. The branch dominated by the flat (non-flat) profiles is marked with solid blue (red) lines. The dashed blue lines indicate the action of the flat profile when it is no longer optimal. (b) The Landau theory $\mathcal{L}_\lambda(m)$ and (c) the optimal density profiles in each regime of $\lambda$.}
\end{figure}

A more interesting behaviour is found when $\bar\sigma' = 0$, which allows the system to exhibit DPTs. To see this, note that when 
\begin{align} \label{eq:elambda_d}
\epsilon_\lambda 
=
\epsilon_\lambda^d
\equiv
\frac{\lambda_d-\lambda_c}{\lambda_c}
= -\frac{128(\bar D \bar\sigma^{(3)}-3\bar D'\bar\sigma'')^2}{81\pi^2\bar D\bar\sigma''(4\bar D''\bar\sigma'' - \bar D\bar\sigma^{(4)})}\,,
\end{align}
the Landau theory has two degenerate minima.
The location of these minima are at $m = 0$ and $m = m_d$, with
\begin{align}
m_d \simeq 
\frac{64(\bar D \bar\sigma^{(3)}-3\bar D'\bar\sigma'')}{9\pi(4\bar D''\bar\sigma'' - \bar D\bar\sigma^{(4)})}\,.
\end{align}
Therefore, at $\lambda=\lambda_d$, there is a first-order DPT where $m$ changes from a zero to a nonzero value $m_d$. Using Eq.~\eqref{eq:lambda_to_Javg}, the former corresponds to a mean current
\begin{align} \label{eq:J_avg_lambda_d_m}
	\langle J \rangle_{\lambda_d}^- \equiv \lim_{\epsilon_\lambda \uparrow \epsilon_\lambda^d} \Psi'(\lambda) = \bar\sigma \lambda_c \left[1 - \frac{128(\bar D \bar\sigma^{(3)}-3\bar D'\bar\sigma'')^2}{81\pi^2\bar D\bar\sigma''(4\bar D''\bar\sigma'' - \bar D\bar\sigma^{(4)})}\right]\,,
\end{align}
and the latter corresponds to a different mean current
\begin{align} \label{eq:J_avg_lambda_d_p}
	\langle J \rangle_{\lambda_d}^+ \equiv \lim_{\epsilon_\lambda \downarrow \epsilon_\lambda^d} \Psi'(\lambda) = \langle J \rangle_{\lambda_d}^- + \frac{1024\lambda_c\bar\sigma''\left(\bar D\bar\sigma^{(3)}-3\bar D'\bar\sigma''\right)^2}{81\pi^2\left(4\bar D''\bar\sigma'' - \bar D\bar\sigma^{(4)}\right)^2}\,.
\end{align}
This results in a jump discontinuity of $\Psi'(\lambda)$
\begin{align} \label{eq:Psi_1st_jump}
\Delta\!\left(\Psi' \right) \equiv 
\lim_{\epsilon_\lambda \downarrow \epsilon_\lambda^d}\Psi' 
- 
\lim_{\epsilon_\lambda \uparrow \epsilon_\lambda^d}\Psi' 
\simeq
\frac{1024\lambda_c\bar\sigma''\left(\bar D\bar\sigma^{(3)}-3\bar D'\bar\sigma''\right)^2}{81\pi^2\left(4\bar D''\bar\sigma'' - \bar D\bar\sigma^{(4)}\right)^2}\,,
\end{align}
which is a standard property of a first-order phase transition. An illustration of $\Psi(\lambda)$ with such first-order DPTs is shown in Fig.~\ref{fig:fig3}, assuming $\bar D\bar\sigma^{(3)} > 3\bar D'\bar\sigma''$.

Below, we analyze the cost of the instanton between the two solutions, in order to study the difference between the possible orderings of the $L\to\infty$ and the $T\to\infty$ limits.
We find that the instanton from the $m = m_d$ to the $m = 0$ solution has a \emph{negative} cost of action, in contrast to the instanton from the $m = 0$ to the $m = m_d$ solution.
This has interesting consequences that we discuss in the next subsection. 

\subsubsection{$T\to\infty$ then $L\to\infty$~~vs.~~$L\to\infty$ then $T\to\infty$}
\label{sssec:lambda_eq_asym_limit_order}

We denote by $(\rho_\lambda^\text{DW},\hat\rho_\lambda^\text{DW})$ an instanton (domain wall) connecting the solutions with $m = 0$ and $m = m_d$ at $\lambda=\lambda_d$. The additional cost of its action can be written as
\begin{align} \label{eq:SDW_1}
\Delta S_\lambda^\text{DW}
&\equiv S_T[\rho_\lambda^\text{DW},\hat\rho_\lambda^\text{DW}] - S_T[\bar\rho + \varphi_m,\lambda x + \hat\varphi_m] \nonumber\\
&= \int_0^T dt\, \int_0^1 dx\, \left[(\hat\rho_\lambda^\text{DW}-\lambda x)\,\partial_t \rho_\lambda^\text{DW} - H(\rho_\lambda^\text{DW},\hat\rho_\lambda^\text{DW}) + H(\bar\rho + \varphi_m,\lambda x + \hat\varphi_m)\right]\,.
\end{align}
where $\varphi_m$ and $\hat\varphi_m$ are given in the form of Eq.~\eqref{eq:phi_m} with $m = 0$ or $m = m_d$ depending on the initial state. For $\Delta S_\lambda^\text{DW}$ to be minimal, $(\rho_\lambda^\text{DW},\hat\rho_\lambda^\text{DW})$ should obey the Hamiltonian field equations~\eqref{eq:rrhl_hist}. Since these equations conserve $H$ along the history, we have 
\begin{align}
\int_0^1 dx\, \left[H(\rho_\lambda^\text{DW},\hat\rho_\lambda^\text{DW})-H(\bar\rho + \varphi_m,\lambda x + \hat\varphi_m)\right] = 0\,.	
\end{align}
Thus Eq.~\eqref{eq:SDW_1} can be rewritten as
\begin{align} \label{eq:SDW_2}
\Delta S_\lambda^\text{DW} = \int_0^T dt\, \int_0^1 dx\, (\hat\rho_\lambda^\text{DW}-\lambda x)\,\partial_t \rho_\lambda^\text{DW}\,.
\end{align}
To proceed, we introduce the notations
\begin{align}
\Delta\rho_\lambda^\text{DW}(x) &\equiv \rho_\lambda^\text{DW}(x,T)-\rho_\lambda^\text{DW}(x,0)\,, \nonumber\\
\Delta f_\lambda^\text{DW}(x) &\equiv f\big(\rho_\lambda^\text{DW}(x,T)\big)-f\big(\rho_\lambda^\text{DW}(x,0)\big)\,,
\end{align}
based on which we can write the identity
\begin{align}
0 = \int_0^1 dx\, \frac{\Delta f_\lambda^\text{DW}-f'(\bar\rho)\Delta\rho_\lambda^\text{DW}}{2} - \int_0^T dt\, \int_0^1 dx\, \left[\frac{f'(\rho_\lambda^\text{DW})-f'(\bar\rho)}{2}\right]\,\partial_t \rho_\lambda^\text{DW}\,.
\end{align}
Adding this side-by-side to Eq.~\eqref{eq:SDW_2}, we obtain
\begin{align} \label{eq:SDW_3}
\Delta S_\lambda^\text{DW}
&= \int_0^1 dx\, \frac{\Delta f_\lambda^\text{DW}-f'(\bar\rho)\Delta\rho_\lambda^\text{DW}}{2}\nonumber\\
&\quad + \int_0^T dt\, \int_0^1 dx\, \left[\hat\rho_\lambda^\text{DW}-\lambda x - \frac{f'(\rho_\lambda^\text{DW})-f'(\bar\rho)}{2}\right]\,\partial_t \rho_\lambda^\text{DW}\,.
\end{align}
Expanding $f(\rho)$ around $\rho = \bar\rho$, the first integral on the rhs of Eq.~\eqref{eq:SDW_3} yields
\begin{align}
\int_0^1 dx\, \frac{\Delta f_\lambda^\text{DW}-f'(\bar\rho)\Delta\rho_\lambda^\text{DW}}{2}
\simeq \frac{f''(\bar\rho)}{4}\int_0^1 dx\, \Delta(m^2) \sin^2(\pi x) = \frac{f''(\bar\rho)}{8}\Delta(m^2)\,,
\end{align}
where $\Delta(m^2)$ is the change of $m^2$ from before to after the instanton. Meanwhile, the second integral of the same equation has an integrand which, according to Eqs.~\eqref{eq:phi_m}, \eqref{eq:sol_10}, and \eqref{eq:sol_20}, satisfies
\begin{align}
	\hat\rho_\lambda^\text{DW}-\lambda x - \frac{f'(\rho_\lambda^\text{DW})-f'(\bar\rho)}{2} = m^2\, \frac{\bar\sigma''\lambda_c}{8\pi\bar\sigma}\sin(2\pi x) + O(m^3)
\end{align}
at $t = 0$ and $t = T$. Assuming that the above quantity stays of order $m^2$ and that $\rho_\lambda^\text{DW}$ stays of order $m$ along the instanton, the contribution from the second integral of Eq.~\eqref{eq:SDW_3} is of order $m^3$, which is higher-order than that of the first integral. Thus we have
\begin{align} \label{eq:SDW_4_appendix}
\Delta S_\lambda^\text{DW} \simeq \frac{f''(\bar\rho)}{8}\left[m(T)^2-m(0)^2\right]\,.
\end{align}
Since $f''(\bar\rho) > 0$, this implies that an instanton starting at $m=m_d$ and ending at $m=0$ costs action, while the opposite leads to a \emph{gain} in the action.
%

As in the symmetry-breaking case, the effect of the instantons can only be accounted for heuristically. We present two possible scenarios for the behavior in the $\lambda$-ensemble and discuss the corresponding behavior in the $J$-ensemble later. As before we assume that the action can be decomposed into contributions from the instantons and those from the saddle-point solutions. The two scenarios differ in the identification of the basic excitation.

\begin{enumerate}
\item Scenario I: The basic excitations are the instantons from $m = 0$ to $m = m_d$ and vice versa. The instanton from $m = 0$ to $m = m_d$ costs action, and occurs on a very slow time scale scaling as $e^{cLm_d^2}$ with $c>0$. In contrast, the instanton from $m = m_d$ to $m = 0$ occurs very quickly. If we identify these as the basic excitations, we have two time scales in the system, one slow and one fast. Then instantons from $m = 0$ to $m = m_d$ are observed only for $T \gg e^{cLm_d^2}$. In contrast, the instantons from $m = m_d$ to $m = 0$ can be observed even with $T \ll e^{cLm_d^2}$, when the initial state is given by $m = m_d$.

\item Scenario II: Here the basic excitation is a pair of instantons from $m = 0$ to $m = m_d$ and from $m = m_d$ to $m = 0$. The order $L m_d^2$ cost of action of the instanton from $m = 0$ to $m = m_d$ is always compensated by a subsequent instanton from $m = m_d$ to $m = 0$. This gives an action cost of the order of $L m_d^3$ for the pair of domain walls. Correspondingly, there is a single time scale $e^{cLm_d^3}$ for the occurrence of the excitation. Then for $T \ll e^{cLm_d^3}$ there are two options. Initial states with $m=0$ stay at $m=0$. In contrast, initial states with $m=m_d$ will switch to $m=0$ at a random times. For $T \gg e^{cLm_d^3}$ histories with an alternating sequence of the two types of instantons are more dominant than those with a static density profile. 
\end{enumerate}

It will be interesting to check numerically which scenario occurs.



\subsection{Generalization to nonequilibrium systems}
\label{ssec:lambda_neq}

The above results can be generalized to nonequilibrium systems with boundary and/or bulk driving. Instead of the equilibrium conditions~\eqref{eq:eq_conditions}, we now assume
\begin{align} \label{eq:neq_conditions}
	\bar\rho_a = \bar\rho - \delta\rho\,,
\quad \bar\rho_b = \bar\rho + \delta\rho\,,
\quad E \neq 0\,,
\end{align}
so that nonzero $\delta\rho$ and $E$ indicate the presence of boundary and bulk driving, respectively. Even then, since the particle--hole exchange operation~\eqref{eq:par2hole} is still applicable, the system can exhibit both symmetry-breaking and first-order DPTs through similar mechanisms. In the following we sketch how the generalization is done.

\subsubsection{Effects of bulk driving}

We first address the case when the system only has nonzero bulk driving ($E \neq 0$) and vanishing boundary driving ($\delta\rho = 0$). In the calculations described in Sec.~\ref{sec:eq_sym} and \ref{ssec:lambda_eq_asym}, this changes all occurrences of $\lambda^2$ to $\lambda(\lambda + 2E)$. Consequently one finds that Eq.~\eqref{eq:lambda_sym_break}, which describes the values of $\lambda$ at which the symmetric profile becomes unstable against a mode of wave number $n\pi$ and frequency $\omega$, is modified to
\begin{align} \label{eq:lambda_sym_break_bulk}
\lambda_{n,\omega} = -E \pm \sqrt{E^2 + \frac{2 \left(n^4 \pi^4 \bar D^2 + \omega^2\right)}{n^2 \pi^2 \bar\sigma \bar\sigma''}}\,.
\end{align}
Using this $\lambda_{n,\omega}$, it is easily seen that the symmetric DPT still occurs due to a time-independent mode with $n = 1$ and $\omega = 0$, so that the transition points are located at
\begin{align}
\lambda_c^\pm = -E \pm \sqrt{E^2 + \frac{2 \pi^2 \bar D^2}{\bar\sigma \bar\sigma''}}\,.
\end{align}
Proceeding with the calculation as before, we obtain the Landau theory
\begin{align} \label{eq:landau_neq_asym}
{\cal L}_\lambda(m) &=
	-\frac{2\pi\bar D^2}{\bar\sigma\bar\sigma''}\,\bar\sigma'\,m
	-\frac{\pi^2 \bar D^2}{2\bar\sigma} \,\epsilon_\lambda \, m^2
	-\frac{2\pi \bar D(\bar D \bar\sigma^{(3)}-3\bar D'\bar\sigma'')}{9\bar\sigma\bar\sigma''}
	\,m^3 \nonumber\\
	&\quad + \left[\frac{\pi^2\bar D\left(4\bar D'' \bar\sigma'' - \bar D \bar\sigma^{(4)}\right)}{64\bar\sigma\bar\sigma''} + \frac{\bar\sigma''^2 E^2}{64\bar\sigma}\right]\, m^4 + O(m^5)\,,
\end{align}
where the only changes are in the coefficient of $m^4$ as well as the shifted $\lambda_c$.

\begin{figure}
\includegraphics[width=0.5\columnwidth]{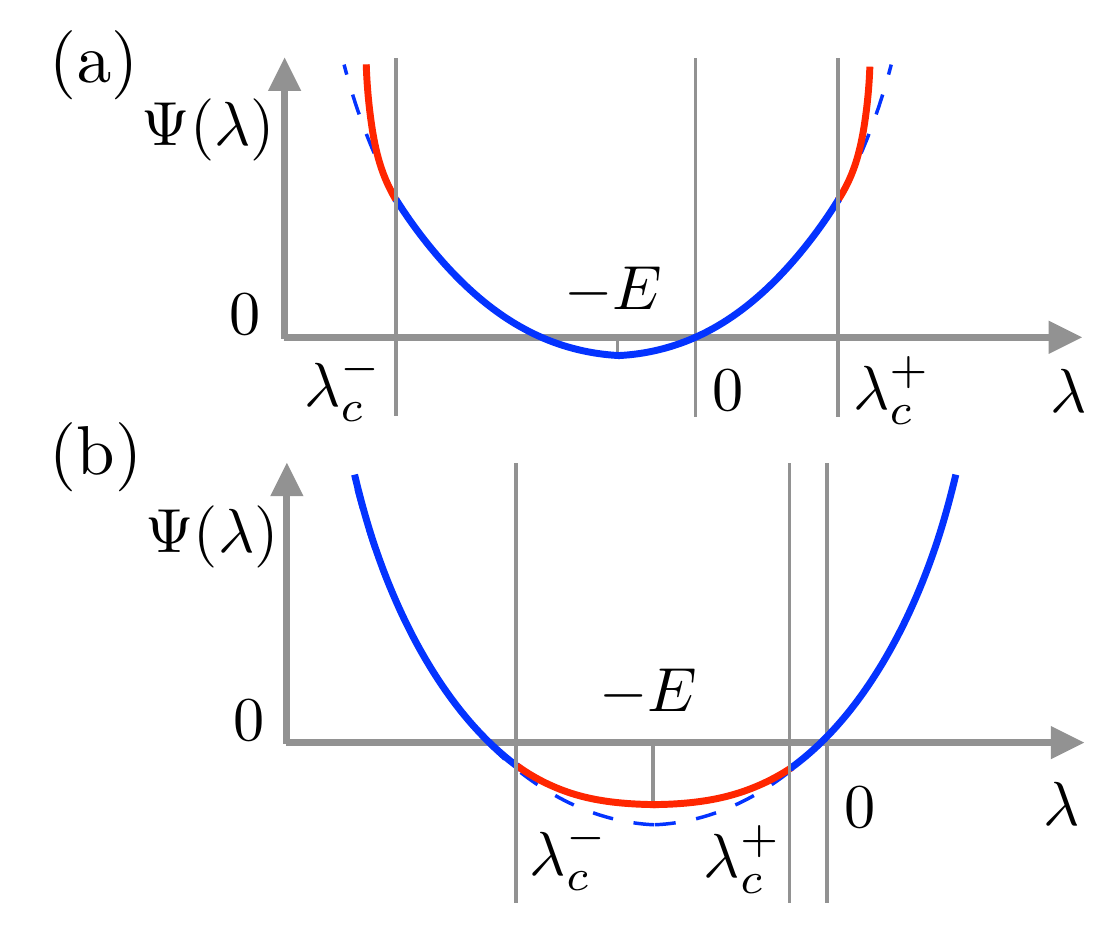}
\caption{\label{fig:fig4} (a) A schematic illustration of the scaled CGF $\Psi(\lambda)$ exhibiting symmetry-breaking DPTs when $E > 0$ and $\bar\sigma'' > 0$. The branch dominated by the symmetric (symmetry-breaking) profile(s) is marked with solid blue (red) lines. The dashed blue lines indicate the action of the symmetric profile when it is no longer dominant. (b) The behavior of $\Psi(\lambda)$ when $E > 0$ and $\bar\sigma'' < 0$.}
\end{figure}

It is notable that, for sufficiently large $E$, DPTs occur for $\bar\sigma'=0$ even when $\bar\sigma'' < 0$. In this case, both values of $\lambda_c$ have the same sign, which implies that the solution with $m \neq 0$ is optimal only for a bounded range of $\lambda$. The singular structures of $\Psi(\lambda)$ for different signs of $\bar\sigma''$ are illustrated in Fig.~\ref{fig:fig4}. We also note that the scaled CGF $\Psi(\lambda)$ satisfies the Gallavotti--Cohen symmetry
$
\Psi(\lambda) = \Psi(- 2E - \lambda)
$~\cite{Gallavotti1995a,*Gallavotti1995b,lebowitz_gallavotticohen-type_1999}.

\subsubsection{Effects of boundary driving}

We now turn to the effects of nonzero boundary driving $\delta\rho \neq 0$. We consider $\bar\rho_a=\bar\rho-\delta\rho$ and $\bar\rho_b=\bar\rho+\delta\rho$ with bulk driving $E$. The first consequence of $\delta\rho \neq 0$ is that the linear profiles shown in Eq.~\eqref{eq:lin_profs} are no longer consistent with the boundary conditions. Treating $\delta\rho$ perturbatively, the symmetric saddle-point profiles are
\begin{align} \label{eq:rho_sym_ueq}
\rho(x) = \bar\rho + \delta\rho\,\rho_1(x) + O(\delta\rho^2) \,, \quad
\hat\rho(x) = \lambda x + \delta\rho\,\hat\rho_1(x) + O(\delta\rho^2) \,.
\end{align}
Using this series expansion to solve the saddle-point equations~\eqref{eq:rrhl_hist} for the steady state, we obtain
\begin{align} \label{eq:rho_sym_ueq_corr}
\rho_1(x) = \csc \frac{F(\lambda)}{2} \sin \left[\frac{F(\lambda)}{2}\left(x-\frac{1}{2}\right)\right] \,, \quad 
\hat\rho_1(x) = \frac{\bar D}{\bar \sigma} \rho_1(x) - \frac{2\bar D}{\bar\sigma} \left( x - \frac{1}{2}\right) \,,
\end{align}
where $F(\lambda)$ denotes
\begin{align} \label{eq:alpha}
  F(\lambda) \equiv \sqrt{\frac{\lambda(\lambda+2E)\bar\sigma\bar\sigma''}{2\bar D^2}} \,.
\end{align}
One can easily verify that the profiles given by Eqs.~\eqref{eq:rho_sym_ueq} and \eqref{eq:rho_sym_ueq_corr} are symmetric under the particle--hole exchange~\eqref{eq:par2hole}.

Using the modified symmetric profiles, we proceed similarly to Sec.~\ref{sec:eq_sym} and~\ref{ssec:lambda_eq_asym}, keeping track of linear corrections in $\delta\rho$. This changes Eq.~\eqref{eq:lambda_sym_break_bulk}, which shows the threshold values of $\lambda$ at which a mode $(n,\omega)$ becomes unstable, to
\begin{align}
\lambda_{n,\omega} \simeq -E \pm \sqrt{E^2 + \frac{2 \left(n^4 \pi^4 \bar D^2 + \omega^2\right)}{n^2 \pi^2 \bar\sigma \bar\sigma''}} + \frac{2 \bar D}{\bar \sigma}\delta\rho \,.
\end{align}
Since $\delta\rho$ shifts every $\lambda_{n,\omega}$ by equal an amount, the symmetry-breaking DPT still occurs due to a time-independent mode, with a transition point shifted by
\begin{align} \label{eq:lambda_c_ueq}
\lambda_c \mapsto \lambda_c + \frac{2 \bar D}{\bar \sigma}\delta\rho \,.
\end{align}
This leads to the same Landau theory given in Eq.~\eqref{eq:landau_neq_asym} with a shifted $\lambda_c$. 
Therefore, to linear order in $\delta\rho$, the physics in this case is identical to that or $\delta\rho=0$.
Finally, we note that the CGF $\Psi(\lambda)$ obeys the Gallavotti--Cohen symmetry
$
	\Psi(\lambda) = \Psi\!\left(-2E+\frac{4\bar D}{\bar\sigma}\delta\rho - \lambda\right)
$~\cite{Gallavotti1995a,*Gallavotti1995b,lebowitz_gallavotticohen-type_1999}.

\section{Transitions in the $J$-ensemble} \label{sec:J_ens}

In this section we analyze the dynamical phase transitions directly in the $J$-ensemble. While in principle, as detailed below, one can directly obtain the results from the $\lambda$-ensemble, the calculation is instructive. In particular, it shows that first-order phase transitions arise in a similar but distinct mechanism from that suggested in \cite{bertini_current_2005,bertini_non_2006}. We begin by giving a quick overview of the formalism, assuming the additivity principle ({\em i.e.} time independence of the optimal histories). After then we turn to discuss the phase transitions and the structure of optimal histories. For simplicity, we focus on equilibrium systems lacking any boundary or bulk driving ($\bar\rho_a = \bar\rho_b = \bar\rho$ and $E = 0$). A generalization to nonequilibrium systems satisfying Eq.~\eqref{eq:neq_conditions} can be done using an approach similar to the one described in Sec.~\ref{ssec:lambda_neq}.

\subsection{Additivity principle and Lagrangian formalism}

We start by noting that the distribution of $J$ can be written in a path-integral form as
\begin{align} \label{eq:PJ_path_integ_full}
P(J) &= \int \mathcal{D}\rho\,\mathcal{D}j\,
	\bigg\langle\delta\Big[\partial_t\rho+\partial_x j\Big]\,
	\delta\!\left[j+D(\rho)\partial_x\rho-\sigma(\rho)E-\sqrt{\sigma(\rho)}\eta\right]\nonumber\\
	&
   \phantom{ \int \mathcal{D}\rho\,\mathcal{D}j\,	\bigg\langle}
   \quad\times\delta\!\left(JT-\int_0^Tdt\,\int_0^1dx\,j\right)\bigg\rangle\,,
\end{align}
where the first two delta functionals impose the Langevin dynamics given by Eqs.~\eqref{eq:continuity} and~\eqref{eq:j}. The third delta function conditions the integral to paths whose time-averaged current is equal to $J$, thus implementing a $J$-ensemble.

The calculation of $\Phi(J)$ is simplified by assuming the additivity principle~\cite{bodineau_current_2004}. This states that the path integral is dominated by histories which are time-independent. Under this assumption, the path integral in Eq.~\eqref{eq:PJ_path_integ_full} is simplified to
\begin{align}
P(J) &= \int \mathcal{D}\rho\,
	\exp\!\left\{-LT\int_0^1dx\,
	\frac{[J+D(\rho)\partial_x\rho-\sigma(\rho)E]^2}{2\sigma(\rho)}\right\}\,.
\end{align}
with $J$ constant.
For large $T$ and $L$, saddle-point asymptotics leads to the LDF
\begin{align}
\Phi^\text{AP}(J) &= \inf_\rho \int_0^1 dx\, \Lambda(\rho,\partial_x\rho)\,,\label{eq:Phi_AP}\\
\Lambda(\rho,\partial_x\rho) &\equiv \frac{[J+D(\rho)\partial_x\rho-\sigma(\rho)E]^2}{2\sigma(\rho)}\,, \label{eq:lagrangian}
\end{align}
where the minimization in the first equation is carried out over all density profiles $\rho = \rho(x)$ satisfying the boundary conditions~\eqref{eq:r_bcs}. Since Eq.~\eqref{eq:Phi_AP} has the form of a least action principle whose Lagrangian is given by Eq.~\eqref{eq:lagrangian}, $\Phi^\text{AP}(J)$ is determined by an optimal profile satisfying the Euler--Lagrange equation
\begin{align}
	\frac{\partial\Lambda}{\partial\rho} - \frac{d}{dx}\frac{\partial\Lambda}{\partial\,(\partial_x\rho)}  = 0\,.
\end{align}
Multiplying both sides of this equation by $\partial_x\rho$ and integrating over $x$, we obtain the saddle-point equation
\begin{align} \label{eq:saddle_J}
\frac{J^2-D(\rho)^2(\partial_x\rho)^2+\sigma(\rho)^2E^2}{2\sigma(\rho)} = K(J)\,,
\end{align}
where $K(J)$ is independent of space and time.

The singularities of $\Phi^\text{AP}(J)$ are found by examining the singular behavior of the solution to Eq.~\eqref{eq:saddle_J} as a function of $J$. In the rest of this section, we discuss how such singularities can be used to identify the DPTs in the $J$-ensemble.

\subsection{Symmetry-breaking transitions at equilibrium}

\subsubsection{A condition for DPT}

Recall that we consider systems with $\bar\rho_a = \bar\rho_b = \bar\rho$, $E = 0$, and particle--hole symmetry. The odd-order derivatives of $D(\rho)$ and $\sigma(\rho)$ at $\rho=\bar\rho$ are zero. Clearly, for such systems the flat profile $\rho(x) = \bar\rho$ satisfies the saddle-point equation~\eqref{eq:saddle_J} with $K(J) = J^2/(2\bar\sigma)$. As we now show, near the mean current $\langle J \rangle = 0$, the flat profile is the optimal profile that minimizes the action in Eq.~\eqref{eq:Phi_AP}. However, this flat profile is unstable against small density modulations for sufficiently large $|J|$.

As stated in Eq.~\eqref{eq:Phi_AP}, the LDF is obtained by minimizing the action
\begin{align} \label{eq:action_def}
	S_J[\rho] \equiv \int_0^1 dx\, \frac{[J+D(\rho)\partial_x\rho]^2}{2\sigma(\rho)}\,.
\end{align} 
Denoting the density modulations around the flat profile by $\varphi = \varphi(x)$, we obtain
\begin{align} \label{eq:dS_int}
	\delta S_J[\varphi] = S_J[\bar\rho+\varphi]-S_J[\bar\rho]
	\simeq \int_0^1 dx\, \frac{2\bar D^2\bar\sigma (\partial_x\varphi)^2-J^2\bar\sigma''\varphi^2}{4\bar\sigma^2}\,,
\end{align}
where we have carried out an integration by parts and used $\varphi(0) = \varphi(1) = 0$. The flat density profile is unstable when $\delta S_J[\varphi] < 0$ for some $\varphi$. When $\bar\sigma'' > 0$, \emph{i.e.}~when $\sigma(\rho)$ has a local minimum at $\rho = \bar\rho$, the two terms in the integrand have opposite signs. While the positive term reflects the propensity of diffusion to flatten out the density profile, the negative term can be attributed to the fact that, for $\bar\sigma'' > 0$, a given current $J$ is easier to carry when $\sigma(\rho)$ is increased by density modulations $\varphi$ moving $\rho$ away from a local minimum of $\sigma$. The prevalence of the latter for sufficiently large $|J|$ destabilizes the flat profile.

Applying a Fourier decomposition
\begin{align}
	\varphi(x) = \sum_{n = 1}^\infty A_n \sin(n \pi x)\,,
\end{align}
we can rewrite $\delta S_J[\varphi]$ as a functional of the amplitudes $A = (A_1,A_2,\ldots)$; namely,
\begin{align}
	\delta S_J[A] \simeq \sum_{n = 1}^\infty \frac{(2 n^2\pi^2\bar D^2\bar\sigma - J^2\bar\sigma'')A_n^2}{8\bar\sigma^2}\,.
\end{align}
This shows that the flat profile becomes unstable against sinusoidal modulations of the form $\varphi \sim \sin(n\pi x)$ when $J^2 \ge 2n^2\pi^2\bar D^2\bar\sigma/\bar\sigma''$. As $|J|$ is increased from zero, the first saddle-point instability occurs due to modulations of wave number $n = 1$. This happens at critical currents $J = J_c$ given by
\begin{align}
J_c = \pm \sqrt{\frac{2\pi^2\bar D^2\bar\sigma}{\bar\sigma''}}\,.
\end{align}

\subsubsection{Derivation of Landau theory}
\label{sssec:J_eq_sym_landau}

In this section we derive a Landau theory for the transition directly in the $J$-ensemble. The above discussions imply that, when $J$ is very close to $J_c$, $\Phi^\text{AP}(J)$ is dominated by a density profile $\rho(x) = \bar\rho + \varphi^m(x)$ with small modulations given by
\begin{align} \label{eq:phi_m_series_J}
	\varphi^m(x) = m \sin(\pi x) + \sum_{l=2}^\infty m^l \varphi_l(x)\,,
\end{align}
where each $\varphi_l$ satisfies the boundary conditions
\begin{align} \label{eq:phi_l_bcs_J}
	\varphi_l(0) = \varphi_l(1) = 0\,.
\end{align}
Taking $m$ to be the order parameter, the Landau theory can be formulated using the additional cost of action due to $\varphi^m(x)$; namely,
\begin{align} \label{eq:landau_J_def}
	\mathcal{L}_J(m) \equiv S_J[\bar\rho + \varphi^m] - S_J[\bar\rho]\,.
\end{align}
From the least action principle~\eqref{eq:Phi_AP}, we get
\begin{align} \label{eq:phi_landau}
	\Phi^\text{AP}(J) = S_J[\bar\rho]+\inf_m \mathcal{L}_J(m)
	= \frac{J^2}{2\bar\sigma}+\inf_m \mathcal{L}_J(m)\,,
\end{align}
so that $\Phi^\text{AP}(J)$ is determined by the minimization of $\mathcal{L}_J(m)$.

\begin{figure}
\includegraphics[width=0.5\columnwidth]{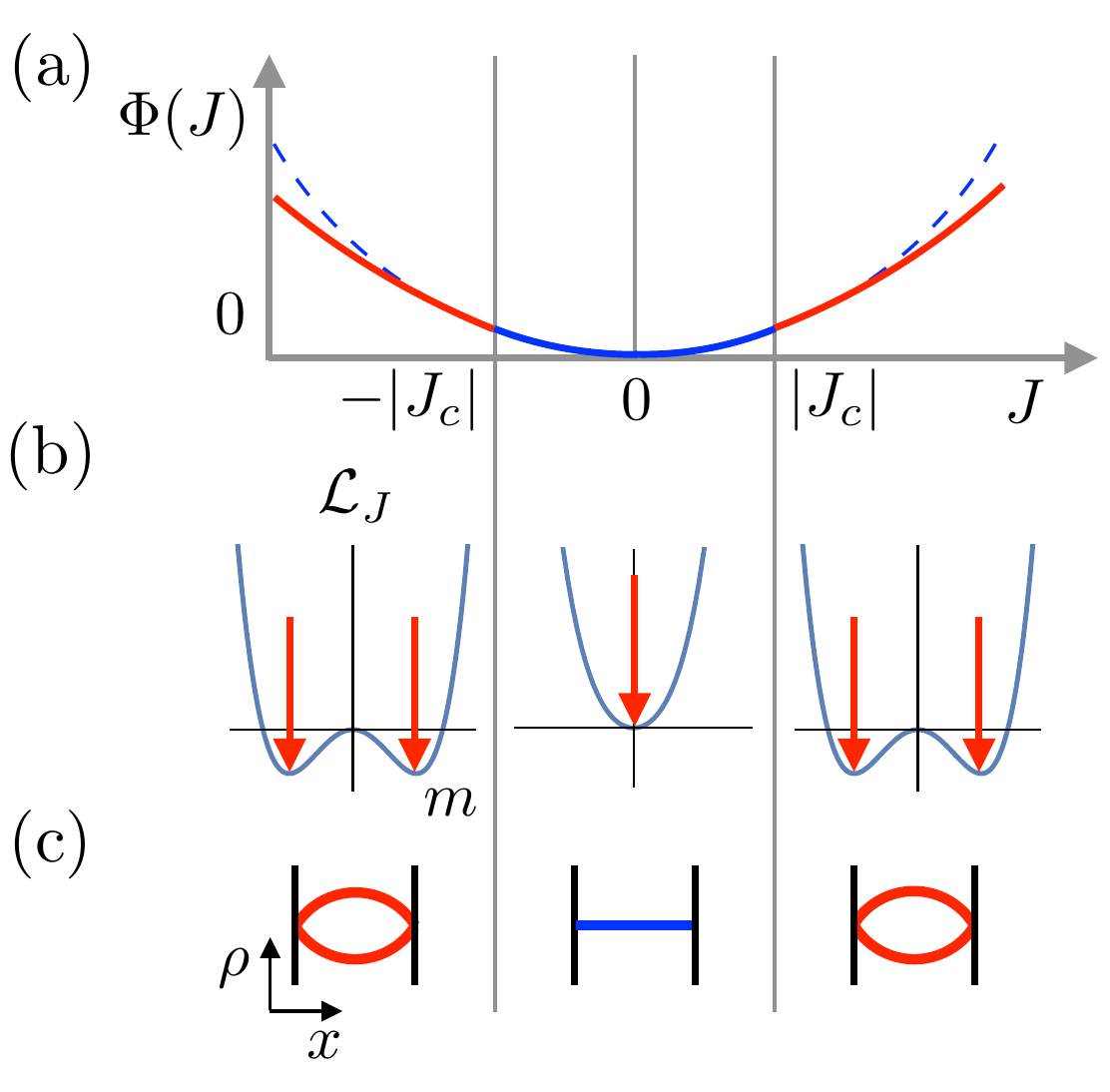}
\caption{\label{fig:fig5} (a) A schematic illustration of the LDF $\Phi(J)$ exhibiting second-order DPTs associated with particle--hole symmetry breaking. The branch dominated by the symmetric (symmetry-breaking) profile(s) is marked with solid blue (red) lines. The dashed blue lines indicate the action of the symmetric profile when it is no longer dominant. (b) The Landau theory $\mathcal{L}_J(m)$ in each regime of $\lambda$. (c) The optimal density profiles in each regime of $\lambda$.}
\end{figure}

In order to calculate $\mathcal{L}_J(m)$, we need to find a non-flat solution $\rho(x) = \bar\rho + \varphi^m(x)$ for the saddle-point equation~\eqref{eq:saddle_J}. For $E = 0$, the equation simplifies to
\begin{align} \label{eq:saddle_J_zero_E}
	J^2 - D(\rho)^2(\partial_x\rho)^2 = 2K(J)\,\sigma(\rho)\,.
\end{align}
It should be noted that $\varphi^m$ can be expanded as in Eq.~\eqref{eq:phi_m_series_J} only when $J$ is sufficiently close to $J_c$. This condition is fulfilled by limiting the range of $J$, so that $\epsilon_J \equiv (J - J_c)/J_c \simeq c_J\, m^2$. Taking $K(J) = J_c^2/(2\bar\sigma) + c_Km^2$, Eq.~\eqref{eq:saddle_J_zero_E} is automatically satisfied up to order $m$. The equation also holds at order $m^2$ if
\begin{align} \label{eq:sol_cK}
	c_K = \frac{2c_J J_c^2 - \pi^2\bar D^2}{2\bar\sigma}\,.
\end{align}
At order $m^3$, Eq.~\eqref{eq:saddle_J_zero_E} implies
\begin{align}
	\pi \sin(\pi x)\varphi_2 + \cos(\pi x)\partial_x\varphi_2 = 0\,.
\end{align}
The only solution for this equation satisfying $\varphi_2(0) = \varphi_2(1) = 0$ is
\begin{align} \label{eq:sol_20_J}
	\varphi_2(x) = 0\,.
\end{align}
Proceeding to the next order, Eq.~\eqref{eq:saddle_J_zero_E} implies
\begin{align}
\partial_x^2 \varphi_3 &= -\pi^2\varphi_3 + 2\pi^2\left(c_J - \frac{\bar D''}{4\bar D} - \frac{\bar\sigma''}{4\bar\sigma} + \frac{\bar\sigma^{(4)}}{16\bar\sigma''}\right)\sin(\pi x)\nonumber\\
&\quad + \left(\frac{\pi^2\bar D''}{2\bar D} - \frac{\pi^2\bar\sigma^{(4)}}{24\bar\sigma''}\right)\sin(3\pi x)\,.
\end{align}
This equation has a solution with $\varphi_3(0) = \varphi_3(1) = 0$ if and only if
\begin{align}
c_J = \frac{\bar D'}{4\bar D} + \frac{\bar\sigma''}{4\bar\sigma} - \frac{\bar\sigma^{(4)}}{16\bar\sigma''}\,,
\end{align}
which leads to
\begin{align} \label{eq:sol_30_J}
\varphi_3(x) = \left(\frac{\bar D''}{16\bar D} - \frac{\bar\sigma^{(4)}}{192\bar\sigma''}\right)\sin(3\pi x)\,.
\end{align}
As expected, Eqs.~\eqref{eq:sol_20_J} and \eqref{eq:sol_30_J} are in agreement with Eqs.~\eqref{eq:sol_20} and \eqref{eq:sol_30}. Finally, using Eqs.~\eqref{eq:action_def}, \eqref{eq:phi_m_series_J}, \eqref{eq:landau_J_def}, \eqref{eq:sol_20_J}, and \eqref{eq:sol_30_J}, $\mathcal{L}_J(m)$ is obtained to order $m^4$ as
\begin{align} \label{eq:landau_J}
\mathcal{L}_J(m) \simeq
	- \frac{\pi^2\bar D^2}{2\bar\sigma} \epsilon_J m^2
	+ \frac{\pi^2 \bar D}{64\bar\sigma^2\bar\sigma''}
	\left(4\bar D''\bar\sigma\bar\sigma''+4\bar D\bar\sigma''^2-\bar D\bar\sigma\bar\sigma^{(4)}\right)m^4\,.
\end{align}
It should be noted that all $\varphi_l(x)$ with $l \ge 3$ do not contribute to $\mathcal{L}_J(m)$ at this order, which was also the case for $\mathcal{L}_\lambda(m)$ in Eq.~\eqref{eq:landau_eq_sym}. If the transport coefficients satisfy
\begin{align}
	\bar\sigma'' > 0 \quad \text{and} \quad
	4\bar D''\bar\sigma\bar\sigma''+4\bar D\bar\sigma''^2-\bar D\bar\sigma\bar\sigma^{(4)} > 0\,,
\end{align}
for $\epsilon_J > 0$ the minimum of $\mathcal{L}_J(m)$ is achieved by
\begin{align} \label{eq:mJpm}
m = m_J^\pm \equiv \pm\left[\frac{16\bar D\bar\sigma\bar\sigma''}{4\bar D''\bar\sigma\bar\sigma''+4\bar D\bar\sigma''^2-\bar D\bar\sigma\bar\sigma^{(4)}}\,\epsilon_J\right]^{1/2}\,.
\end{align}
Using Eqs.~\eqref{eq:phi_landau}, \eqref{eq:landau_J}, and \eqref{eq:mJpm}, we therefore obtain 
\begin{align} \label{eq:Phi_AP_eq_sym}
\Phi^\text{AP}(J) = \begin{cases}
 	\frac{J^2}{2\bar\sigma} &\text{if $\epsilon_J < 0$\,,} \\
 	\frac{J^2}{2\bar\sigma} - \frac{4\pi^2\bar D^3\bar\sigma''}{4\bar D''\bar\sigma\bar\sigma''+4\bar D\bar\sigma''^2-\bar D\bar\sigma\bar\sigma^{(4)}}\,\epsilon_J^2 &\text{if $\epsilon_J \ge 0$\,.}
 \end{cases}
\end{align}
According to Eq.~\eqref{eq:Phi_AP_eq_sym}, the second derivative of $\Phi^\text{AP}(J)$ jumps from $\lim_{\epsilon_J \uparrow 0}\partial_J^2\Phi^\text{AP}(J) = 1/(2\bar\sigma)$ to
\begin{align}
\lim_{\epsilon_J \downarrow 0}\partial_J^2\Phi^\text{AP}(J) = \frac{4\bar D''\bar\sigma\bar\sigma''-\bar D\bar\sigma\bar\sigma^{(4)}}{\bar\sigma(4\bar D''\bar\sigma\bar\sigma''+4\bar D\bar\sigma''^2-\bar D\bar\sigma\bar\sigma^{(4)})}
\end{align}
as $\epsilon_J$ crosses zero from below. Thus, if the transport coefficients satisfy $4\bar D''\bar\sigma\bar\sigma''-\bar D\bar\sigma\bar\sigma^{(4)} > 0$, $\Phi^\text{AP}(J)$ is convex on both sides of $\epsilon_J = 0$. We recall that, as stated by Eq.~\eqref{eq:sym_break_criterion}, systems satisfying this condition as well as $\bar\sigma'' > 0$ have symmetry-breaking DPTs in the $\lambda$-ensemble, which are described by the Landau theory at order $m^4$. Indeed, applying the inverse Legendre transform
\begin{align} \label{eq:psi_to_phi}
\Phi(J) = \inf_\lambda [\lambda J - \Psi(\lambda)]
\end{align}
to $\Psi(\lambda)$ given by Eq.~\eqref{eq:Psi_eq_sym}, one obtains $\Phi^\text{AP}(J) = \Phi(J)$. This, together with the Legendre transform~\eqref{eq:phi_to_psi}, implies that the $\lambda$- and $J$-ensembles yield equivalent descriptions of the symmetry-breaking DPTs. An illustration of $\Phi(J)$ exhibiting such singular features is given in Fig.~\ref{fig:fig5}.

\subsubsection{$T\to\infty$ then $L\to\infty$~~vs.~~$L\to\infty$ then $T\to\infty$}
\label{sssec:J_eq_sym_limit_order}

Based on the equivalence of ensembles discussed above, we can apply the theory of instantons in the $\lambda$-ensemble to predict the shape of typically observed histories for $\epsilon_J > 0$. Using Eq.~\eqref{eq:SDW_4} and the scaling $m_J \sim \epsilon_J^{1/2}$ obtained in Eq.~\eqref{eq:mJpm}, the cost of each instanton can be written as $\Delta S_J^\text{DW} \sim \epsilon_J^{3/2}$. Thus the typical time between an adjacent pair of instantons scales as $\tau_\text{dom} \sim e^{cL\epsilon_J^{3/2}}$ with $c > 0$. If $T \gg \tau_\text{dom}$, the histories contain multiple excitations of instantons. In contrast, if $T \ll \tau_\text{dom}$, only one of the two optimal profiles is observed with equal probability.


\subsection{Effects of weak particle--hole asymmetry at equilibrium}

We now turn to address the effects of weak particle--hole asymmetry on the DPTs of systems at equilibrium ($\bar\rho_a = \bar\rho_b = \bar\rho$, $E = 0$) in the $J$-ensemble. Here, as in the $\lambda$-ensemble, we treat the odd-order derivatives $\bar D^{(2n+1)}$ and $\bar\sigma^{(2n+1)}$ perturbatively to obtain a Landau theory for the transition. The Landau theory produces a nonconvex cusp singularity of $\Phi^\text{AP}(J)$. Recalling that the additivity principle assumes optimal profiles with a uniform and time-independent current, this implies phase coexistence in time between two possible values of the current within an interval around the cusp singularity.

\subsubsection{Derivation of the Landau theory}
\label{sssec:J_eq_asym_landau}

As done in Eq.~\eqref{eq:scalings_asym} for the $\lambda$-ensemble, we assume that the odd-order derivatives are small, with the lowest-order ones scaling as $\bar D' \simeq c_D m$, $\bar\sigma' \simeq c_1 m^3$, and $\bar\sigma^{(3)} \simeq c_3 m$. Then we can solve the saddle-point equation~\eqref{eq:saddle_J_zero_E} order by order for density profiles of the form $\rho = \bar\rho + \varphi^m$, where $\varphi^m$ satisfies Eqs.~\eqref{eq:phi_m_series_J} and \eqref{eq:phi_l_bcs_J}. Proceeding as in Sec.~\ref{sssec:J_eq_sym_landau}, for $c_K$ given by Eq.~\eqref{eq:sol_cK} and
\begin{align}
c_J = -\frac{2c_1}{\pi\bar\sigma''} - \frac{2c_3}{3\pi\bar\sigma''} +\frac{2c_D}{\pi\bar D} + \frac{\bar D'}{4\bar D} + \frac{\bar\sigma''}{4\bar\sigma} - \frac{\bar\sigma^{(4)}}{16\bar\sigma''}\,,
\end{align}
one obtains nonzero solutions for $\varphi_l$ with $1 \le  l \le 3$. Using these results in Eq.~\eqref{eq:landau_J_def}, $\mathcal{L}_J(m)$ is obtained to order $m^4$ as
\begin{align} \label{eq:landau_J_asym}
\mathcal{L}_J(m) \simeq
	&-\frac{2\pi \bar D^2}{\bar\sigma\bar\sigma''}\bar\sigma'\,m
	- \frac{\pi^2\bar D^2}{2\bar\sigma} \epsilon_J m^2
	- \frac{2\pi \bar D}{9\bar\sigma\bar\sigma''}\left(\bar D\bar\sigma^{(3)}-3\bar D'\bar\sigma''\right)m^3 \nonumber\\
	&+ \frac{\pi^2 \bar D}{64\bar\sigma^2\bar\sigma''}
	\left(4\bar D''\bar\sigma\bar\sigma''+4\bar D\bar\sigma''^2-\bar D\bar\sigma\bar\sigma^{(4)}\right)m^4\,.
\end{align}

When $\bar\sigma' \neq 0$, the linear term of $\mathcal{L}_J(m)$ destroys the DPT in the vicinity of $J_c$ --- a result that was also seen in the $\lambda$-ensemble. 
If $\bar\sigma' = 0$, $\mathcal{L}_J(m)$ has two degenerate minima when
\begin{align} \label{eq:eJ_star}
	\epsilon_J = \epsilon_J^* \equiv \frac{J_* - J_c}{J_c}
	= -\frac{128\bar\sigma\left(\bar D\bar\sigma^{(3)}-3\bar D'\bar\sigma''\right)^2}{81\pi^2\bar D\bar\sigma''\left(4\bar D''\bar\sigma\bar\sigma''+4\bar D\bar\sigma''^2-\bar D\bar\sigma\bar\sigma^{(4)}\right)}\,,
\end{align}
with one minima located at $m = 0$ and another at
\begin{align} \label{eq:m_star}
	m = m_* \equiv \frac{64\bar\sigma\left(\bar D\bar\sigma^{(3)}-3\bar D'\bar\sigma''\right)}{9\pi\left(4\bar D''\bar\sigma\bar\sigma''+4\bar D\bar\sigma''^2-\bar D\bar\sigma\bar\sigma^{(4)}\right)}\,.
\end{align}
Using Eqs.~\eqref{eq:landau_J_asym}, \eqref{eq:eJ_star}, and \eqref{eq:m_star} in Eq.~\eqref{eq:phi_landau}, one observes that $\Phi^\text{AP}(J)$ has a jump discontinuity in its first derivative
\begin{align}
\Delta\!\left(\partial_J\Phi^\text{AP}\right) \equiv \lim_{\epsilon_J\downarrow\epsilon_J^*}\partial_J\Phi^\text{AP} - \lim_{\epsilon_J\uparrow\epsilon_J^*}\partial_J\Phi^\text{AP}
\simeq -\frac{1024J_c\bar\sigma''\left(\bar D\bar\sigma^{(3)}-3\bar D'\bar\sigma''\right)^2}{81\pi^2\left(4\bar D''\bar\sigma\bar\sigma''+4\bar D\bar\sigma''^2-\bar D\bar\sigma\bar\sigma^{(4)}\right)^2}\,.
\end{align}
As shown in Fig.~\ref{fig:fig6}, the sign of this jump discontinuity is such that $\Phi^\text{AP}(J)$ has a cusp pointing upward at $J = J_*$. The resulting shape of $\Phi^\text{AP}(J)$ is nonconvex. The implications of this nonconvexity is discussed below.

\subsubsection{Implications of the nonconvex $\Phi^\text{AP}(J)$}

\begin{figure}
\includegraphics[width=0.5\columnwidth]{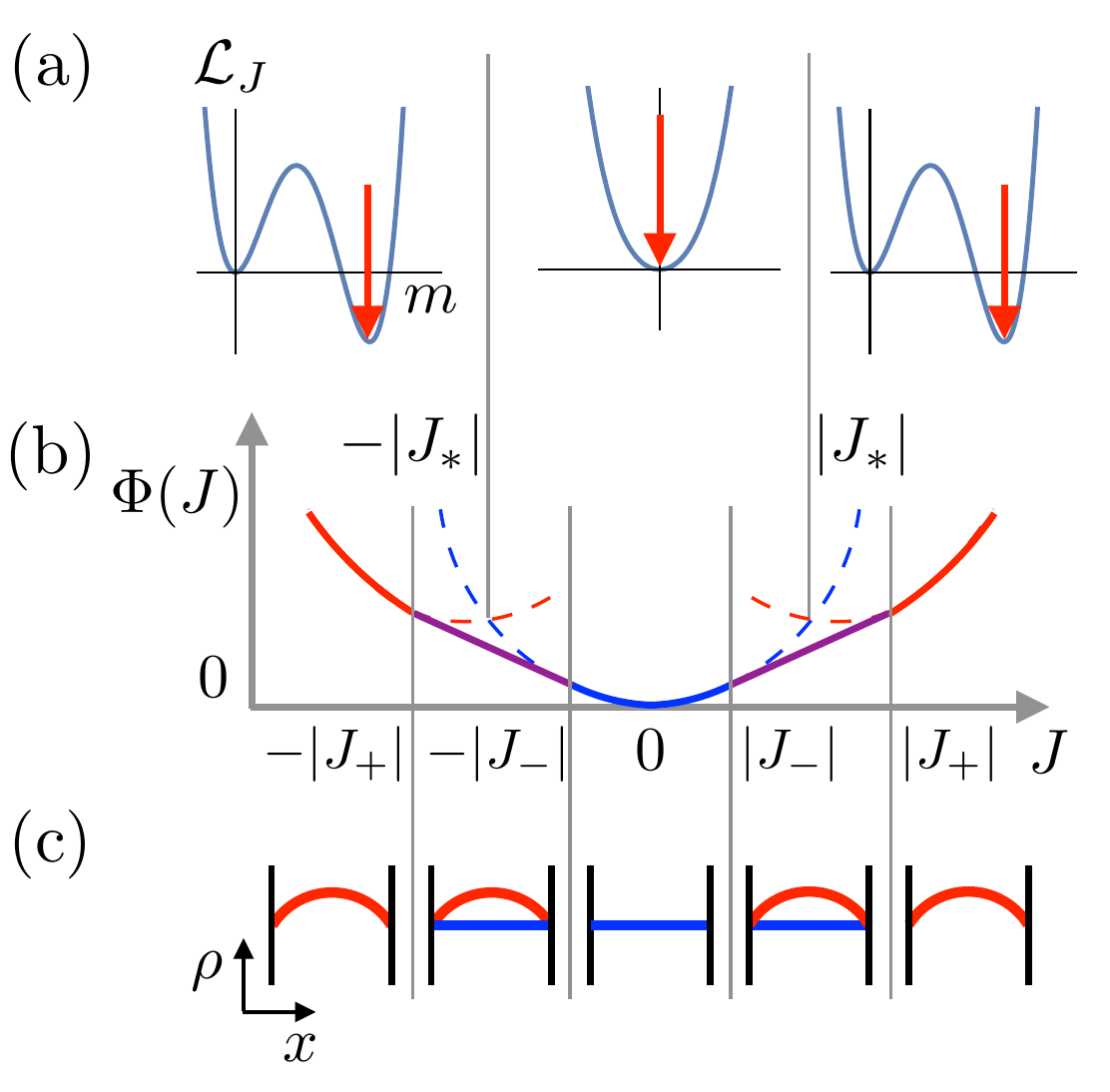}
\caption{\label{fig:fig6} (a) The Landau theory $\mathcal{L}_J(m)$, obtained by assuming the additivity principle, in different regimes of $J$ demarcated by $J = \pm |J_*|$. (b) A schematic illustration of the LDF $\Phi(J)$ exhibiting first-order DPTs. The branch dominated by the flat (non-flat) profile(s) is marked with solid blue (red) lines, and the coexistence regimes of both profiles are marked with solid purple lines. The dashed blue (red) lines indicate the action of the flat (non-flat) profile when it occupies the entire history. (c) The optimal density profiles in each regime of $\lambda$.}
\end{figure}

Following the ideas from equilibrium phase coexistence, we consider the convex envelope $\Phi^\text{env}(J)$ of the nonconvex $\Phi^\text{AP}(J)$ derived above (see Fig.~\ref{fig:fig6} for a schematic illustration). Examining the behavior of $\Phi^\text{AP}(J)$ in the vicinity of $J_c$ given by Eqs.~\eqref{eq:phi_landau} and \eqref{eq:landau_J_asym}, $\Phi^\text{env}(J)$ is obtained as
\begin{align}
\Phi^\text{env}(J) = \begin{cases}
 \Phi^\text{AP}(J_-) + \frac{\Phi^\text{AP}(J_+)-\Phi^\text{AP}(J_-)}{J_+ - J_-}(J-J_-) &\text{ if $\epsilon_J^- \le \epsilon_J \le \epsilon_J^+$,}\\
 \Phi^\text{AP}(J)	&\text{ otherwise,}
 \end{cases}
\end{align}
where the endpoints of the linear regime ($\epsilon_J^- \le \epsilon_J \le \epsilon_J^+$) are obtained by a common-tangent construction and given by 
\begin{align}
\epsilon_J^{-} &\equiv \frac{J_- - J_c}{J_c} \simeq -\frac{128(\bar D \bar\sigma^{(3)}-3\bar D'\bar\sigma'')^2}{81\pi^2\bar D\bar\sigma''(4\bar D''\bar\sigma'' - \bar D\bar\sigma^{(4)})}\,,\label{eq:eJm}\\
\epsilon_J^{+} &\equiv \frac{J_+ - J_c}{J_c} \simeq \epsilon_J^{-} + \frac{1024\bar\sigma''\left(\bar D\bar\sigma^{(3)}-3\bar D'\bar\sigma''\right)^2}{81\pi^2\bar\sigma\left(4\bar D''\bar\sigma'' - \bar D\bar\sigma^{(4)}\right)^2}\,. \label{eq:eJp}
\end{align}
Within this regime there is a coexistence between time-independent solutions corresponding to $J = J_-$ and $J = J_+$ with instantons (domain walls) connecting them.
In the $T \to \infty$ limit, the contribution of the instantons to $\Phi(J)$ is negligible. 
For a current $J=pJ_- + (1-p)J_+$ with $0\leq p\leq 1$, the system spends a total sojourn time of $pT$ in the $J = J_-$ solution and time $(1-p)T$ in the $J = J_+$ solution.
Clearly, in this region, $\Phi^\text{env}\big(pJ_- + (1-p)J_+\big) \leq \Phi^\text{AP}\big(pJ_- + (1-p)J_+\big)$. Hence,  $\Phi^\text{AP}(J)$ fails to give the correct description of $\Phi(J)$. Instead we have $\Phi(J) = \Phi^\text{env}(J)$, which describes the phase coexistence for $\epsilon_J^- \le \epsilon_J \le \epsilon_J^+$ (see Fig.~\ref{fig:fig6}).

One can check that such behavior of $\Phi(J)$ is consistent with that of $\Psi(\lambda)$ in the vicinity of a first-order DPT (see the discussion in Sec.~\ref{ssec:lambda_eq_asym}): using Eqs.~\eqref{eq:lambda_c_def}, \eqref{eq:elambda_d}, \eqref{eq:J_avg_lambda_d_m}, and \eqref{eq:J_avg_lambda_d_p}, one can show $\lambda_d = \Phi'(J_-) = \Phi'(J_+)$ and $J_\pm = \left\langle J \right\rangle^\pm_{\lambda_d}$. These equations reflect the validity of the inverse Legendre transform~\eqref{eq:psi_to_phi} and the one-to-one correspondence between $\Psi(\lambda)$ and $\Phi(J)$ in the vicinity of $J_c$. Hence, at the level of large deviations, the first-order DPTs in the $J$-ensemble are equivalent to those in the $\lambda$-ensemble.

\subsubsection{$T\to\infty$ then $L\to\infty$~~vs.~~$L\to\infty$ then $T\to\infty$}

Based on the equivalence between the $\lambda$- and $J$-ensembles discussed above, here we use the results obtained in Sec.~\ref{sssec:lambda_eq_asym_limit_order}  for the cost of instantons to understand how the order of the limits $T \to \infty$ and $L \to \infty$ affects the structure typical histories for $\epsilon_J^- \le \epsilon_J \le \epsilon_J^+$. Noting that $\lambda_d = \Phi'(J)$ in this regime, we can take $\Delta S_{\lambda_d}^\text{DW}$, given in Eq.~\eqref{eq:SDW_4_appendix}, to be the cost of each instanton. As in Sec.~\ref{sssec:lambda_eq_asym_limit_order}, there are two possible scenarios depending on the basic excitation.
\begin{enumerate}
\item Scenario I: If the basic excitation is an instanton from $J = J_-$ ($m = 0$) to $J = J_+$ ($m = m_d \sim |\epsilon_J^d|^{1/2}$) and vice versa, for $T \gg e^{cLm_d^2}$ there are multiple instantons between the two profiles, which obey the constraint that the total time spent in the $J = J_-$ ($J = J_+$) profile is given by $\Delta t_- = pT$ ($\Delta t_+ = (1-p)T$). For $T \ll e^{cLm_d^2}$, there is only a single instanton from $J = J_+$ to $J = J_-$, which occurs at $t = (1-p)T$. Note that this instanton gains action while going from $J = J_-$ to $J = J_+$ costs actions. This sets the order of appearance of $J_+$ and $J_-$.
\item Scenario II: If the basic excitation is a pair of instantons from one profile to another followed by the reverse process, multiple instantons are observed for $T \gg e^{cLm_d^3}$, with the constraint on $\Delta t_-$ and $\Delta t_+$ discussed above. If $T \ll e^{cLm_d^3}$, there is only a single instanton from $J = J_+$ to $J = J_-$ occurring at $t = (1-p)T$.
\end{enumerate}
We note here that, in contrast to the case of the $\lambda$-ensemble, the strict constraint on the value of $J$ enforces the existence of at least a single instanton.

\section{Exactly solvable model}
\label{sec:exact_model}

In this section we analyze the symmetry-breaking transition in an exactly solvable which we study both in the $J$ and $\lambda$-ensembles. The model we consider is defined through
\begin{equation}
  \label{eq:Dsigmaquad}
  D(\rho)= 1\,,\qquad 
  \sigma(\rho)=1+\rho^2
\end{equation}
and we consider the boundary conditions $\bar \rho_a = \bar \rho_b=0$.

\subsection{Symmetry breaking in the $\lambda$-ensemble}
\label{sec:symm-break-lambda}

Assuming that the additivity principle holds, the saddle-point equations~(\ref{eq:rrhl_hist}) in the $\lambda$-ensemble reduce to their time-independent forms
\begin{align} \label{eq:saddle_traj_time_indept}
\partial_x \left[\partial_x \rho - (1+\rho^2)\partial_x \hat\rho_\lambda \right] = 0 \,, \quad
\partial_x ^2 \hat\rho_\lambda + \rho\,(\partial_x \hat\rho_\lambda)^2 = 0 \,.
\end{align}
The first equation implies
\begin{equation}
  \label{eq:K1}
  \partial_x \hat\rho_\lambda=\frac{\partial_x \rho +K_1}{1+\rho^2}\,,
\end{equation}
where $K_1$ is an integration constant.
Substituting this into the second equations yields (after an integration)
\begin{equation}
  \label{eq:K2}
  \frac{K_1^2 -(\partial_x \rho)^2}{1+\rho^2}=K_2\,,
\end{equation}
where $K_2$ is another integration constant.
%
This implies by differentiation
\begin{equation}
  \label{eq:diffK2}
  -\partial_x \rho\:\partial_x ^2\rho=K_2\,\rho\,\partial_x \rho\:.
\end{equation}
Therefore, either $\partial_x \rho=0$ so that $\rho(x)=0$, or:
\begin{equation}
  \label{eq:eqrholambda}
  \partial_x ^2\rho+K_2\,\rho(x)=0\:.
\end{equation}
Note that, as seen from~(\ref{eq:K1}) and the boundary conditions~(\ref{eq:rhl_bcs}) on $\hat\rho_\lambda(x)$, $\lambda$ is related to $K_1$ through:
\begin{align}
\lambda
&=
\int_0^1\!\!\! dx \: \frac{\partial_x \rho +K_1}{1+\rho(x)^2}
= -\Big[ \arctan \rho(x)\Big]_{x=0}^{x=1} \!\!+ K_1 \int_0^1  \!\!\!dx \: \frac{1}{1+\rho(x)^2}
  \label{eq:K1lambda}
=
K_1 \int_0^1 \!\!\!dx \: \frac{1}{1+\rho(x)^2}\:.
\end{align}
Substituting~(\ref{eq:K1}) into the action, one obtains
\begin{equation}
\psi(\lambda)
= -\frac 1{T} S_\lambda[\rho,\hat\rho]
 = \frac{K_1^2 -(\partial_x \rho)^2}{2(1+\rho^2)}\stackrel{(\ref{eq:K2})}=\frac 12 K_2 \:.
\label{eq:psiK2}
\end{equation}
With the above results we can now obtain an expression for the CGF.
When $\rho(x)$ is flat, one has
\begin{equation}
  \label{eq:lambdarhoconst}
  \lambda \stackrel{\eqref{eq:K1lambda}}= K_1
\qquad
\text{and}
\quad
  \psi(\lambda)\stackrel{\eqref{eq:psiK2}}=\frac 12  K_2\stackrel{(\ref{eq:K2})}=\frac 12 K_1^2=\frac{\lambda^2}{2}
\:.
\end{equation}
On the other hand, Eq.~(\ref{eq:eqrholambda}) implies that a non-flat profile verifying the boundary conditions can exist only if $K_2>0$.
This gives
\begin{align}
  \label{eq:rho_nm_lambda}
  \rho_{n,m}(x) = m \sin (n\pi x)  \qquad\text{with}\quad K_2 = n^2 \pi^2,\ n\in\mathbb N^\star \quad\text{and}\quad m\neq 0\:,
\end{align}
from which one infers from~(\ref{eq:K2}) that
\begin{equation}
  \label{eq:K1nm}
  K_1^2=n^2\pi^2(1+m^2)\:.
\end{equation}
Substituting~(\ref{eq:rho_nm_lambda}) into~(\ref{eq:K1lambda}) gives
 $K_1^2=\lambda^2\,(1+m^2)$, which together with Eq.~(\ref{eq:K1nm}) implies that $|\lambda|=n\pi$.
From~$\psi(\lambda)=\frac 12  K_2$ one finds that the lowest action is obtained for $n=1$ and given by
\begin{equation}
|\lambda|=\pi\,.\label{eq:lambdapi}
\end{equation}
The flat profile $\rho(x)$ describes the solution for $|\lambda|\leq\lambda_c$ with $\lambda_c=\pi$, while at $|\lambda|=\lambda_c$ all profiles $\rho_{n=1,m}(x)$ given by~(\ref{eq:rho_nm_lambda}) are solutions to the saddle-point equations and give the same CGF $\psi(\pm\lambda_c)=\pi^2/2$.

Combining all the above results we find that the CGF is given by
\begin{equation}
  \label{eq:despsilambda}
  \psi(\lambda) = \frac{\lambda^2}2  \qquad\text{for}\quad |\lambda|\leq\pi
\end{equation}
with the values of $\lambda$ bounded between $-\pi$ and $\pi$ (\emph{i.e.}~the CGF is defined on a compact domain).
This saturation of the values of $\lambda$ is reminiscent of the saturation of the chemical potential in a condensation of an ideal Bose gas~\cite{huang_statistical_1987}.
Indeed, as we show below, to change to the $J$-ensemble one needs to consider finite-$L$ corrections to $\psi(\lambda)$. These can be obtained as detained in Appendix.~\ref{sec:appFScorrLDFs} and are given by
\begin{align}
  \psi_L(\lambda) 
&= \frac{\lambda^2}{2} - \frac{1}{L^2} \frac 12 \sum_{n\geq 1}\big\{n\pi \sqrt{n^2\pi^2-\lambda^2}-n^2\pi^2+\tfrac 12 \lambda^2\big\}
  \label{eq:psiLdirectsum}
\\
&= \frac{\lambda^2}{2}  + \frac{1}{L^2} \frac 18 \mathcal F\big(\tfrac 12\lambda^2\big)\,,
  \label{eq:psiLdirect}
\end{align}
with $\mathcal F(u)$ denoting the universal function~\cite{bodineau_long_2008}
\begin{equation}
  \label{eq:defF}
 \mathcal F(u) = -4 \sum_{n\geq 1}\big\{n\pi \sqrt{n^2\pi^2-2u}-n^2\pi^2+u \big\} \:.
\end{equation}

To perform the Legendre transform, one has to find $\lambda=\lambda(J)$ which solves the relation $J=\psi'_L(\lambda)$.
Taking advantage of the parity symmetry, we focus on the domain $J>0$.
Using Eq.~\eqref{eq:psiLdirectsum}, one finds  that for $\lambda\uparrow\lambda_c$
\begin{equation}
  \label{eq:bhvpsiLprimetr}
  J= \psi'_L(\lambda)=\lambda+\frac{\lambda}{2L^2}\frac{1}{\sqrt{\pi^2-\lambda^2}}\,,
\end{equation}
or:
\begin{equation}
  \label{eq:bhvpsiLprimetr2}
  \psi'_L(\lambda_c-\epsilon)=\pi+\frac{1}{L^2}\sqrt{\frac \pi 8}\frac{1}{\sqrt{\epsilon}} \:,
\end{equation}
where only the $n=1$ mode in~(\ref{eq:psiLdirectsum}) is accounted for in the $\lambda\uparrow\lambda_c$ asymptotics.
Therefore, in the large $L$ asymptotics, choosing $\lambda(J)=\pi-\epsilon$ with $\epsilon\sim L^{-4}$ solves for values of $J$ larger than $\pi$.
Specifically we solve Eq.~(\ref{eq:bhvpsiLprimetr}) to obtain
\begin{equation}
  \label{eq:lambdastarhighC}
  \lambda(J)=\pi\,\Big(1-\frac1{8L^4J^2}\Big)
  + O(L^{-5}) \:.
\end{equation}
Then, for $J>\pi$, we find the rate function for the current distribution:
\begin{align}
  \label{eq:PhiL-from-psiL-large-currents}
\Phi_L(J)
&
=
J\,\lambda(J)-\psi_L\big(\lambda(J)\big)
\\
&
\stackrel{\eqref{eq:psiLdirect}}=
J\,\pi\,\Big(1-\frac1{8L^4J^2}\Big)
  -\frac 12 \pi^2\,\Big(1-\frac1{8L^4J^2}\Big)^2
  -\frac 1{L^2} \frac 18 \mathcal F\big(\tfrac 12\pi^2\big)
  +o(L^{-2})
\\
&
\stackrel{\phantom{\eqref{eq:psiLdirect}}}=
\frac 12 \pi (2J-\pi)
-\frac 1{L^2} \frac 18 \mathcal F\big(\tfrac 12\pi^2\big)
+
o(L^{-2})
\:.
\label{eq:devPhilargeJlargeL}
\end{align}

Thus, we obtain that for $|J|>J_c$ (with $J_c=\pi$) the rate function $\Phi(J)=\lim_{L\to\infty} \Phi_L(J)$ has two affine branches --~which could not be retrieved from the infinite-$L$ CGF 
$\psi(\lambda)$.
In fact, it is rather straightforward to directly perform the calculation of $\Phi(J)$ and to obtain the same second-order phase transition in the MFT settings as we detail below.


\subsection{Symmetry breaking in the $J$-ensemble}
\label{sec:symmetry-breaking-j}

Assuming time-independent optimal profiles, the action in the $J$-ensemble is given by
\begin{align}
  \label{eq:actionJ}
  S[\rho,J]
  &=
  T \int d x\:
  \frac{\big[J+D(\rho(x))\,\partial_x \rho(x)\big]^2}{2\sigma(\rho(x))
} \:.
\end{align}
At dominant order, $\Phi(J) = \frac 1T S[\rho^\star,J]$ where $\rho^\star(x)$ is the dominant solution of the saddle-point equation:
\begin{equation}
  \label{eq:saddlepoint1}
  \frac{\rho (x) \left[J^2-(\partial_x \rho(x))^2\right]+\rho (x)^2 \,\partial_x^2 \rho(x)+\partial_x^2 \rho(x)}{\left(\rho (x)^2+1\right)^2}=0
\:.
\end{equation}
The only flat solution to this equation is given by $\rho(x)=0$. Then trivially $\Phi(J)=J^2/2$. 

We now look for possibles non-flat profiles with a lower action.
Multiplying by $\partial_x \rho(x)$, one finds that there is a conserved quantity given by
\begin{align}
  \label{eq:constK}
  k=\frac{J^2-(\partial_x \rho(x))^2}{1+\rho (x)^2}
\:.
\end{align}
Multiplying the previous equality by $1+\rho (x)^2$ and differentiating with respect to~$x$, one obtains that
\begin{align}
  \label{eq:eqmotrho}
  \partial_x^2 \rho(x)+ k \rho (x)=0
\:.
\end{align}
The shape of the solution depends on the sign of $k$.
This equation has non-constant solutions of the form
\begin{align}
  \label{eq:rho_nm}
  \rho_{n,m}(x) = m \sin (n\pi x)\,,  \qquad\text{with}\quad k = n^2 \pi^2,\ n\in\mathbb N^\star \quad\text{and}\quad m\neq 0
\:.
\end{align}
Using Eq.~\eqref{eq:constK}, we obtain
\begin{align}
  \label{eq:3}
\frac{J^2-\pi ^2 \left(m^2+1\right) n^2}{m^2 \sin ^2(\pi  n x)+1} = 0\,,
\end{align}
which gives
\begin{equation}
\label{eq:relationJnm}
J^2= n^2 \pi^2 (1+m^2)
\:.
\end{equation}
For a given value of $n$, this implies that non-flat profiles exist only for
\begin{equation}
  \label{eq:Jcn}
  |J| \geq J_c^{(n)}\equiv n\pi
\:.
\end{equation}
The corresponding optimal profiles read
\begin{equation}
  \label{eq:rhon}
  \rho_n(x)=\pm\sqrt{\frac{J^2}{n^2\pi^2}-1}\  \sin (n \pi x)
\:.
\end{equation}
Inserting this expression into~(\ref{eq:actionJ}), one finds the corresponding value of the LDF
\begin{equation}
  \label{eq:Phin}
  \Phi_n(J) = \frac 12 n\pi (2J-n\pi)\,,   \qquad\text{with}\quad |J| \geq J_c^{(n)}\equiv n\pi
\:.
\end{equation}
Comparing to the constant profile solution ($\rho(x)=0$), one checks directly that the final LDF reads
\begin{equation}
  \label{eq:finalPhi}
  \Phi(J)=
  \begin{cases}
    \frac 12 J^2          & |J| \leq \pi\,,\\
    \frac 12 \pi (2J-\pi) & |J| \geq \pi\,.\\
  \end{cases}
\end{equation}
The transition corresponds to a symmetry breaking at $|J|=J_c\equiv \pi$ from a flat profile to a pair of non-flat profiles given by~(\ref{eq:rhon}) for $n=1$.

In Appendix~\ref{sec:appFScorrLDFs}, we show that 
for $|J|< J_c$, the finite-size corrections to this result can be computed by the MFT and are given by
\begin{equation}
  \label{eq:devPhi}
  \Phi_L(J) = \frac {J^2} {2} - \frac 1{8L^2}  \mathcal F \Big(\tfrac 12 J^2\Big) + o(L^{-2})
\:.
\end{equation}
As we have seen at the end of the last Section, these are important in order to understand the changes between the $J$- and the $\lambda$-ensembles.
In the $J$-ensemble considered in the present Section, the picture of the symmetry breaking is more direct since two opposite profiles appear at $|J|>J_c$; in contrast, due to the linearity of $\Phi(|J|>J_c)$, the full symmetry-broken phases are reduced to the points $|\lambda|=\lambda_c$ in the $\lambda$-ensemble.

\section{Conclusions}
\label{sec:conclusions}

In this paper, we have studied symmetry-breaking and first-order dynamical phase transitions in one-dimensional diffusive systems connecting a pair of reservoirs. Based on the macroscopic fluctuation theory, we showed that the transitions are induced by time-independent unstable modes and can be described by Landau theories. We also showed that the order of the large-time ($T\to\infty$) and the large-system limits ($L\to\infty$) plays an important role in the structure of dominating observed histories. We proposed two possible scenarios which distinguish a regime of static histories from that of multiple instantons (domain walls in time).
These scenarios are based on arguments beyond the macroscopic fluctuation theory and thus remain to be checked numerically.
Finally, we analyzed the symmetry-breaking DPT in an exactly solvable model and studied its leading finite-size corrections, providing an explicit non-perturbative example.
In this model, the transition bears similarities with Bose--Einstein condensation.

We also note that the transitions in diffusive systems with periodic boundaries, which are induced by time-dependent modes, can be similarly described by the Landau theory derived in this paper~\cite{Bodineau:2007iq}. It still remains to be clarified whether diffusive systems with open boundaries can have the transitions driven by time-dependent modes~\cite{Shpielberg2017}.
In addition, it would be interesting to identify DPTs that may occur for currents beyond the critical one that we identified~\cite{Shpielberg2017}.

\acknowledgments{
We also thank Giovanni Jona-Lasinio for helpful suggestions. 
Y.B. and Y.K. are supported by an ISF grant, and Y.B. is supported in part at the Technion by a fellowship from the Lady Davis Foundation. 
V.L. is supported by the ANR-15-CE40-0020-03 Grant LSD and also acknowledges support by the ERC Starting Grant No. 680275 MALIG and UGA IRS PHEMIN project.
}

\appendix*

\section{Finite-size corrections to the LDFs in the $1+\rho^2$ model}
\label{sec:appFScorrLDFs}

\subsection{The $\lambda$-ensemble}

In the domain of definition $|\lambda|\leq\pi$ of $\psi(\lambda)$, we consider space-time fluctuations around the saddle-point as follows:
\begin{equation}
  \label{eq:saddlerhorhohatdevphi}
  \rho(x,t)=\rho^\star(x)+\phi(x,t) = \phi(x,t)\,,
  \qquad
  \hat\rho(x,t)=\hat\rho^\star(x)+\hat\phi(x,t)=\lambda x +\hat\phi(x,t)
\:.
\end{equation}
Expanding the action in powers of the small fields $\phi(x,t)$ and $\hat\phi(x,t)$, one finds the quadratic form
\begin{equation}
  \label{eq:S2lambda}
  \delta^2S[\phi,\hat\phi]=
  \int_0^T\mathrm{d}t\,
  \int_0^1\mathrm{d}x\,
   \left[\hat\phi  \, \partial_t{\phi} + \partial_x  \phi \, \partial_x \hat\phi
   -\frac 12 (\partial_x  \hat\phi)^2 - \frac {\lambda^2}2 \phi^2 \right] 
\:.
\end{equation}
In order to integrate the corresponding Gaussian fluctuations, we introduce the following spatial Fourier decomposition, which obeys the spatial boundary conditions $\phi(0,t) = \phi(1,t) = \hat\phi(0,t) = \hat\phi(1,t) = 0$:
\begin{align}
  \label{eq:defFphihatphi}
  \phi(x,t)& = \sum_{n\geq 1} a_n(t) \sin(n\pi x)\,,
\\
  \hat\phi(x,t)& = \sum_{n\geq 1} \hat a_n(t) \sin(n\pi x)
\,.
\end{align}
Using then the trigonometric identities
\begin{align}
  \label{eq:scalarp}
  \int_0^1 d x \sin (n\pi x) \sin (m\pi x) &=\frac 12\delta_{nm}\,,
\\
  \int_0^1 d x \cos (n\pi x) \cos (m\pi x) &=\frac 12\delta_{nm}\,,
\label{eq:scalarp2}
\end{align}
the quadratic action becomes a sum over independent modes
\begin{equation}
  \label{eq:S2lambdaa}
  \delta^2S[\phi,\hat\phi]=
\frac 12
  \int_0^T\mathrm{d}t\,
  \sum_{n\geq 1}
   \left[\hat a_n(t)  \partial_t{a_n(t) } + n^2\pi^2 a_n(t) \hat a_n(t)
   -\frac 12 n^2\pi^2\hat a_n^2 - \frac {\lambda^2}2 a_n^2 \right] 
\:.
\end{equation}
Using the rescaling $a_n(t)\mapsto\sqrt{2}a_n(t)$ and $\hat a_n(t)\mapsto\sqrt{2}\hat a_n(t)$, one obtains
\begin{equation}
  \label{eq:S2lambda2}
  \delta^2S[\phi,\hat\phi]=
  \int_0^T\mathrm{d}t\,
  \sum_{n\geq 1}
   \left[\hat a_n(t)  \partial_t{a_n(t) } + n^2\pi^2 a_n(t) \hat a_n(t)
   -\frac 12 n^2\pi^2\hat a_n^2 - \frac {\lambda^2}2 a_n^2 \right] 
\:.
\end{equation}
The large-time behavior of this quadratic action is given by the sum of the ground-state eigenvalue of the following independent harmonic oscillators (see \emph{e.g.}~\cite{baek_singularities_2015})
\begin{equation}
  \label{eq:Hoh}
  \mathbb H_n = n^2\pi^2 a^\dag a - \tfrac 12 n^2\pi^2{a^\dag}^2 - \tfrac 12 \lambda^2 a^2 
\:.
\end{equation}
Summing over the individual ground states, one finds the finite-size corrections to CGF from  $-\delta^2S$ as
\begin{align}
  \psi_L(\lambda) 
&= \frac{\lambda^2}{2} - \frac{1}{L^2} \frac 12 \sum_{n\geq 1}\big\{n\pi \sqrt{n^2\pi^2-\lambda^2}-n^2\pi^2+\tfrac 12 \lambda^2\big\}+o(L^{-2})
\\
&= \frac{\lambda^2}{2}  + \frac{1}{L^2} \frac 18 \mathcal F\big(\tfrac 12\lambda^2\big)+o(L^{-2})
  \label{eq:psiLdirect_app}
\:,
\end{align}
as announced in the main text.

\subsection{The $J$-ensemble}

The space-time fluctuations around the flat solution (\emph{i.e.}~for $|J|\leq J_c$) are
\begin{align}
  \label{eq:deffl}
  \rho(x,t)& = \sum_{n\geq 1} a_n(t) \sin(n\pi x)\equiv\delta \rho(x,t)\,,
\\
  j(x,t)& = J+  \sum_{n\geq 1} \partial_t a_n(t) \frac{\cos(n\pi x)}{n\pi} \equiv J+\delta j(x,t)
\,,
\end{align}
which satisfy the continuity equation $\partial_t\rho+\partial_xj=0$.
Expanding the action up to second order for small $\delta\rho$ and $\delta\phi$, one gets
\begin{align}
  \label{eq:actionJtot2}
  \delta^2S[\rho]
  &=
  \frac 12 \int_0^{t_f} d t\int_0^1 d x\:
   \Big\{
    \big(\delta j(x,t)+\partial_x\delta \rho(x,t)\big)^2-J^2
   \delta \rho (x,t)^2
    \Big\}  
\:.
\end{align}
Using then \eqref{eq:scalarp}-\eqref{eq:scalarp2}, one finds
\begin{align}
  \label{eq:integrals_TF}
  \int_0^1d x \ \delta \rho^2 &= \frac 12 \sum_{n\geq 1} a_n(t)^2\,,
\\
  \int_0^1d x \ \delta j^2    &= \frac 12 \sum_{n\geq 1} \frac 1{n^2\pi^2}\big(\partial_t a_n(t)\big)^2\,,
\\
  \int_0^1d x \        (\partial_x\delta\rho)^2    &= \frac 12 \sum_{n\geq 1} {n^2\pi^2} a_n(t)^2\,,
\\
  \int_0^1d x \ \delta j\partial_x\delta\rho       &=  \sum_{n\geq 1} \partial_t(a_n(t)^2)
\label{eq:intjrhoprime}
\,.
\end{align}
%
We note that terms in the form of the Eq.~\eqref{eq:intjrhoprime}, when integrated over time, only contribute temporal boundary terms to the action.
Bearing this in mind, one finally obtains
\begin{align}
  \label{eq:actionJtot2TF}
  \delta^2S[\rho]
  &=
  \frac 12 \int_0^{t_f} d t \sum_{n\geq 1}
\frac 12
   \Big\{
    \big(n^2\pi^2-J^2\big) a_n(t)^2
+
    \frac{1}{n^2\pi^2} \big(\partial_t a_n(t)\big)^2
    \Big\}  
\\
  &=
  \frac 12 \int_0^{t_f} d t \sum_{n\geq 1}
   \Big\{
    M_n \Omega_n^2 a_n(t)^2
+
    M_n \big(\partial_t a_n(t)\big)^2
    \Big\}  
\:.
\end{align}
This expression represents a collection of independent harmonic oscillators indexed by $n$ with parameters
\begin{equation}
  \label{eq:paramsOH}
  M_n \Omega^2_n = \frac 12 ( n^2\pi^2-J^2)\,,
\qquad
  M_n =\frac{1}{2n^2\pi^2}
\:.
\end{equation}
Assuming for simplicity that $a_n(0)=a_n(t_f)=0$ (which should not be important at large $t_f$) one can use the standard results on the Euclidean harmonic oscillator
\begin{align}
  \label{eq:OH}
  \int \mathcal D \mathbf a \:e^{-\delta^2 S} 
  &= 
  \prod_{n\geq 1} \sqrt{\frac{M_n\Omega_n}{2\pi\sinh (\Omega_n t_f)}}
\\
  &= 
  \exp\bigg[ {\frac 12 \sum_{n\geq 1}\log {\frac{M_n\Omega_n}{2\pi\sinh (\Omega_n t_f)}}} \bigg]
  \label{eq:OH2}
\:.
\end{align}
At large times, the leading-order behavior arises from the $\sinh$ component; taking into account the effect of cut-offs (\emph{e.g.} as in \cite{bodineau_long_2008}), then from Eqs.~\eqref{eq:OH2} and $\Omega_n= n\pi \sqrt{n^2\pi^2-J^2}$, one obtains
\begin{align}
  \delta^2\Phi 
  &= \frac 12 \sum_{n\geq 1} \Big[\Omega_n-\Omega_n|_{J=0} + \text{correction terms from cut-offs}\Big]
\\
  &= \frac 12 \sum_{n\geq 1}\big\{n\pi \sqrt{n^2\pi^2-J^2}-n^2\pi^2+\tfrac 12 J^2\big\}
  \label{eq:corPhi}
\:.
\end{align}
Using the definition of the universal function
\begin{equation}
  \label{eq:defF_app}
 \mathcal F(u) = -4 \sum_{n\geq 1}\big\{n\pi \sqrt{n^2\pi^2-2u}-n^2\pi^2+u \big\}\,,
\end{equation}
one obtains for $|J|< J_c$
\begin{equation}
  \label{eq:devPhi_app}
  \Phi_L(J) = \frac {J^2} {2} - \frac 1{8L^2}  \mathcal F \Big(\tfrac 12 J^2\Big) + o(L^{-2})
\:.
\end{equation}
(Note that the saddle-point term can also yield order $L^{-2}$ corrections, which depend on the microscopic model~\cite{bodineau_long_2008,imparato_equilibriumlike_2009} and are not taken into account here).
This is the result announced in~\eqref{eq:devPhi},
which is compatible with the result obtained by Legendre transform from the finite-size corrections to the CGF.


\bibliography{symmetry-breaking-long}

\begin{thebibliography}{74}%
\makeatletter
\providecommand \@ifxundefined [1]{%
 \@ifx{#1\undefined}
}%
\providecommand \@ifnum [1]{%
 \ifnum #1\expandafter \@firstoftwo
 \else \expandafter \@secondoftwo
 \fi
}%
\providecommand \@ifx [1]{%
 \ifx #1\expandafter \@firstoftwo
 \else \expandafter \@secondoftwo
 \fi
}%
\providecommand \natexlab [1]{#1}%
\providecommand \enquote  [1]{``#1''}%
\providecommand \bibnamefont  [1]{#1}%
\providecommand \bibfnamefont [1]{#1}%
\providecommand \citenamefont [1]{#1}%
\providecommand \href@noop [0]{\@secondoftwo}%
\providecommand \href [0]{\begingroup \@sanitize@url \@href}%
\providecommand \@href[1]{\@@startlink{#1}\@@href}%
\providecommand \@@href[1]{\endgroup#1\@@endlink}%
\providecommand \@sanitize@url [0]{\catcode `\\12\catcode `\$12\catcode
  `\&12\catcode `\#12\catcode `\^12\catcode `\_12\catcode `\%12\relax}%
\providecommand \@@startlink[1]{}%
\providecommand \@@endlink[0]{}%
\providecommand \url  [0]{\begingroup\@sanitize@url \@url }%
\providecommand \@url [1]{\endgroup\@href {#1}{\urlprefix }}%
\providecommand \urlprefix  [0]{URL }%
\providecommand \Eprint [0]{\href }%
\providecommand \doibase [0]{http://dx.doi.org/}%
\providecommand \selectlanguage [0]{\@gobble}%
\providecommand \bibinfo  [0]{\@secondoftwo}%
\providecommand \bibfield  [0]{\@secondoftwo}%
\providecommand \translation [1]{[#1]}%
\providecommand \BibitemOpen [0]{}%
\providecommand \bibitemStop [0]{}%
\providecommand \bibitemNoStop [0]{.\EOS\space}%
\providecommand \EOS [0]{\spacefactor3000\relax}%
\providecommand \BibitemShut  [1]{\csname bibitem#1\endcsname}%
\let\auto@bib@innerbib\@empty
\bibitem [{\citenamefont {Levitov}\ and\ \citenamefont
  {Lesovik}(1993)}]{levitov1993pis}%
  \BibitemOpen
  \bibfield  {author} {\bibinfo {author} {\bibfnamefont {L.~S.}\ \bibnamefont
  {Levitov}}\ and\ \bibinfo {author} {\bibfnamefont {G.~B.}\ \bibnamefont
  {Lesovik}},\ }\href
  {http://www.jetpletters.ac.ru/ps/1186/article_17907.shtml} {\bibfield
  {journal} {\bibinfo  {journal} {JETP Lett.}\ }\textbf {\bibinfo {volume}
  {58}},\ \bibinfo {pages} {230} (\bibinfo {year} {1993})}\BibitemShut
  {NoStop}%
\bibitem [{\citenamefont {Levitov}\ \emph {et~al.}(1996)\citenamefont
  {Levitov}, \citenamefont {Lee},\ and\ \citenamefont
  {Lesovik}}]{levitov_electron_1996}%
  \BibitemOpen
  \bibfield  {author} {\bibinfo {author} {\bibfnamefont {L.~S.}\ \bibnamefont
  {Levitov}}, \bibinfo {author} {\bibfnamefont {H.}~\bibnamefont {Lee}}, \ and\
  \bibinfo {author} {\bibfnamefont {G.~B.}\ \bibnamefont {Lesovik}},\ }\href
  {\doibase 10.1063/1.531672} {\bibfield  {journal} {\bibinfo  {journal} {J.
  Math. Phys.}\ }\textbf {\bibinfo {volume} {37}},\ \bibinfo {pages} {4845}
  (\bibinfo {year} {1996})}\BibitemShut {NoStop}%
\bibitem [{\citenamefont {Pilgram}\ \emph {et~al.}(2003)\citenamefont
  {Pilgram}, \citenamefont {Jordan}, \citenamefont {Sukhorukov},\ and\
  \citenamefont {B{\"u}ttiker}}]{pilgram_stochastic_2003}%
  \BibitemOpen
  \bibfield  {author} {\bibinfo {author} {\bibfnamefont {S.}~\bibnamefont
  {Pilgram}}, \bibinfo {author} {\bibfnamefont {A.~N.}\ \bibnamefont {Jordan}},
  \bibinfo {author} {\bibfnamefont {E.~V.}\ \bibnamefont {Sukhorukov}}, \ and\
  \bibinfo {author} {\bibfnamefont {M.}~\bibnamefont {B{\"u}ttiker}},\ }\href
  {\doibase 10.1103/PhysRevLett.90.206801} {\bibfield  {journal} {\bibinfo
  {journal} {Phys. Rev. Lett.}\ }\textbf {\bibinfo {volume} {90}},\ \bibinfo
  {pages} {206801} (\bibinfo {year} {2003})}\BibitemShut {NoStop}%
\bibitem [{\citenamefont {Jordan}\ \emph {et~al.}(2004)\citenamefont {Jordan},
  \citenamefont {Sukhorukov},\ and\ \citenamefont
  {Pilgram}}]{jordan_fluctuation_2004}%
  \BibitemOpen
  \bibfield  {author} {\bibinfo {author} {\bibfnamefont {A.~N.}\ \bibnamefont
  {Jordan}}, \bibinfo {author} {\bibfnamefont {E.~V.}\ \bibnamefont
  {Sukhorukov}}, \ and\ \bibinfo {author} {\bibfnamefont {S.}~\bibnamefont
  {Pilgram}},\ }\href {\doibase 10.1063/1.1803927} {\bibfield  {journal}
  {\bibinfo  {journal} {J. Math. Phys.}\ }\textbf {\bibinfo {volume} {45}},\
  \bibinfo {pages} {4386} (\bibinfo {year} {2004})}\BibitemShut {NoStop}%
\bibitem [{\citenamefont {Esposito}\ \emph {et~al.}(2009)\citenamefont
  {Esposito}, \citenamefont {Harbola},\ and\ \citenamefont
  {Mukamel}}]{Esposito2009}%
  \BibitemOpen
  \bibfield  {author} {\bibinfo {author} {\bibfnamefont {M.}~\bibnamefont
  {Esposito}}, \bibinfo {author} {\bibfnamefont {U.}~\bibnamefont {Harbola}}, \
  and\ \bibinfo {author} {\bibfnamefont {S.}~\bibnamefont {Mukamel}},\ }\href
  {http://link.aps.org/doi/10.1103/RevModPhys.81.1665} {\bibfield  {journal}
  {\bibinfo  {journal} {Rev. Mod. Phys.}\ }\textbf {\bibinfo {volume} {81}},\
  \bibinfo {pages} {1665} (\bibinfo {year} {2009})}\BibitemShut {NoStop}%
\bibitem [{\citenamefont {Flindt}\ and\ \citenamefont
  {Garrahan}(2013)}]{flindt_trajectory_2013}%
  \BibitemOpen
  \bibfield  {author} {\bibinfo {author} {\bibfnamefont {C.}~\bibnamefont
  {Flindt}}\ and\ \bibinfo {author} {\bibfnamefont {J.~P.}\ \bibnamefont
  {Garrahan}},\ }\href {\doibase 10.1103/PhysRevLett.110.050601} {\bibfield
  {journal} {\bibinfo  {journal} {Phys. Rev. Lett.}\ }\textbf {\bibinfo
  {volume} {110}},\ \bibinfo {pages} {050601} (\bibinfo {year}
  {2013})}\BibitemShut {NoStop}%
\bibitem [{\citenamefont {Genway}\ \emph {et~al.}(2014)\citenamefont {Genway},
  \citenamefont {Hickey}, \citenamefont {Garrahan},\ and\ \citenamefont
  {Armour}}]{genway_trajectory_2014}%
  \BibitemOpen
  \bibfield  {author} {\bibinfo {author} {\bibfnamefont {S.}~\bibnamefont
  {Genway}}, \bibinfo {author} {\bibfnamefont {J.~M.}\ \bibnamefont {Hickey}},
  \bibinfo {author} {\bibfnamefont {J.~P.}\ \bibnamefont {Garrahan}}, \ and\
  \bibinfo {author} {\bibfnamefont {A.~D.}\ \bibnamefont {Armour}},\ }\href
  {\doibase 10.1088/1751-8113/47/50/505001} {\bibfield  {journal} {\bibinfo
  {journal} {J. Phys. A}\ }\textbf {\bibinfo {volume} {47}},\ \bibinfo {pages}
  {505001} (\bibinfo {year} {2014})}\BibitemShut {NoStop}%
\bibitem [{\citenamefont {Derrida}(2007)}]{Derrida2007}%
  \BibitemOpen
  \bibfield  {author} {\bibinfo {author} {\bibfnamefont {B.}~\bibnamefont
  {Derrida}},\ }\href {http://stacks.iop.org/1742-5468/2007/i=07/a=P07023}
  {\bibfield  {journal} {\bibinfo  {journal} {J. Stat. Mech.}\ }\textbf
  {\bibinfo {volume} {2007}},\ \bibinfo {pages} {P07023} (\bibinfo {year}
  {2007})}\BibitemShut {NoStop}%
\bibitem [{\citenamefont {Derrida}\ and\ \citenamefont
  {Appert}(1999)}]{derrida_universal_1999}%
  \BibitemOpen
  \bibfield  {author} {\bibinfo {author} {\bibfnamefont {B.}~\bibnamefont
  {Derrida}}\ and\ \bibinfo {author} {\bibfnamefont {C.}~\bibnamefont
  {Appert}},\ }\href {\doibase 10.1023/A:1004599526997} {\bibfield  {journal}
  {\bibinfo  {journal} {J. Stat. Phys.}\ }\textbf {\bibinfo {volume} {94}},\
  \bibinfo {pages} {1} (\bibinfo {year} {1999})}\BibitemShut {NoStop}%
\bibitem [{\citenamefont {Bodineau}\ and\ \citenamefont
  {Derrida}(2005)}]{bodineau_distribution_2005}%
  \BibitemOpen
  \bibfield  {author} {\bibinfo {author} {\bibfnamefont {T.}~\bibnamefont
  {Bodineau}}\ and\ \bibinfo {author} {\bibfnamefont {B.}~\bibnamefont
  {Derrida}},\ }\href {\doibase 10.1103/PhysRevE.72.066110} {\bibfield
  {journal} {\bibinfo  {journal} {Phys. Rev. E}\ }\textbf {\bibinfo {volume}
  {72}},\ \bibinfo {pages} {066110} (\bibinfo {year} {2005})}\BibitemShut
  {NoStop}%
\bibitem [{\citenamefont {Bodineau}\ and\ \citenamefont
  {Derrida}(2007)}]{Bodineau:2007iq}%
  \BibitemOpen
  \bibfield  {author} {\bibinfo {author} {\bibfnamefont {T.}~\bibnamefont
  {Bodineau}}\ and\ \bibinfo {author} {\bibfnamefont {B.}~\bibnamefont
  {Derrida}},\ }\href {\doibase 10.1016/j.crhy.2007.04.014} {\bibfield
  {journal} {\bibinfo  {journal} {C. R. Physique}\ }\textbf {\bibinfo {volume}
  {8}},\ \bibinfo {pages} {540} (\bibinfo {year} {2007})}\BibitemShut {NoStop}%
\bibitem [{\citenamefont {Prolhac}\ and\ \citenamefont
  {Mallick}(2008)}]{prolhac_current_2008}%
  \BibitemOpen
  \bibfield  {author} {\bibinfo {author} {\bibfnamefont {S.}~\bibnamefont
  {Prolhac}}\ and\ \bibinfo {author} {\bibfnamefont {K.}~\bibnamefont
  {Mallick}},\ }\href {\doibase 10.1088/1751-8113/41/17/175002} {\bibfield
  {journal} {\bibinfo  {journal} {J. Phys. A}\ }\textbf {\bibinfo {volume}
  {41}},\ \bibinfo {pages} {175002} (\bibinfo {year} {2008})}\BibitemShut
  {NoStop}%
\bibitem [{\citenamefont {Appert-Rolland}\ \emph {et~al.}(2008)\citenamefont
  {Appert-Rolland}, \citenamefont {Derrida}, \citenamefont {Lecomte},\ and\
  \citenamefont {van Wijland}}]{appert-rolland_universal_2008}%
  \BibitemOpen
  \bibfield  {author} {\bibinfo {author} {\bibfnamefont {C.}~\bibnamefont
  {Appert-Rolland}}, \bibinfo {author} {\bibfnamefont {B.}~\bibnamefont
  {Derrida}}, \bibinfo {author} {\bibfnamefont {V.}~\bibnamefont {Lecomte}}, \
  and\ \bibinfo {author} {\bibfnamefont {F.}~\bibnamefont {van Wijland}},\
  }\href {\doibase 10.1103/PhysRevE.78.021122} {\bibfield  {journal} {\bibinfo
  {journal} {Phys. Rev. E}\ }\textbf {\bibinfo {volume} {78}},\ \bibinfo
  {pages} {021122} (\bibinfo {year} {2008})}\BibitemShut {NoStop}%
\bibitem [{\citenamefont {Prolhac}\ and\ \citenamefont
  {Mallick}(2009)}]{prolhac_cumulants_2009}%
  \BibitemOpen
  \bibfield  {author} {\bibinfo {author} {\bibfnamefont {S.}~\bibnamefont
  {Prolhac}}\ and\ \bibinfo {author} {\bibfnamefont {K.}~\bibnamefont
  {Mallick}},\ }\href {\doibase 10.1088/1751-8113/42/17/175001} {\bibfield
  {journal} {\bibinfo  {journal} {J. Phys. A}\ }\textbf {\bibinfo {volume}
  {42}},\ \bibinfo {pages} {175001} (\bibinfo {year} {2009})}\BibitemShut
  {NoStop}%
\bibitem [{\citenamefont {Baek}\ \emph {et~al.}(2016)\citenamefont {Baek},
  \citenamefont {Kafri},\ and\ \citenamefont {Lecomte}}]{baek_large_n_2016}%
  \BibitemOpen
  \bibfield  {author} {\bibinfo {author} {\bibfnamefont {Y.}~\bibnamefont
  {Baek}}, \bibinfo {author} {\bibfnamefont {Y.}~\bibnamefont {Kafri}}, \ and\
  \bibinfo {author} {\bibfnamefont {V.}~\bibnamefont {Lecomte}},\ }\href
  {http://stacks.iop.org/1742-5468/2016/i=5/a=053203} {\bibfield  {journal}
  {\bibinfo  {journal} {J. Stat. Mech.}\ }\textbf {\bibinfo {volume} {2016}},\
  \bibinfo {pages} {053203} (\bibinfo {year} {2016})}\BibitemShut {NoStop}%
\bibitem [{\citenamefont {Hurtado}\ and\ \citenamefont
  {Garrido}(2009)}]{hurtado_test_2009}%
  \BibitemOpen
  \bibfield  {author} {\bibinfo {author} {\bibfnamefont {P.~I.}\ \bibnamefont
  {Hurtado}}\ and\ \bibinfo {author} {\bibfnamefont {P.~L.}\ \bibnamefont
  {Garrido}},\ }\href {\doibase 10.1103/PhysRevLett.102.250601} {\bibfield
  {journal} {\bibinfo  {journal} {Phys. Rev. Lett.}\ }\textbf {\bibinfo
  {volume} {102}},\ \bibinfo {pages} {250601} (\bibinfo {year}
  {2009})}\BibitemShut {NoStop}%
\bibitem [{\citenamefont {Prados}\ \emph {et~al.}(2011)\citenamefont {Prados},
  \citenamefont {Lasanta},\ and\ \citenamefont {Hurtado}}]{prados_large_2011}%
  \BibitemOpen
  \bibfield  {author} {\bibinfo {author} {\bibfnamefont {A.}~\bibnamefont
  {Prados}}, \bibinfo {author} {\bibfnamefont {A.}~\bibnamefont {Lasanta}}, \
  and\ \bibinfo {author} {\bibfnamefont {P.~I.}\ \bibnamefont {Hurtado}},\
  }\href {\doibase 10.1103/PhysRevLett.107.140601} {\bibfield  {journal}
  {\bibinfo  {journal} {Phys. Rev. Lett.}\ }\textbf {\bibinfo {volume} {107}},\
  \bibinfo {pages} {140601} (\bibinfo {year} {2011})}\BibitemShut {NoStop}%
\bibitem [{\citenamefont {de~Gier}\ and\ \citenamefont
  {Essler}(2011)}]{de_gier_large_2011}%
  \BibitemOpen
  \bibfield  {author} {\bibinfo {author} {\bibfnamefont {J.}~\bibnamefont
  {de~Gier}}\ and\ \bibinfo {author} {\bibfnamefont {F.~H.~L.}\ \bibnamefont
  {Essler}},\ }\href {\doibase 10.1103/PhysRevLett.107.010602} {\bibfield
  {journal} {\bibinfo  {journal} {Phys. Rev. Lett.}\ }\textbf {\bibinfo
  {volume} {107}},\ \bibinfo {pages} {010602} (\bibinfo {year}
  {2011})}\BibitemShut {NoStop}%
\bibitem [{\citenamefont {Lazarescu}\ and\ \citenamefont
  {Mallick}(2011)}]{lazarescu_exact_2011}%
  \BibitemOpen
  \bibfield  {author} {\bibinfo {author} {\bibfnamefont {A.}~\bibnamefont
  {Lazarescu}}\ and\ \bibinfo {author} {\bibfnamefont {K.}~\bibnamefont
  {Mallick}},\ }\href {\doibase 10.1088/1751-8113/44/31/315001} {\bibfield
  {journal} {\bibinfo  {journal} {J. Phys. A}\ }\textbf {\bibinfo {volume}
  {44}},\ \bibinfo {pages} {315001} (\bibinfo {year} {2011})}\BibitemShut
  {NoStop}%
\bibitem [{\citenamefont {Hurtado}\ and\ \citenamefont
  {Garrido}(2011{\natexlab{a}})}]{hurtado_spontaneous_2011}%
  \BibitemOpen
  \bibfield  {author} {\bibinfo {author} {\bibfnamefont {P.~I.}\ \bibnamefont
  {Hurtado}}\ and\ \bibinfo {author} {\bibfnamefont {P.~L.}\ \bibnamefont
  {Garrido}},\ }\href {\doibase 10.1103/PhysRevLett.107.180601} {\bibfield
  {journal} {\bibinfo  {journal} {Phys. Rev. Lett.}\ }\textbf {\bibinfo
  {volume} {107}},\ \bibinfo {pages} {180601} (\bibinfo {year}
  {2011}{\natexlab{a}})}\BibitemShut {NoStop}%
\bibitem [{\citenamefont {Derrida}(2011)}]{derrida_microscopic_2011}%
  \BibitemOpen
  \bibfield  {author} {\bibinfo {author} {\bibfnamefont {B.}~\bibnamefont
  {Derrida}},\ }\href {\doibase 10.1088/1742-5468/2011/01/P01030} {\bibfield
  {journal} {\bibinfo  {journal} {J. Stat. Mech.}\ }\textbf {\bibinfo {volume}
  {2011}},\ \bibinfo {pages} {P01030} (\bibinfo {year} {2011})}\BibitemShut
  {NoStop}%
\bibitem [{\citenamefont {Gorissen}\ and\ \citenamefont
  {Vanderzande}(2012)}]{gorissen_current_2012}%
  \BibitemOpen
  \bibfield  {author} {\bibinfo {author} {\bibfnamefont {M.}~\bibnamefont
  {Gorissen}}\ and\ \bibinfo {author} {\bibfnamefont {C.}~\bibnamefont
  {Vanderzande}},\ }\href {\doibase 10.1103/PhysRevE.86.051114} {\bibfield
  {journal} {\bibinfo  {journal} {Phys. Rev. E}\ }\textbf {\bibinfo {volume}
  {86}},\ \bibinfo {pages} {051114} (\bibinfo {year} {2012})}\BibitemShut
  {NoStop}%
\bibitem [{\citenamefont {Gorissen}\ \emph {et~al.}(2012)\citenamefont
  {Gorissen}, \citenamefont {Lazarescu}, \citenamefont {Mallick},\ and\
  \citenamefont {Vanderzande}}]{gorissen_exact_2012}%
  \BibitemOpen
  \bibfield  {author} {\bibinfo {author} {\bibfnamefont {M.}~\bibnamefont
  {Gorissen}}, \bibinfo {author} {\bibfnamefont {A.}~\bibnamefont {Lazarescu}},
  \bibinfo {author} {\bibfnamefont {K.}~\bibnamefont {Mallick}}, \ and\
  \bibinfo {author} {\bibfnamefont {C.}~\bibnamefont {Vanderzande}},\ }\href
  {\doibase 10.1103/PhysRevLett.109.170601} {\bibfield  {journal} {\bibinfo
  {journal} {Phys. Rev. Lett.}\ }\textbf {\bibinfo {volume} {109}},\ \bibinfo
  {pages} {170601} (\bibinfo {year} {2012})}\BibitemShut {NoStop}%
\bibitem [{\citenamefont {Krapivsky}\ and\ \citenamefont
  {Meerson}(2012)}]{krapivsky_fluctuations_2012}%
  \BibitemOpen
  \bibfield  {author} {\bibinfo {author} {\bibfnamefont {P.~L.}\ \bibnamefont
  {Krapivsky}}\ and\ \bibinfo {author} {\bibfnamefont {B.}~\bibnamefont
  {Meerson}},\ }\href {\doibase 10.1103/PhysRevE.86.031106} {\bibfield
  {journal} {\bibinfo  {journal} {Phys. Rev. E}\ }\textbf {\bibinfo {volume}
  {86}},\ \bibinfo {pages} {031106} (\bibinfo {year} {2012})}\BibitemShut
  {NoStop}%
\bibitem [{\citenamefont {Meerson}\ and\ \citenamefont
  {Sasorov}(2013)}]{meerson_extreme_2013}%
  \BibitemOpen
  \bibfield  {author} {\bibinfo {author} {\bibfnamefont {B.}~\bibnamefont
  {Meerson}}\ and\ \bibinfo {author} {\bibfnamefont {P.~V.}\ \bibnamefont
  {Sasorov}},\ }\href {\doibase 10.1088/1742-5468/2013/12/P12011} {\bibfield
  {journal} {\bibinfo  {journal} {J. Stat. Mech.}\ }\textbf {\bibinfo {volume}
  {2013}},\ \bibinfo {pages} {P12011} (\bibinfo {year} {2013})}\BibitemShut
  {NoStop}%
\bibitem [{\citenamefont {Meerson}\ and\ \citenamefont
  {Sasorov}(2014)}]{meerson_extreme_2014}%
  \BibitemOpen
  \bibfield  {author} {\bibinfo {author} {\bibfnamefont {B.}~\bibnamefont
  {Meerson}}\ and\ \bibinfo {author} {\bibfnamefont {P.~V.}\ \bibnamefont
  {Sasorov}},\ }\href {\doibase 10.1103/PhysRevE.89.010101} {\bibfield
  {journal} {\bibinfo  {journal} {Phys. Rev. E}\ }\textbf {\bibinfo {volume}
  {89}},\ \bibinfo {pages} {010101} (\bibinfo {year} {2014})}\BibitemShut
  {NoStop}%
\bibitem [{\citenamefont {Žnidarič}(2014)}]{znidaric_exact_2014}%
  \BibitemOpen
  \bibfield  {author} {\bibinfo {author} {\bibfnamefont {M.}~\bibnamefont
  {Žnidarič}},\ }\href {\doibase 10.1103/PhysRevLett.112.040602} {\bibfield
  {journal} {\bibinfo  {journal} {Phys. Rev. Lett.}\ }\textbf {\bibinfo
  {volume} {112}},\ \bibinfo {pages} {040602} (\bibinfo {year}
  {2014})}\BibitemShut {NoStop}%
\bibitem [{\citenamefont {Hurtado}\ \emph {et~al.}(2014)\citenamefont
  {Hurtado}, \citenamefont {Espigares}, \citenamefont {del Pozo},\ and\
  \citenamefont {Garrido}}]{Hurtado:2014bn}%
  \BibitemOpen
  \bibfield  {author} {\bibinfo {author} {\bibfnamefont {P.~I.}\ \bibnamefont
  {Hurtado}}, \bibinfo {author} {\bibfnamefont {C.~P.}\ \bibnamefont
  {Espigares}}, \bibinfo {author} {\bibfnamefont {J.~J.}\ \bibnamefont {del
  Pozo}}, \ and\ \bibinfo {author} {\bibfnamefont {P.~L.}\ \bibnamefont
  {Garrido}},\ }\href {\doibase 10.1007/s10955-013-0894-6} {\bibfield
  {journal} {\bibinfo  {journal} {J. Stat. Phys.}\ }\textbf {\bibinfo {volume}
  {154}},\ \bibinfo {pages} {214} (\bibinfo {year} {2014})}\BibitemShut
  {NoStop}%
\bibitem [{\citenamefont {Lazarescu}(2015)}]{Lazarescu2015}%
  \BibitemOpen
  \bibfield  {author} {\bibinfo {author} {\bibfnamefont {A.}~\bibnamefont
  {Lazarescu}},\ }\href {\doibase 10.1088/1751-8113/48/50/503001} {\bibfield
  {journal} {\bibinfo  {journal} {J. Phys. A}\ }\textbf {\bibinfo {volume}
  {48}},\ \bibinfo {pages} {503001} (\bibinfo {year} {2015})}\BibitemShut
  {NoStop}%
\bibitem [{\citenamefont {Shpielberg}\ and\ \citenamefont
  {Akkermans}(2016)}]{Shpielberg2016}%
  \BibitemOpen
  \bibfield  {author} {\bibinfo {author} {\bibfnamefont {O.}~\bibnamefont
  {Shpielberg}}\ and\ \bibinfo {author} {\bibfnamefont {E.}~\bibnamefont
  {Akkermans}},\ }\href {\doibase 10.1103/PhysRevLett.116.240603} {\bibfield
  {journal} {\bibinfo  {journal} {Phys. Rev. Lett.}\ }\textbf {\bibinfo
  {volume} {116}},\ \bibinfo {pages} {240603} (\bibinfo {year}
  {2016})}\BibitemShut {NoStop}%
\bibitem [{\citenamefont {Zarfaty}\ and\ \citenamefont
  {Meerson}(2016)}]{Zarfaty:2016dv}%
  \BibitemOpen
  \bibfield  {author} {\bibinfo {author} {\bibfnamefont {L.}~\bibnamefont
  {Zarfaty}}\ and\ \bibinfo {author} {\bibfnamefont {B.}~\bibnamefont
  {Meerson}},\ }\href {\doibase 10.1088/1742-5468/2016/03/033304} {\bibfield
  {journal} {\bibinfo  {journal} {J. Stat. Mech.}\ }\textbf {\bibinfo {volume}
  {2016}},\ \bibinfo {pages} {033304} (\bibinfo {year} {2016})}\BibitemShut
  {NoStop}%
\bibitem [{\citenamefont {Hirschberg}\ \emph {et~al.}(2015)\citenamefont
  {Hirschberg}, \citenamefont {Mukamel},\ and\ \citenamefont
  {Sch{\"u}tz}}]{hirschberg_zrp_2015}%
  \BibitemOpen
  \bibfield  {author} {\bibinfo {author} {\bibfnamefont {O.}~\bibnamefont
  {Hirschberg}}, \bibinfo {author} {\bibfnamefont {D.}~\bibnamefont {Mukamel}},
  \ and\ \bibinfo {author} {\bibfnamefont {G.~M.}\ \bibnamefont {Sch{\"u}tz}},\
  }\href {http://stacks.iop.org/1742-5468/2015/i=11/a=P11023} {\bibfield
  {journal} {\bibinfo  {journal} {J. Stat. Mech.}\ }\textbf {\bibinfo {volume}
  {2015}},\ \bibinfo {pages} {P11023} (\bibinfo {year} {2015})}\BibitemShut
  {NoStop}%
\bibitem [{\citenamefont {Shpielberg}\ \emph {et~al.}(2017)\citenamefont
  {Shpielberg}, \citenamefont {Don},\ and\ \citenamefont
  {Akkermans}}]{Shpielberg2017}%
  \BibitemOpen
  \bibfield  {author} {\bibinfo {author} {\bibfnamefont {O.}~\bibnamefont
  {Shpielberg}}, \bibinfo {author} {\bibfnamefont {Y.}~\bibnamefont {Don}}, \
  and\ \bibinfo {author} {\bibfnamefont {E.}~\bibnamefont {Akkermans}},\ }\href
  {https://link.aps.org/doi/10.1103/PhysRevE.95.032137} {\bibfield  {journal}
  {\bibinfo  {journal} {Phys. Rev. E}\ }\textbf {\bibinfo {volume} {95}},\
  \bibinfo {pages} {032137} (\bibinfo {year} {2017})}\BibitemShut {NoStop}%
\bibitem [{\citenamefont {Touchette}(2009)}]{touchette_large_2009}%
  \BibitemOpen
  \bibfield  {author} {\bibinfo {author} {\bibfnamefont {H.}~\bibnamefont
  {Touchette}},\ }\href {\doibase 10.1016/j.physrep.2009.05.002} {\bibfield
  {journal} {\bibinfo  {journal} {Phys. Rep.}\ }\textbf {\bibinfo {volume}
  {478}},\ \bibinfo {pages} {1} (\bibinfo {year} {2009})}\BibitemShut {NoStop}%
\bibitem [{\citenamefont {Derrida}\ and\ \citenamefont
  {Lebowitz}(1998)}]{derrida_exact_1998}%
  \BibitemOpen
  \bibfield  {author} {\bibinfo {author} {\bibfnamefont {B.}~\bibnamefont
  {Derrida}}\ and\ \bibinfo {author} {\bibfnamefont {J.~L.}\ \bibnamefont
  {Lebowitz}},\ }\href {\doibase 10.1103/PhysRevLett.80.209} {\bibfield
  {journal} {\bibinfo  {journal} {Phys. Rev. Lett.}\ }\textbf {\bibinfo
  {volume} {80}},\ \bibinfo {pages} {209} (\bibinfo {year} {1998})}\BibitemShut
  {NoStop}%
\bibitem [{\citenamefont {Mallick}(2011)}]{mallick_exact_2011}%
  \BibitemOpen
  \bibfield  {author} {\bibinfo {author} {\bibfnamefont {K.}~\bibnamefont
  {Mallick}},\ }\href {\doibase 10.1088/1742-5468/2011/01/P01024} {\bibfield
  {journal} {\bibinfo  {journal} {J. Stat. Mech.}\ }\textbf {\bibinfo {volume}
  {2011}},\ \bibinfo {pages} {P01024} (\bibinfo {year} {2011})}\BibitemShut
  {NoStop}%
\bibitem [{\citenamefont {Lazarescu}(2013)}]{lazarescu_matrix_2013}%
  \BibitemOpen
  \bibfield  {author} {\bibinfo {author} {\bibfnamefont {A.}~\bibnamefont
  {Lazarescu}},\ }\href {\doibase 10.1088/1751-8113/46/14/145003} {\bibfield
  {journal} {\bibinfo  {journal} {J. Phys. A}\ }\textbf {\bibinfo {volume}
  {46}},\ \bibinfo {pages} {145003} (\bibinfo {year} {2013})}\BibitemShut
  {NoStop}%
\bibitem [{\citenamefont {Ayyer}(2016)}]{ayyer_full_2015}%
  \BibitemOpen
  \bibfield  {author} {\bibinfo {author} {\bibfnamefont {A.}~\bibnamefont
  {Ayyer}},\ }\href {\doibase 10.1088/1751-8113/49/15/155003} {\bibfield
  {journal} {\bibinfo  {journal} {J. Phys. A}\ }\textbf {\bibinfo {volume}
  {49}},\ \bibinfo {pages} {155003} (\bibinfo {year} {2016})}\BibitemShut
  {NoStop}%
\bibitem [{\citenamefont {Spohn}(1983)}]{spohn_long_1983}%
  \BibitemOpen
  \bibfield  {author} {\bibinfo {author} {\bibfnamefont {H.}~\bibnamefont
  {Spohn}},\ }\href {\doibase 10.1088/0305-4470/16/18/029} {\bibfield
  {journal} {\bibinfo  {journal} {J. Phys. A}\ }\textbf {\bibinfo {volume}
  {16}},\ \bibinfo {pages} {4275} (\bibinfo {year} {1983})}\BibitemShut
  {NoStop}%
\bibitem [{\citenamefont {Spohn}(1991)}]{Spohn_Book}%
  \BibitemOpen
  \bibfield  {author} {\bibinfo {author} {\bibfnamefont {H.}~\bibnamefont
  {Spohn}},\ }\href@noop {} {\emph {\bibinfo {title} {Large scale dynamics of
  interacting particles}}}\ (\bibinfo  {publisher} {Springer},\ \bibinfo
  {address} {Berlin},\ \bibinfo {year} {1991})\BibitemShut {NoStop}%
\bibitem [{\citenamefont {Bertini}\ \emph {et~al.}(2002)\citenamefont
  {Bertini}, \citenamefont {De~Sole}, \citenamefont {Gabrielli}, \citenamefont
  {Jona-Lasinio},\ and\ \citenamefont {Landim}}]{bertini_macroscopic_2002}%
  \BibitemOpen
  \bibfield  {author} {\bibinfo {author} {\bibfnamefont {L.}~\bibnamefont
  {Bertini}}, \bibinfo {author} {\bibfnamefont {A.}~\bibnamefont {De~Sole}},
  \bibinfo {author} {\bibfnamefont {D.}~\bibnamefont {Gabrielli}}, \bibinfo
  {author} {\bibfnamefont {G.}~\bibnamefont {Jona-Lasinio}}, \ and\ \bibinfo
  {author} {\bibfnamefont {C.}~\bibnamefont {Landim}},\ }\href {\doibase
  10.1023/A:1014525911391} {\bibfield  {journal} {\bibinfo  {journal} {J. Stat.
  Phys.}\ }\textbf {\bibinfo {volume} {107}},\ \bibinfo {pages} {635} (\bibinfo
  {year} {2002})}\BibitemShut {NoStop}%
\bibitem [{\citenamefont {Bertini}\ \emph {et~al.}(2015)\citenamefont
  {Bertini}, \citenamefont {De~Sole}, \citenamefont {Gabrielli}, \citenamefont
  {Jona-Lasinio},\ and\ \citenamefont {Landim}}]{bertini_macroscopic_2015}%
  \BibitemOpen
  \bibfield  {author} {\bibinfo {author} {\bibfnamefont {L.}~\bibnamefont
  {Bertini}}, \bibinfo {author} {\bibfnamefont {A.}~\bibnamefont {De~Sole}},
  \bibinfo {author} {\bibfnamefont {D.}~\bibnamefont {Gabrielli}}, \bibinfo
  {author} {\bibfnamefont {G.}~\bibnamefont {Jona-Lasinio}}, \ and\ \bibinfo
  {author} {\bibfnamefont {C.}~\bibnamefont {Landim}},\ }\href {\doibase
  10.1103/RevModPhys.87.593} {\bibfield  {journal} {\bibinfo  {journal} {Rev.
  Mod. Phys.}\ }\textbf {\bibinfo {volume} {87}},\ \bibinfo {pages} {593}
  (\bibinfo {year} {2015})}\BibitemShut {NoStop}%
\bibitem [{\citenamefont {Derrida}\ \emph {et~al.}(2004)\citenamefont
  {Derrida}, \citenamefont {Douçot},\ and\ \citenamefont
  {Roche}}]{derrida_current_2004}%
  \BibitemOpen
  \bibfield  {author} {\bibinfo {author} {\bibfnamefont {B.}~\bibnamefont
  {Derrida}}, \bibinfo {author} {\bibfnamefont {B.}~\bibnamefont {Douçot}}, \
  and\ \bibinfo {author} {\bibfnamefont {P.-E.}\ \bibnamefont {Roche}},\ }\href
  {\doibase 10.1023/B:JOSS.0000022379.95508.b2} {\bibfield  {journal} {\bibinfo
   {journal} {J. Stat. Phys.}\ }\textbf {\bibinfo {volume} {115}},\ \bibinfo
  {pages} {717} (\bibinfo {year} {2004})}\BibitemShut {NoStop}%
\bibitem [{\citenamefont {Bodineau}\ and\ \citenamefont
  {Derrida}(2004)}]{bodineau_current_2004}%
  \BibitemOpen
  \bibfield  {author} {\bibinfo {author} {\bibfnamefont {T.}~\bibnamefont
  {Bodineau}}\ and\ \bibinfo {author} {\bibfnamefont {B.}~\bibnamefont
  {Derrida}},\ }\href {\doibase 10.1103/PhysRevLett.92.180601} {\bibfield
  {journal} {\bibinfo  {journal} {Phys. Rev. Lett.}\ }\textbf {\bibinfo
  {volume} {92}},\ \bibinfo {pages} {180601} (\bibinfo {year}
  {2004})}\BibitemShut {NoStop}%
\bibitem [{\citenamefont {Bertini}\ \emph {et~al.}(2005)\citenamefont
  {Bertini}, \citenamefont {De~Sole}, \citenamefont {Gabrielli}, \citenamefont
  {Jona-Lasinio},\ and\ \citenamefont {Landim}}]{bertini_current_2005}%
  \BibitemOpen
  \bibfield  {author} {\bibinfo {author} {\bibfnamefont {L.}~\bibnamefont
  {Bertini}}, \bibinfo {author} {\bibfnamefont {A.}~\bibnamefont {De~Sole}},
  \bibinfo {author} {\bibfnamefont {D.}~\bibnamefont {Gabrielli}}, \bibinfo
  {author} {\bibfnamefont {G.}~\bibnamefont {Jona-Lasinio}}, \ and\ \bibinfo
  {author} {\bibfnamefont {C.}~\bibnamefont {Landim}},\ }\href {\doibase
  10.1103/PhysRevLett.94.030601} {\bibfield  {journal} {\bibinfo  {journal}
  {Phys. Rev. Lett.}\ }\textbf {\bibinfo {volume} {94}},\ \bibinfo {pages}
  {030601} (\bibinfo {year} {2005})}\BibitemShut {NoStop}%
\bibitem [{\citenamefont {Bertini}\ \emph {et~al.}(2006)\citenamefont
  {Bertini}, \citenamefont {De~Sole}, \citenamefont {Gabrielli}, \citenamefont
  {Jona-Lasinio},\ and\ \citenamefont {Landim}}]{bertini_non_2006}%
  \BibitemOpen
  \bibfield  {author} {\bibinfo {author} {\bibfnamefont {L.}~\bibnamefont
  {Bertini}}, \bibinfo {author} {\bibfnamefont {A.}~\bibnamefont {De~Sole}},
  \bibinfo {author} {\bibfnamefont {D.}~\bibnamefont {Gabrielli}}, \bibinfo
  {author} {\bibfnamefont {G.}~\bibnamefont {Jona-Lasinio}}, \ and\ \bibinfo
  {author} {\bibfnamefont {C.}~\bibnamefont {Landim}},\ }\href {\doibase
  10.1007/s10955-006-9056-4} {\bibfield  {journal} {\bibinfo  {journal} {J.
  Stat. Phys.}\ }\textbf {\bibinfo {volume} {123}},\ \bibinfo {pages} {237}
  (\bibinfo {year} {2006})}\BibitemShut {NoStop}%
\bibitem [{\citenamefont {Harris}\ \emph {et~al.}(2005)\citenamefont {Harris},
  \citenamefont {R{\'a}kos},\ and\ \citenamefont {Sch{\"u}tz}}]{Harris2005}%
  \BibitemOpen
  \bibfield  {author} {\bibinfo {author} {\bibfnamefont {R.~J.}\ \bibnamefont
  {Harris}}, \bibinfo {author} {\bibfnamefont {A.}~\bibnamefont {R{\'a}kos}}, \
  and\ \bibinfo {author} {\bibfnamefont {G.~M.}\ \bibnamefont {Sch{\"u}tz}},\
  }\href {http://stacks.iop.org/1742-5468/2005/i=08/a=P08003} {\bibfield
  {journal} {\bibinfo  {journal} {J. Stat. Mech.}\ }\textbf {\bibinfo {volume}
  {2005}},\ \bibinfo {pages} {P08003} (\bibinfo {year} {2005})}\BibitemShut
  {NoStop}%
\bibitem [{\citenamefont {Imparato}\ \emph {et~al.}(2009)\citenamefont
  {Imparato}, \citenamefont {Lecomte},\ and\ \citenamefont {van
  Wijland}}]{imparato_equilibriumlike_2009}%
  \BibitemOpen
  \bibfield  {author} {\bibinfo {author} {\bibfnamefont {A.}~\bibnamefont
  {Imparato}}, \bibinfo {author} {\bibfnamefont {V.}~\bibnamefont {Lecomte}}, \
  and\ \bibinfo {author} {\bibfnamefont {F.}~\bibnamefont {van Wijland}},\
  }\href {\doibase 10.1103/PhysRevE.80.011131} {\bibfield  {journal} {\bibinfo
  {journal} {Phys. Rev. E}\ }\textbf {\bibinfo {volume} {80}},\ \bibinfo
  {pages} {011131} (\bibinfo {year} {2009})}\BibitemShut {NoStop}%
\bibitem [{\citenamefont {Lecomte}\ \emph {et~al.}(2010)\citenamefont
  {Lecomte}, \citenamefont {Imparato},\ and\ \citenamefont {van
  Wijland}}]{Lecomte2010}%
  \BibitemOpen
  \bibfield  {author} {\bibinfo {author} {\bibfnamefont {V.}~\bibnamefont
  {Lecomte}}, \bibinfo {author} {\bibfnamefont {A.}~\bibnamefont {Imparato}}, \
  and\ \bibinfo {author} {\bibfnamefont {F.}~\bibnamefont {van Wijland}},\
  }\href {http://ptps.oxfordjournals.org/content/184/276.abstract} {\bibfield
  {journal} {\bibinfo  {journal} {Progress of Theoretical Physics Supplement}\
  }\textbf {\bibinfo {volume} {184}},\ \bibinfo {pages} {276} (\bibinfo {year}
  {2010})}\BibitemShut {NoStop}%
\bibitem [{\citenamefont {Bodineau}\ \emph
  {et~al.}(2008{\natexlab{a}})\citenamefont {Bodineau}, \citenamefont
  {Derrida},\ and\ \citenamefont {Lebowitz}}]{bodineau_vortices_2008}%
  \BibitemOpen
  \bibfield  {author} {\bibinfo {author} {\bibfnamefont {T.}~\bibnamefont
  {Bodineau}}, \bibinfo {author} {\bibfnamefont {B.}~\bibnamefont {Derrida}}, \
  and\ \bibinfo {author} {\bibfnamefont {J.~L.}\ \bibnamefont {Lebowitz}},\
  }\href {\doibase 10.1007/s10955-008-9518-y} {\bibfield  {journal} {\bibinfo
  {journal} {J. Stat. Phys.}\ }\textbf {\bibinfo {volume} {131}},\ \bibinfo
  {pages} {821} (\bibinfo {year} {2008}{\natexlab{a}})}\BibitemShut {NoStop}%
\bibitem [{\citenamefont {Akkermans}\ \emph {et~al.}(2013)\citenamefont
  {Akkermans}, \citenamefont {Bodineau}, \citenamefont {Derrida},\ and\
  \citenamefont {Shpielberg}}]{Akkermans2013}%
  \BibitemOpen
  \bibfield  {author} {\bibinfo {author} {\bibfnamefont {E.}~\bibnamefont
  {Akkermans}}, \bibinfo {author} {\bibfnamefont {T.}~\bibnamefont {Bodineau}},
  \bibinfo {author} {\bibfnamefont {B.}~\bibnamefont {Derrida}}, \ and\
  \bibinfo {author} {\bibfnamefont {O.}~\bibnamefont {Shpielberg}},\ }\href
  {http://stacks.iop.org/0295-5075/103/i=2/a=20001} {\bibfield  {journal}
  {\bibinfo  {journal} {Europhys. Lett.}\ }\textbf {\bibinfo {volume} {103}},\
  \bibinfo {pages} {20001} (\bibinfo {year} {2013})}\BibitemShut {NoStop}%
\bibitem [{\citenamefont {Baek}\ and\ \citenamefont
  {Kafri}(2015)}]{baek_singularities_2015}%
  \BibitemOpen
  \bibfield  {author} {\bibinfo {author} {\bibfnamefont {Y.}~\bibnamefont
  {Baek}}\ and\ \bibinfo {author} {\bibfnamefont {Y.}~\bibnamefont {Kafri}},\
  }\href {\doibase 10.1088/1742-5468/2015/08/P08026} {\bibfield  {journal}
  {\bibinfo  {journal} {J. Stat. Mech.}\ }\textbf {\bibinfo {volume} {2015}},\
  \bibinfo {pages} {P08026} (\bibinfo {year} {2015})}\BibitemShut {NoStop}%
\bibitem [{\citenamefont {Derrida}\ \emph {et~al.}(2002)\citenamefont
  {Derrida}, \citenamefont {Lebowitz},\ and\ \citenamefont
  {Speer}}]{derrida_lebowitz_speer_2002}%
  \BibitemOpen
  \bibfield  {author} {\bibinfo {author} {\bibfnamefont {B.}~\bibnamefont
  {Derrida}}, \bibinfo {author} {\bibfnamefont {J.~L.}\ \bibnamefont
  {Lebowitz}}, \ and\ \bibinfo {author} {\bibfnamefont {E.~R.}\ \bibnamefont
  {Speer}},\ }\href {\doibase 10.1103/PhysRevLett.89.030601} {\bibfield
  {journal} {\bibinfo  {journal} {Phys. Rev. Lett.}\ }\textbf {\bibinfo
  {volume} {89}},\ \bibinfo {pages} {030601} (\bibinfo {year}
  {2002})}\BibitemShut {NoStop}%
\bibitem [{\citenamefont {Lecomte}\ \emph {et~al.}(2007)\citenamefont
  {Lecomte}, \citenamefont {Appert-Rolland},\ and\ \citenamefont
  {Wijland}}]{lecomte_thermodynamic_2007}%
  \BibitemOpen
  \bibfield  {author} {\bibinfo {author} {\bibfnamefont {V.}~\bibnamefont
  {Lecomte}}, \bibinfo {author} {\bibfnamefont {C.}~\bibnamefont
  {Appert-Rolland}}, \ and\ \bibinfo {author} {\bibfnamefont {F.}~\bibnamefont
  {Wijland}},\ }\href {\doibase 10.1007/s10955-006-9254-0} {\bibfield
  {journal} {\bibinfo  {journal} {J. Stat. Phys.}\ }\textbf {\bibinfo {volume}
  {127}},\ \bibinfo {pages} {51} (\bibinfo {year} {2007})}\BibitemShut
  {NoStop}%
\bibitem [{\citenamefont {Garrahan}\ \emph {et~al.}(2007)\citenamefont
  {Garrahan}, \citenamefont {Jack}, \citenamefont {Lecomte}, \citenamefont
  {Pitard}, \citenamefont {van Duijvendijk},\ and\ \citenamefont {van
  Wijland}}]{garrahan_dynamical_2007}%
  \BibitemOpen
  \bibfield  {author} {\bibinfo {author} {\bibfnamefont {J.~P.}\ \bibnamefont
  {Garrahan}}, \bibinfo {author} {\bibfnamefont {R.~L.}\ \bibnamefont {Jack}},
  \bibinfo {author} {\bibfnamefont {V.}~\bibnamefont {Lecomte}}, \bibinfo
  {author} {\bibfnamefont {E.}~\bibnamefont {Pitard}}, \bibinfo {author}
  {\bibfnamefont {K.}~\bibnamefont {van Duijvendijk}}, \ and\ \bibinfo {author}
  {\bibfnamefont {F.}~\bibnamefont {van Wijland}},\ }\href {\doibase
  10.1103/PhysRevLett.98.195702} {\bibfield  {journal} {\bibinfo  {journal}
  {Phys. Rev. Lett.}\ }\textbf {\bibinfo {volume} {98}},\ \bibinfo {pages}
  {195702} (\bibinfo {year} {2007})}\BibitemShut {NoStop}%
\bibitem [{\citenamefont {Bunin}\ \emph {et~al.}(2012)\citenamefont {Bunin},
  \citenamefont {Kafri},\ and\ \citenamefont
  {Podolsky}}]{bunin_non-differentiable_2012}%
  \BibitemOpen
  \bibfield  {author} {\bibinfo {author} {\bibfnamefont {G.}~\bibnamefont
  {Bunin}}, \bibinfo {author} {\bibfnamefont {Y.}~\bibnamefont {Kafri}}, \ and\
  \bibinfo {author} {\bibfnamefont {D.}~\bibnamefont {Podolsky}},\ }\href
  {\doibase 10.1088/1742-5468/2012/10/L10001} {\bibfield  {journal} {\bibinfo
  {journal} {J. Stat. Mech.}\ }\textbf {\bibinfo {volume} {2012}},\ \bibinfo
  {pages} {L10001} (\bibinfo {year} {2012})}\BibitemShut {NoStop}%
\bibitem [{\citenamefont {Tiz{\'o}n-Escamilla}\ \emph
  {et~al.}(2017)\citenamefont {Tiz{\'o}n-Escamilla}, \citenamefont {Hurtado},\
  and\ \citenamefont {Garrido}}]{tizon-escamilla_order_2016}%
  \BibitemOpen
  \bibfield  {author} {\bibinfo {author} {\bibfnamefont {N.}~\bibnamefont
  {Tiz{\'o}n-Escamilla}}, \bibinfo {author} {\bibfnamefont {P.~I.}\
  \bibnamefont {Hurtado}}, \ and\ \bibinfo {author} {\bibfnamefont {P.~L.}\
  \bibnamefont {Garrido}},\ }\href
  {https://link.aps.org/doi/10.1103/PhysRevE.95.032119} {\bibfield  {journal}
  {\bibinfo  {journal} {Phys. Rev. E}\ }\textbf {\bibinfo {volume} {95}},\
  \bibinfo {pages} {032119} (\bibinfo {year} {2017})}\BibitemShut {NoStop}%
\bibitem [{\citenamefont {Shpielberg}()}]{shpielberg_geometrical_2017}%
  \BibitemOpen
  \bibfield  {author} {\bibinfo {author} {\bibfnamefont {O.}~\bibnamefont
  {Shpielberg}},\ }\href@noop {} {\ }\Eprint {http://arxiv.org/abs/1706.04126}
  {arXiv:1706.04126 [cond-mat.stat-mech]} \BibitemShut {NoStop}%
\bibitem [{\citenamefont {Hurtado}\ and\ \citenamefont
  {Garrido}(2011{\natexlab{b}})}]{Hurtado2011}%
  \BibitemOpen
  \bibfield  {author} {\bibinfo {author} {\bibfnamefont {P.~I.}\ \bibnamefont
  {Hurtado}}\ and\ \bibinfo {author} {\bibfnamefont {P.~L.}\ \bibnamefont
  {Garrido}},\ }\href {http://link.aps.org/doi/10.1103/PhysRevLett.107.180601}
  {\bibfield  {journal} {\bibinfo  {journal} {Phys. Rev. Lett.}\ }\textbf
  {\bibinfo {volume} {107}},\ \bibinfo {pages} {180601} (\bibinfo {year}
  {2011}{\natexlab{b}})}\BibitemShut {NoStop}%
\bibitem [{\citenamefont {Espigares}\ \emph {et~al.}(2013)\citenamefont
  {Espigares}, \citenamefont {Garrido},\ and\ \citenamefont
  {Hurtado}}]{Espigares:2013dx}%
  \BibitemOpen
  \bibfield  {author} {\bibinfo {author} {\bibfnamefont {C.~P.}\ \bibnamefont
  {Espigares}}, \bibinfo {author} {\bibfnamefont {P.~L.}\ \bibnamefont
  {Garrido}}, \ and\ \bibinfo {author} {\bibfnamefont {P.~I.}\ \bibnamefont
  {Hurtado}},\ }\href {\doibase 10.1103/PhysRevE.87.032115} {\bibfield
  {journal} {\bibinfo  {journal} {Phys. Rev. E}\ }\textbf {\bibinfo {volume}
  {87}},\ \bibinfo {pages} {032115} (\bibinfo {year} {2013})}\BibitemShut
  {NoStop}%
\bibitem [{\citenamefont {Baek}\ \emph {et~al.}(2017)\citenamefont {Baek},
  \citenamefont {Kafri},\ and\ \citenamefont {Lecomte}}]{baek_dynamical_2016}%
  \BibitemOpen
  \bibfield  {author} {\bibinfo {author} {\bibfnamefont {Y.}~\bibnamefont
  {Baek}}, \bibinfo {author} {\bibfnamefont {Y.}~\bibnamefont {Kafri}}, \ and\
  \bibinfo {author} {\bibfnamefont {V.}~\bibnamefont {Lecomte}},\ }\href
  {\doibase 10.1103/PhysRevLett.118.030604} {\bibfield  {journal} {\bibinfo
  {journal} {Phys. Rev. Lett.}\ }\textbf {\bibinfo {volume} {118}},\ \bibinfo
  {pages} {030604} (\bibinfo {year} {2017})}\BibitemShut {NoStop}%
\bibitem [{\citenamefont {Katz}\ \emph {et~al.}(1983)\citenamefont {Katz},
  \citenamefont {Lebowitz},\ and\ \citenamefont
  {Spohn}}]{katz_nonequilibrium_1983}%
  \BibitemOpen
  \bibfield  {author} {\bibinfo {author} {\bibfnamefont {S.}~\bibnamefont
  {Katz}}, \bibinfo {author} {\bibfnamefont {J.~L.}\ \bibnamefont {Lebowitz}},
  \ and\ \bibinfo {author} {\bibfnamefont {H.}~\bibnamefont {Spohn}},\ }\href
  {\doibase 10.1007/BF01018556} {\bibfield  {journal} {\bibinfo  {journal} {J.
  Stat. Phys.}\ }\textbf {\bibinfo {volume} {34}},\ \bibinfo {pages} {497}
  (\bibinfo {year} {1983})}\BibitemShut {NoStop}%
\bibitem [{\citenamefont {Adam}\ \emph {et~al.}(2007)\citenamefont {Adam},
  \citenamefont {Hwang}, \citenamefont {Galitski},\ and\ \citenamefont
  {Das~Sarma}}]{Adam2007}%
  \BibitemOpen
  \bibfield  {author} {\bibinfo {author} {\bibfnamefont {S.}~\bibnamefont
  {Adam}}, \bibinfo {author} {\bibfnamefont {E.~H.}\ \bibnamefont {Hwang}},
  \bibinfo {author} {\bibfnamefont {V.~M.}\ \bibnamefont {Galitski}}, \ and\
  \bibinfo {author} {\bibfnamefont {S.}~\bibnamefont {Das~Sarma}},\ }\href
  {http://www.pnas.org/content/104/47/18392.full} {\bibfield  {journal}
  {\bibinfo  {journal} {Proc. Natl. Acad. Sci.}\ }\textbf {\bibinfo {volume}
  {104}},\ \bibinfo {pages} {18392} (\bibinfo {year} {2007})}\BibitemShut
  {NoStop}%
\bibitem [{\citenamefont {Tan}\ \emph {et~al.}(2007)\citenamefont {Tan},
  \citenamefont {Zhang}, \citenamefont {Bolotin}, \citenamefont {Zhao},
  \citenamefont {Adam}, \citenamefont {Hwang}, \citenamefont {Das~Sarma},
  \citenamefont {Stormer},\ and\ \citenamefont {Kim}}]{Tan2007}%
  \BibitemOpen
  \bibfield  {author} {\bibinfo {author} {\bibfnamefont {Y.~W.}\ \bibnamefont
  {Tan}}, \bibinfo {author} {\bibfnamefont {Y.}~\bibnamefont {Zhang}}, \bibinfo
  {author} {\bibfnamefont {K.}~\bibnamefont {Bolotin}}, \bibinfo {author}
  {\bibfnamefont {Y.}~\bibnamefont {Zhao}}, \bibinfo {author} {\bibfnamefont
  {S.}~\bibnamefont {Adam}}, \bibinfo {author} {\bibfnamefont {E.~H.}\
  \bibnamefont {Hwang}}, \bibinfo {author} {\bibfnamefont {S.}~\bibnamefont
  {Das~Sarma}}, \bibinfo {author} {\bibfnamefont {H.~L.}\ \bibnamefont
  {Stormer}}, \ and\ \bibinfo {author} {\bibfnamefont {P.}~\bibnamefont
  {Kim}},\ }\href {http://link.aps.org/doi/10.1103/PhysRevLett.99.246803}
  {\bibfield  {journal} {\bibinfo  {journal} {Phys. Rev. Lett.}\ }\textbf
  {\bibinfo {volume} {99}},\ \bibinfo {pages} {246803} (\bibinfo {year}
  {2007})}\BibitemShut {NoStop}%
\bibitem [{\citenamefont {Chen}\ \emph {et~al.}(2008)\citenamefont {Chen},
  \citenamefont {Jang}, \citenamefont {Adam}, \citenamefont {Fuhrer},
  \citenamefont {Williams},\ and\ \citenamefont {Ishigami}}]{Chen2008}%
  \BibitemOpen
  \bibfield  {author} {\bibinfo {author} {\bibfnamefont {J.~H.}\ \bibnamefont
  {Chen}}, \bibinfo {author} {\bibfnamefont {C.}~\bibnamefont {Jang}}, \bibinfo
  {author} {\bibfnamefont {S.}~\bibnamefont {Adam}}, \bibinfo {author}
  {\bibfnamefont {M.~S.}\ \bibnamefont {Fuhrer}}, \bibinfo {author}
  {\bibfnamefont {E.~D.}\ \bibnamefont {Williams}}, \ and\ \bibinfo {author}
  {\bibfnamefont {M.}~\bibnamefont {Ishigami}},\ }\href
  {http://www.nature.com/doifinder/10.1038/nphys935} {\bibfield  {journal}
  {\bibinfo  {journal} {Nat. Phys.}\ }\textbf {\bibinfo {volume} {4}},\
  \bibinfo {pages} {377} (\bibinfo {year} {2008})}\BibitemShut {NoStop}%
\bibitem [{\citenamefont {Martin}\ \emph {et~al.}(1973)\citenamefont {Martin},
  \citenamefont {Siggia},\ and\ \citenamefont
  {Rose}}]{martin_statistical_1973}%
  \BibitemOpen
  \bibfield  {author} {\bibinfo {author} {\bibfnamefont {P.~C.}\ \bibnamefont
  {Martin}}, \bibinfo {author} {\bibfnamefont {E.~D.}\ \bibnamefont {Siggia}},
  \ and\ \bibinfo {author} {\bibfnamefont {H.~A.}\ \bibnamefont {Rose}},\
  }\href {\doibase 10.1103/PhysRevA.8.423} {\bibfield  {journal} {\bibinfo
  {journal} {Phys. Rev. A}\ }\textbf {\bibinfo {volume} {8}},\ \bibinfo {pages}
  {423} (\bibinfo {year} {1973})}\BibitemShut {NoStop}%
\bibitem [{\citenamefont {{de Dominicis}}(1976)}]{dominicis_technics_1976}%
  \BibitemOpen
  \bibfield  {author} {\bibinfo {author} {\bibfnamefont {C.}~\bibnamefont {{de
  Dominicis}}},\ }\href
  {http://inis.iaea.org/Search/search.aspx?orig_q=RN:7266465} {\bibfield
  {journal} {\bibinfo  {journal} {J. Phys. Colloques}\ }\textbf {\bibinfo
  {volume} {37}},\ \bibinfo {pages} {247} (\bibinfo {year} {1976})}\BibitemShut
  {NoStop}%
\bibitem [{\citenamefont {Janssen}(1976)}]{janssen_lagrangean_1976}%
  \BibitemOpen
  \bibfield  {author} {\bibinfo {author} {\bibfnamefont {H.-K.}\ \bibnamefont
  {Janssen}},\ }\href {\doibase 10.1007/BF01316547} {\bibfield  {journal}
  {\bibinfo  {journal} {Z. Physik. B}\ }\textbf {\bibinfo {volume} {23}},\
  \bibinfo {pages} {377} (\bibinfo {year} {1976})}\BibitemShut {NoStop}%
\bibitem [{\citenamefont {Tailleur}\ \emph {et~al.}(2008)\citenamefont
  {Tailleur}, \citenamefont {Kurchan},\ and\ \citenamefont
  {Lecomte}}]{Tailleur2008}%
  \BibitemOpen
  \bibfield  {author} {\bibinfo {author} {\bibfnamefont {J.}~\bibnamefont
  {Tailleur}}, \bibinfo {author} {\bibfnamefont {J.}~\bibnamefont {Kurchan}}, \
  and\ \bibinfo {author} {\bibfnamefont {V.}~\bibnamefont {Lecomte}},\ }\href
  {http://stacks.iop.org/1751-8121/41/i=50/a=505001} {\bibfield  {journal}
  {\bibinfo  {journal} {J. Phys. A}\ }\textbf {\bibinfo {volume} {41}},\
  \bibinfo {pages} {505001} (\bibinfo {year} {2008})}\BibitemShut {NoStop}%
\bibitem [{\citenamefont {Gallavotti}\ and\ \citenamefont
  {Cohen}(1995{\natexlab{a}})}]{Gallavotti1995a}%
  \BibitemOpen
  \bibfield  {author} {\bibinfo {author} {\bibfnamefont {G.}~\bibnamefont
  {Gallavotti}}\ and\ \bibinfo {author} {\bibfnamefont {E.~G.~D.}\ \bibnamefont
  {Cohen}},\ }\href {http://link.aps.org/doi/10.1103/PhysRevLett.74.2694}
  {\bibfield  {journal} {\bibinfo  {journal} {Phys. Rev. Lett.}\ }\textbf
  {\bibinfo {volume} {74}},\ \bibinfo {pages} {2694} (\bibinfo {year}
  {1995}{\natexlab{a}})}\BibitemShut {NoStop}%
\bibitem [{\citenamefont {Gallavotti}\ and\ \citenamefont
  {Cohen}(1995{\natexlab{b}})}]{Gallavotti1995b}%
  \BibitemOpen
  \bibfield  {author} {\bibinfo {author} {\bibfnamefont {G.}~\bibnamefont
  {Gallavotti}}\ and\ \bibinfo {author} {\bibfnamefont {E.~G.~D.}\ \bibnamefont
  {Cohen}},\ }\href {http://dx.doi.org/10.1007/BF02179860} {\bibfield
  {journal} {\bibinfo  {journal} {J. Stat. Phys.}\ }\textbf {\bibinfo {volume}
  {80}},\ \bibinfo {pages} {931} (\bibinfo {year}
  {1995}{\natexlab{b}})}\BibitemShut {NoStop}%
\bibitem [{\citenamefont {Lebowitz}\ and\ \citenamefont
  {Spohn}(1999)}]{lebowitz_gallavotticohen-type_1999}%
  \BibitemOpen
  \bibfield  {author} {\bibinfo {author} {\bibfnamefont {J.~L.}\ \bibnamefont
  {Lebowitz}}\ and\ \bibinfo {author} {\bibfnamefont {H.}~\bibnamefont
  {Spohn}},\ }\href {\doibase 10.1023/A%3A1004589714161} {\bibfield  {journal}
  {\bibinfo  {journal} {J. Stat. Phys.}\ }\textbf {\bibinfo {volume} {95}},\
  \bibinfo {pages} {333} (\bibinfo {year} {1999})}\BibitemShut {NoStop}%
\bibitem [{\citenamefont {Huang}(1987)}]{huang_statistical_1987}%
  \BibitemOpen
  \bibfield  {author} {\bibinfo {author} {\bibfnamefont {K.}~\bibnamefont
  {Huang}},\ }\href@noop {} {\emph {\bibinfo {title} {Statistical
  mechanics}}},\ \bibinfo {edition} {2nd}\ ed.\ (\bibinfo  {publisher}
  {Wiley},\ \bibinfo {address} {New York},\ \bibinfo {year} {1987})\BibitemShut
  {NoStop}%
\bibitem [{\citenamefont {Bodineau}\ \emph
  {et~al.}(2008{\natexlab{b}})\citenamefont {Bodineau}, \citenamefont
  {Derrida}, \citenamefont {Lecomte},\ and\ \citenamefont
  {Wijland}}]{bodineau_long_2008}%
  \BibitemOpen
  \bibfield  {author} {\bibinfo {author} {\bibfnamefont {T.}~\bibnamefont
  {Bodineau}}, \bibinfo {author} {\bibfnamefont {B.}~\bibnamefont {Derrida}},
  \bibinfo {author} {\bibfnamefont {V.}~\bibnamefont {Lecomte}}, \ and\
  \bibinfo {author} {\bibfnamefont {F.}~\bibnamefont {Wijland}},\ }\href
  {\doibase 10.1007/s10955-008-9647-3} {\bibfield  {journal} {\bibinfo
  {journal} {J. Stat. Phys.}\ }\textbf {\bibinfo {volume} {133}},\ \bibinfo
  {pages} {1013} (\bibinfo {year} {2008}{\natexlab{b}})}\BibitemShut {NoStop}%
\end{thebibliography}%

\end{document}